\documentclass[12pt,draftcls,onecolumn]{IEEEtran}

\IEEEoverridecommandlockouts

\usepackage{algorithm,algorithmic,nicefrac}
\usepackage{amsmath,amsfonts,amssymb}
\usepackage{txfonts}
\usepackage{psfrag}
\usepackage{epsfig,graphicx,subfigure}
\usepackage{epstopdf}
\usepackage{stfloats}
\usepackage{epsf}
\usepackage{hyperref}
\usepackage{placeins}

\usepackage{tikz}
\usetikzlibrary{shapes.misc}
\usetikzlibrary{decorations.pathreplacing}

\newcommand{\T}{\mathcal{T}}
\newcommand{\V}{\mathcal{V}}
\newcommand{\E}{\mathcal{E}}
\newcommand{\I}{\mathcal{I}}

\title{\LARGE \bf
Limits of Friendship Networks in Predicting Epidemic Risk
}

\author{Lorenzo Coviello, Massimo Franceschetti, Manuel Garc\'ia-Herranz, Iyad Rahwan
\thanks{LC and IR: MIT Media Lab. MGH: UNICEF Innovation Unit. MF: University of California San Diego. Email:
        {\tt\small lorenzoc@mit.edu}}%
}

\date{}

\begin{document}

\maketitle

\begin{abstract}
The spread of an infection on a real-world social network is determined by the interplay of two processes --
the dynamics of the network, whose structure changes over time according to the encounters between individuals,
and the dynamics on the network, whose nodes can infect each other after an encounter.
Physical encounter is the most common vehicle for the spread of infectious diseases, but detailed information about said encounters is often unavailable because expensive, unpractical to collect or privacy sensitive.
The present work asks whether the friendship ties between the individuals in a social network successfully predict who is at risk.
Using a dataset from a popular online review service, we build a time-varying network that is a proxy of physical encounter between users and a static network based on their reported friendship.
Through computer simulations, we compare infection processes on the resulting networks and show that,
whereas distance on the friendship network is correlated to epidemic risk, friendship provides a poor identification of the individuals at risk if the infection is driven by physical encounter.
Our analyses suggest that such limit is not due to the randomness of the infection process, but to the structural differences of the two networks.
In addition, we argue that our results are not driven by the static nature of the friendship network as opposed to the time-varying nature of the encounter network,
as a static version of the encounter network provides more accurate prediction of risk than the friendship network.
In contrast to the macroscopic similarity between processes spreading on different networks -- confirmed by our simulations, the differences in local connectivity determined by the two definitions of edges result in striking differences between the dynamics at a microscopic level, preventing the identification of the nodes at risk.
Despite the limits highlighted by our analyses, we show that periodical and relatively infrequent monitoring of the real infection on the encounter network allows to correct the predicted infection on the friendship network and to achieve satisfactory prediction accuracy.
In addition, the friendship network contains valuable information to effectively contain epidemic outbreaks when a limited budget is available for immunization.
\end{abstract}

\tableofcontents

\section{Introduction}

The forecast and mitigation of epidemics is a central theme in public health~\cite{colizza2007predictability,ferguson2005strategies,ferguson2006strategies,ferguson2003planning,halloran2002containing,hufnagel2004forecast,longini2005containing},
and events such as the recent ebola epidemic constantly drive the attention and resources of governments, institutions such as the World Health Organization, and the research community~\cite{gomes2014assessing,halloran2014ebola,kupferschmidt2014estimating,lofgren2014opinion,merler2015spatiotemporal,poletto2014assessing}.
The study of infectious processes on real-world networks is of interests to diverse disciplines, and similar models have been proposed to characterize the spread of information, behaviors, cultural norms, innovation, as well as the diffusion of computer viruses~\cite{goffman1964generalization,lloyd2001viruses,pastor2001epidemic,rapoport1953spread,valente1995network}.
Therefore, epidemiologists, computer scientists and social scientist have joint forces in the study of contagion phenomena.
Due to the impossibility to study the spread of infectious diseases through controlled experiments, modeling efforts have prevailed
~\cite{heesterbeek2000mathematical,keeling2008modeling, koopman2004modeling,pastor2014epidemic,pastor2001epidemic}.
Recently, advancements in computation tools determined the emergence of data-driven simulations in the study of epidemic outbreaks and dynamical processes in general~\cite{vespignani2012modelling}.

The spread of an infection over a real-world network is determined by the interplay of two processes:
the dynamics \emph{of} the network, whose edges change over time according to the encounters between individuals;
and the dynamics \emph{on} the network, whose nodes can infect each other after they encounter.
When the two dynamics operate at comparable time scales, their interdependence appears particularly relevant,
the time-varying nature of the network cannot be ignored~\cite{gross2008adaptive,holme2014information,kivela2012multiscale,perra2012random,sayama2013modeling} and specifically devised control strategies are necessary~\cite{liu2014controlling}.
Aggregating the dynamics of the edges into a static version of the network can provide useful insights~\cite{eubank2004modelling} but it can introduce bias~\cite{hoffmann2012generalized,perra2012random}.
Empirical work suggests that the bursty activity patterns of individuals slow down spreading~\cite{karsai2011small, stehle2011simulation, vazquez2007impact}, but temporal correlations seem to accelerate the early phase of an epidemic~\cite{jo2014analytically,rocha2011simulated}

Physical encounter is the most common vehicle for the spread of infectious diseases (as in the case of airborne diseases), and detailed information about said encounters is fundamental for monitoring and containing outbreaks. 
Various sources of data can serve as a proxy of physical encounter -- 
checkins on social networking platforms~\cite{cheng2011exploring,noulas2012tale,noulas2011empirical},
traffic records~\cite{balcan2009multiscale,sun2014efficient,sun2013understanding},
phone call records~\cite{gonzalez2008understanding,hidalgo2008dynamics,onnela2007structure},
wifi and RFID wearable sensors data~\cite{cattuto2010dynamics,heibeck2010honest,hui2005pocket,o2006instrumenting,salathe2010high,stehle2011simulation},
geographical and non-geographical information shared online~\cite{backstrom2010find,cheng2010you}, surveys and diaries of daily contact~\cite{edmunds1997mixes,mikolajczyk2008social,mossong2008social}, and recently multiplex data~\cite{stopczynski2014measuring}.

However, pervasive and detailed information is rarely available and might be expensive and unpractical to collect (as in the case of sensor technologies~\cite{cattuto2010dynamics,salathe2010high,stehle2011simulation}), prone to errors (as in the case of survey data~\cite{eagle2009inferring,read2008dynamic}), and privacy-sensitive~\cite{altshuler2013stealing,barkhuus2012mismeasurement,barkhuus2003location,de2013unique,klasnja2009exploring,lederer2003wants,shokri2011quantifying,zhou2008preserving}.
In general, researchers have to rely on the information in their possession, and in this work we consider the case of self-reported relationships between individuals, such as friendship between the users of an online social network.
Recent research has shown that communication records and social ties are useful to explain and predict human dynamics.
Location data from cell phone records and online social networks has shown that social relationships can partially explain the patterns of human mobility~\cite{cho2011friendship}.
Contact tracing based on phone communication activity has been proposed as a proxy of physical interaction in the context of a method to reduce the final size of epidemic outbreaks~\cite{farrahi2014epidemic}.
Both real-word social relationships (e.g., family, professional, friendship ties) and online social relationships (e.g. Facebook friendship, follower-followee relationships on Twitter) predict the diffusion of behaviors~\cite{aral2011creating, aral2012identifying,centola2010spread, christakis2007spread, christakis2008collective}.
At a structural level, there is evidence that networks generated from wearable sensor measurements, diaries of daily contacts, online links and self-reported friendship present similar structural properties~\cite{mastrandrea2015contact}, but contacts recorded by wearable sensors might not be reported in surveys, especially when the contact's duration is short~\cite{smieszek2014should}.
In general, it is not clear whether and within which limits friendship can be considered a valid proxy of physical encounter,
as a process spreading from an initial seed, or ``patient zero'', can reach only the nodes in its \emph{set of influence} through paths that respect time ordering~\cite{holme2005network}.

Given an infection transmitted by physical encounter on a social network, the present work asks whether the friendship ties between the same individuals successfully predict who is at risk. 

Using the Yelp Dataset Challenge dataset (\url{www.yelp.com/dataset_challenge}), we build a time-varying network that is a proxy of physical encounter between users and a static network based on their reported friendship. We refer to these networks as the encounter network and the friendship network, respectively.
Through computer simulations, we compare the evolution of Susceptible-Infected (SI) processes~\cite{anderson1991infectious} on the two networks, in terms of the sets of infected individuals.
Given a seed, is the set of nodes infected on the friendship network a good approximation of those infected in the encounter network?


%

Our contribution is twofold.
First, we propose similarity measures to quantify how precisely the set of individuals predicted to be at risk according to a given spreading model (e.g., friendship) approximates the set of individuals at risk according to a different underlying spreading model (e.g., physical encounter).
Given a target infection size and a seed present in both networks, we separately simulate infections starting at that seed in both networks and compare the sets of infected nodes.
The proposed measures allows disentangling between the randomness of the infection process and the effect of the structural differences between the networks. 
Given this measure, we show that despite friendship networks produce similar epidemic dynamics at the macro level, friendship provides a poor identification of the individuals at risk if the infection is driven by physical encounter.
That is, the sets of individuals infected on the friendship network are in general very different from the corresponding ones on the encounter network.
This is true even after controlling for the fact that certain individuals might be connected in one network and not in the other.
Our analyses suggest that such difference is primarily determined by the structural differences of the two networks, and due only in part to the randomness of the infection process.
Despite the randomness of the infection increases the unpredictability of the set of infected individuals (between independent processes initiated at the same seed on the same network), topological characteristics amplify such unpredictability when considering the two different networks.
In addition, our results are not driven by the static nature of the friendship network as opposed to the time-varying nature of the encounter network, 
as a static version of the encounter network provides more accurate prediction of risk than the friendship network.
Also, similar conclusions hold if we compare the encounter network to a time-varying version of the friendship network.
The limits of the friendship network in predicting epidemic risk are not simply due to the time ordering of the influence sets determined by physical encounter.

Despite the limits highlighted by our analyses, we show that periodical and relatively infrequent monitoring of the real infection on the encounter network allows to correct the predicted infection on the friendship network and to achieve satisfactory prediction accuracy.
This corresponds to a less extreme scenario in which the researcher has still knowledge of the friendship network, but, in addition, is able to monitor the infected population (on the encounter network) at given times.
In particular, we compare the sets of infected individuals on the two networks right before each correction and show that a good level of prediction accuracy is established early in the process and maintained over time.
Our results suggest that the ability to periodically monitor the infection on the encounter network is the key to overcome the limits of the friendship network in predicting epidemic risk.

In addition, we show that the friendship network encodes useful information for the containment of epidemic outbreaks.
We consider scenarios in which a fixed budget is available for immunization (e.g., limited amount of vaccine) and must be effectively allocated in order to contain the epidemic.
In contrast to the simple method of purely random immunization, we consider a strategy that selects random friends of randomly chosen individuals for immunization, a method already proposed to predict the peak of an epidemic outbreak~\cite{christakis2010social} and the spread of information online~\cite{garcia2014using}.
This strategy is motivated by the ``friendship paradox'', the network property for which the average friend of an individual is more connected than the average individual~\cite{feld1991your}, and is simple, only requiring individuals to name a friend.
The strategy allows effective use of the immunization budget, substantially increasing the probability that an infection dies out in its early stage, and strongly reducing the expected final infection size with respect to purely random immunization.
Moreover, it only requires a small additional cost to obtain the same effect of an ideal strategy that administers immunization to future encounters rather than friends.

Since seminal work on the structure and growth of complex networks~\cite{barabasi1999emergence,faloutsos1999power,watts1998collective},
interdisciplinary research has shown that biological networks, social networks and the Internet are governed by similar rules
~\cite{albert2002statistical,boccaletti2006complex,jeong2000large,newman2003structure}, and share similar structure~\cite{girvan2002community,newman2004finding,palla2005uncovering}.
In particular, very similar models have been proposed to characterize the spread of epidemics, information, behaviors, and cultural norms.
Despite the macroscopic similarity between processes spreading on different networks (confirmed by our simulations), our work shows that the differences in local connectivity determined by the two definitions of edges result in striking differences between the dynamics at a microscopic level, which prevent the identification of the nodes at risk.

\subsection{Outline}
Section~\ref{yelp:sec:dataset} describes the dataset and introduces the friendship network and the encounter network, as well as a static version of the encounter network and a time-varying version of the friendship network that will be considered in the analyses.
Section~\ref{yelp:sec:infection} introduces the epidemic process, defines the metrics to measure its spread, and describes the sensor selection mechanisms considered in the analysis of the process at the macro level.
Section~\ref{yelp:sec:friend_distance_risk} shows that, given an infection initiated at a single seed and spreading on the encounter network, nodes at shorter distance from the seed on the friendship network have higher risk of infection.
Section~\ref{yelp:sec:encounter_friendship_statics} compares processes initiated at the same seed but spreading separately on the encounter network, on the friendship network and on a static version of the encounter network, and highlights the limits of the friendship network in predicting epidemic risk (if the epidemic spreads via physical encounter).
To further support that our results are not driven by the static nature of the friendship network as opposed to the time-varying nature of the encounter network, Section~\ref{yelp:sec:time_population} compares epidemic risk on the encounter network and on the time-varying version of the friendship network,
whereas \ref{yelp:sec:static_population} compares epidemic risk on the friendship network and on the static version of the encounter network.
Section~\ref{yelp:sec:window} considers the situation in which the estimated set of infected individuals on the friendship network can be periodically corrected to match the set of infected individuals on the encounter network, and shows that even relatively infrequent correction overcomes the limits of the friendship networks in predicting epidemic risk.
Section~\ref{yelp:sec:immunization} considers the problem of effectively containing epidemic outbreaks when a limited budget is available for immunization, and shows that the friendship network contains valuable information to effectively allocate this budget.
Section~\ref{yelp:sec:time_infection} and Section~\ref{yelp:sec:static_infection} provide a characterization of the epidemics at a macroscopic level.
In particular, Section~\ref{yelp:sec:time_infection} consider processes spreading on the encounter network and on the time-varying version of the friendship network, and Section~\ref{yelp:sec:static_infection} consider processes spreading on the friendship network and on the static version of the encounter network.
We conclude in Section~\ref{yelp:sec:conclusion}.

\section{The friendship network and the encounter network}
\label{yelp:sec:dataset}
The Yelp Dataset Challenge dataset (\url{www.yelp.com/dataset_challenge})
consists in $1,569,264$ reviews and $495,107$ tips to $61,184$ businesses (in $10$ cities around the world) posted by $366,715$ users over a period spanning over than $10$ years.
Within this period, we consider $1,469$ consecutive days ranging from 1/1/2011 to 1/8/2015, as reviews before 2011 are less numerous.
Each review and tip includes the user who posted it, the reviewed business, and the date it was posted.
Yelp users can form friendship ties between each other, and the list of friends of each user is included in the dataset.
Time information about the formation of friendship ties is not available.
Using the dataset, we define two networks, called the friendship network and the encounter network respectively.

Let $U$ be the set of users, $F\subseteq U\times U$ be the set of friendship ties, $B$ the set of businesses, $T$ be the set of days, $R\subseteq U\times B\times T$ be the set of reviews and tips (which we will refer to as reviews).
For each user $u\in U$ let $F_u\subset U$ be the set of friends of $u$.
Therefore $F=\cup_{u\in U}\{(u,v):v\in F_u\}$.
Each review (or tip) $r\in R$ is a triple $(u,b,t)$ where $u\in U, b\in B, t\in T$.

\subsection{The friendship network}
\label{yelp:sec:friendship_network}
\begin{figure}
\centering
\includegraphics[scale=0.35]{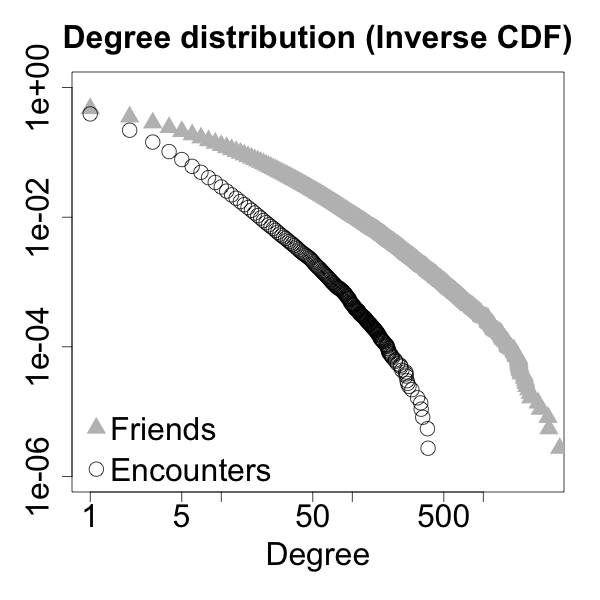}
\caption{Inverse Cumulative Distribution Function of friend degree (grey triangles) and encounter degree (white circles). The friend degree of a user is defined as the her number of friendship ties. The encounter degree of a user is defined as her number of encounters during all period of observation.}
\label{fig:degree_distribution}
\end{figure}

Of all users, $174,100$ have at least one friend, with an average number of friends per user, or friend degree, $14.8$.
The friend degree distribution is shown in Figure~\ref{fig:degree_distribution} (triangles).

Let $N_F=(U,F)$ be the static friendship network.
As we consider processes spreading between connected nodes, connectedness is the key property of the networks.
Therefore, we restrict our attention to the giant component, as users outside giant components form small components whose dynamics are not relevant.
The giant component defined by friendship includes $168,923$ users (whereas the second largest component has $8$ users).
In what follows, we will identify $N_F$ with its giant component.
Observe that this network is static, as its edges do not change over time. 

\subsection{The encounter network}
\label{yelp:sec:encounter_network}

The most common vehicle for the spread of infectious diseases is physical contact (rather than friendship) between individuals.
Strictly speaking, two users in $U$ encountered on a given day $t$ if they visit the same business on day $t$ at the same time.
In the present work, we use reviews as a proxy of physical encounter: an edge is active between two users in $U$ on day $t$ if they posted a review to the same business on day $t$.
This constitutes an approximation to real physical encounter, which requires users to \emph{visit} (rather than review) a business at about the same time.
This approximation is justified as the time of a review is a proxy of the time of the visit to a business, and the element that spreads over a network (e.g., a virus or an opinion) does not necessarily require direct physical contact.
For example, in the case of airborne transmission, particles can remain suspended in the air for hours after an infected individuals has occupied a room~\cite{brankston2007transmission}.
In the context of our dataset, after an infected user visits a business, the infection might spread to customers who visit the business later in the day.
Also, the virus can infect customers which are not included in the dataset, and from them can infect another user who visits the business in a later moment.

\begin{figure}
\centering
\includegraphics[scale=0.50]{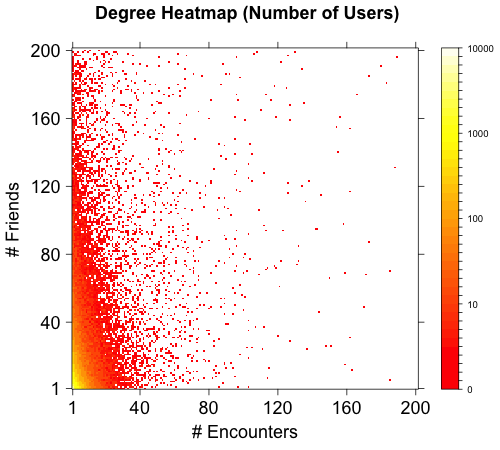}
\caption{Heat map of friend degree and encounter degree of all users with at least one friend and one encounter (friend degree and encounter degree are limited to $200$ in the plot).}
\label{fig:degree_heatmap}
\end{figure}

In the dataset, $143,780$ users have at least one encounter, with an average number of encounters, or encounter degree, of $3.9$.
The distribution is shown in Figure~\ref{fig:degree_distribution} (circles).
Figure~\ref{fig:degree_heatmap} shows a heat map of friend degree and encounter degree of users.
Despite friend degree and encounter degree are correlated (Pearson product-moment correlation $0.3416$, p-value $< 2.2\cdot10^{-16}$), the similarity of the sets of the friends and encounters of an individual is low.
Considering the $72,786$ users with at least one friend and one encounter, the average Jaccard similarity of their encounter and friend sets is $0.01716$, with only $9,527$ of them with a value different than zero.
Despite epidemic processes spreading on the friendship and on the encounter network evolve in a qualitatively similar way, the differences in local connectivity determined by the two definitions of edges result in very different sets of nodes predicted to be at risk.

For each $t\in T$, $U(t)=\{u\in U: (u,b,t)\in R \text{ for some } b\in B\}$ is the set of users who wrote a review on day $t$.
We refer to $U(t)$ as the active users on day $t$.

For each $t\in T$ and $u\in U(t)$, $E_u(t)=\{v\in U(t), v\neq u: (u,b,t)\in R \text{ and } (v,b,t)\in R \text{ for some } b\in B\} \subseteq U$ is the set of encounters of user $u$ on day $t$ (i.e., users who visited at least one of the businesses visited by $u$).
$E(t)=\cup_{u\in U}\{(u,v):v\in E_u(t)\} \subseteq U\times U$ is the set of encounters on day $t$.

For each $t\in T$, let $N_E(t)=(U,E(t))$ be the network defined by the encounters on day $t$.
Observe that the node set in the definition is $U$ rather than $U(t)$.
The \emph{encounter network} is the sequence $\{N_E(t)\}_{t\in T}$.
As connectedness is the key property in a spreading process, we consider the $133,038$ users who had at least one encounter during $T$.

\subsection{The static encounter network and the time-varying friendship network}
\label{yelp:sec:other_networks}

To argue that our results are not driven by the static nature of the friendship network as opposed to the time-varying nature of the encounter network, we define a static version of the encounter network and a time-varying version of the friendship network and we will show that similar conclusions hold.

For each $t\in T$, $F(t)=\{(u,v)\in F: u,v\in U(t)\}$ is the set of friendship ties between active users on day $t$.
Observe that friendship ties are not associated to temporal information (i.e., the time in which the edge formed is unknown).
For each $t\in T$, let $N_F(t)=(U,F(t))$ be the friendship network between active users.
Observe that the node set in the definition is $U$ rather than $U(t)$.
The \emph{friendship time-varying network} is the sequence $\{N_F(t)\}_{t\in T}$.
We consider the $41,664$ users who, during $T$, had at least an active friend on a day in which they were active.

Let $E_u=\cup_{t\in T}E_u(t) \subseteq U$ be the set of encounters of $u$ during $T$, and $E= \cup_{t\in T} \cup_{u\in U}\{(u,v):v\in E(t)\} \subseteq U\times U$ be the set of encounters between users in $U$.
The \emph{static encounter network} is $N_E=(U,E)$.
We restrict our attention to the giant component of the static encounter network, which includes $113,187$ users (whereas the second largest component has $23$ users).

\section{Infection dynamics}
\label{yelp:sec:infection}

To model the spread of an infectious disease, we consider a Susceptible-Infected (SI) process~\cite{anderson1991infectious} , in which nodes never recover after being infected.
Here, we give a general definition of the process that applies to both the static and the time-varying networks defined above.
Given a set of nodes $\V$, a set of edges $\E\subseteq \V\times \V$ and a set of time indices $\T$, let $\{N(t)\}_{t\in \T}$ be a sequence of networks, where $N(t)=(\V,\E(t))$ with $\E(t)\subseteq \E$.
For a static network, $\E(t)=\E$ for all of $t\in\T$.

Let $\I(t)$ denote the set of infected nodes at time $t$, of cardinality $I(t)$.
The infection starts at time $t=0$ from a set $\I(0)$ of infected seeds.

Consider any $t>0$. 
The infection spreads from the set of already infected nodes $\I(t-1)$ as follows.
For each non-infected node $v\in \V\backslash\I(t-1)$, let $d_v(t)=\vert\{u\in \I(t-1): (u,v)\in \E(t) \vert\}$, that is, the number of neighbors of $v$ at time $t$ which are infected at time $t-1$.
Let $B(t)=\{v\in\V\backslash\I(t-1): d_v(t)>0\}$, that is, the set of susceptible nodes at time $t$.
Each node $v\in B(t)$ gets infected with probability $\beta d_v(t)$, where $\beta\in[0,1]$ is the rate of infection.

When $\beta=1$ the infection process is deterministic and, at time $t$, all non-infected neighbors of the nodes infected by time $t-1$ become infected.
For finite values of $\beta$, the infection spreads in a stochastic way.

For the time-varying networks defined above (i.e., the encounter network and the time-varying friendship network), $\T=T$.
The infection will propagate for $\vert T\vert$ time steps, resulting in an infected population $\I(\vert T\vert)$.
For static networks  (i.e., the friendship network and static encounter network), $\T = [0,\infty)$ and the infection propagates until $\I(t)=\V$ (i.e., until the entire population is infected).

\subsection{Infection time}
\label{yelp:sec:time}
Given a realization of the infection process, for each $\alpha\in[0,1]$ let 
$$
\tau(\alpha) = \min\{t:I(t)/\vert\V\vert\ge \alpha\}.
$$
$\tau(\alpha)$ is a random variable and represents the first time in which an $\alpha$-fraction of the nodes $\V$ are infected (once $\I(0)$ is fixed, $\tau(\alpha)$ is a degenerate random variable for $\beta=1$).
Given a realization of the SI process on a time-varying network, let $\tau(\alpha)=\infty$ for $\alpha>\I(\vert T\vert)/\vert\V\vert$.


We also consider the number, rather than the fraction, of infected nodes.
Given a realization of the infection process, for each $M\in[0,\vert\V\vert]$, let
$$
t(M) = \min\{t:I(t)\ge M\},
$$
The random variable $t(M)$ denotes the first time in which at least $M$ nodes are infected.
Given a realization of the SI process on a time-varying network, let $t(M)=\infty$ for $M>\I(\vert T\vert)$.

\subsection{Seed selection}
\label{yelp:sec:seed}
In a static network, seeds are chosen at random and without replacement.
In a time-varying network, the infection can start propagating at the first time $t$ in which there is an edge between an infected seed and a non-infected node, that is, at time
$$
t_0(\I(0)) = \min\{t:\exists (u,v)\in\E(t) \text{ for some } u\in\I(0), v\in\V\backslash\I(0)\}.
$$
As a remark, for $\beta<1$, it is possible that no node is infected at time $t_0$.
Seeds are selected uniformly at random and without replacement among all nodes $v\in \V$ such that $t_0(\{v\})\le 500$, that is, nodes that have a neighbor in the time-varying network by time $t=500$.

\subsection{Detection time with sensors}
\label{yelp:sec:sensors}
In real scenarios, it might be infeasible to monitor all nodes in the network.
Constraints of different nature (e.g., budget, physical, privacy) might limit the researchers to monitor a subset $S\subset\V$ of all nodes, referred to as sensors.
At each time $t$, let $\I_S(t)=\I(t)\cap S$ be the set of infected sensors, and $I_S(t)$ be its cardinality.
Assuming as before that the network and the set of seeds are given, for each $\alpha\in[0,1]$ let 
$$
\tau_S(\alpha) = \min\{t:I_S(t)/\vert S\vert\ge \alpha\}.
$$
That is, $\tau_S(\alpha)$ represents the first time in which an $\alpha$-fraction of the sensors $S$ are infected.
Given a realization of the SI process on a time-varying network, let $\tau_S(\alpha)=\infty$ for $\alpha>I_S(\vert T\vert)/\vert S\vert$.


We consider two types of sensor selection, \emph{random} sensors and \emph{friend} sensors, defined as follows.
Let $m$ be a fixed parameter.
A set $S$ of random sensors is obtained by selecting $m$ nodes from $\V$ uniformly at random and without replacement.
A set $S$ of friend sensors is obtained in two steps.
First, $S$ is initialized as the empty set, and a set $S_0$ of random nodes is obtained by selecting $m$ users from $\V$ uniformly at random and without replacement.
Then, for each node $u\in S_0$, a \emph{friend} $v\in\V$ is selected uniformly at random from $F_u$ (i.e., from the set of friends of $u$) and added to $S$. We require each friend sensor to be in $\V$ and to be friend of a node in $S_0$.
We remark that, even for encounter networks, friend sensors are selected on the basis of friendship rather than encounter.
We make this assumption because explicit relationships (such as friendship, family or professional ties) might be accessible or inferable in a real setting in which the researcher has to select a set of sensors.
Observe that, in the case of friend sensors, the size of the resulting set $S$ might be smaller than $m$.

Given the fact that, on average, people have fewer friends than their friends have (also know and the friendship paradox~\cite{feld1991your}), randomly sampled friends are more connected than randomly sampled individuals and are shown to provide earlier detection of phenomena spreading over complex networks~\cite{christakis2010social,garcia2014using}.

\FloatBarrier
\section{Friendship distance and epidemic risk}
\label{yelp:sec:friend_distance_risk}

In this section we show that distance on the friendship network is correlated to epidemic rick.
Given and infection initiated at a single seed and spreading on the encounter network, nodes at a shorter distance from the seed on the encounter network have a higher probability of becoming infected.
In the rest of the section, we always consider infections spreading on the encounter network and distance defined on the friendship network.

Given nodes $s$ and $s'$ in the friendship network, let $d(s,s')$ denote their distance (i.e., the length of the shortest path connecting them).
Given node $s$ and an integer $d>0$, let
$$
N_d(s) = \{s: d(s,s')=d\}
$$
be the set of nodes at distance $d$ from $s$, and let $n_d(s)$ be its cardinality.
$N_1(s)$ and $n_d(s)$ denote the set of neighbors and the degree of $s$, respectively. 

Let $i$ denote an infection process, and $s_i$ the selected seed.
Given an infection initiated at a seed $s_i$ until time $T$, let $\I(s_i)$ be the set of infected nodes at time $T$.
For each $d>0$ let
$$
\I_d(s_i) = \I(s_i) \cap N_d(s)
$$
be the set of infected nodes that are at distance $d$ from $s_i$ on the encounter network.
The infection rate of nodes at distance $d$ from $s_i$ is defined as
$$
r_d(s_i) = \frac{\vert \I_d(s_i)\vert}{n_d(s)}.
$$
The empirical average of $r_d(s_i)$ over $S$ simulations is given by
$$
\bar r_d = \frac{1}{S} \sum_{i=1}^{S}r_d(s_i),
$$
and represents the risk of becoming infected if the seed is at distance $d$.

\begin{figure}
\centering
\includegraphics[scale=0.35]{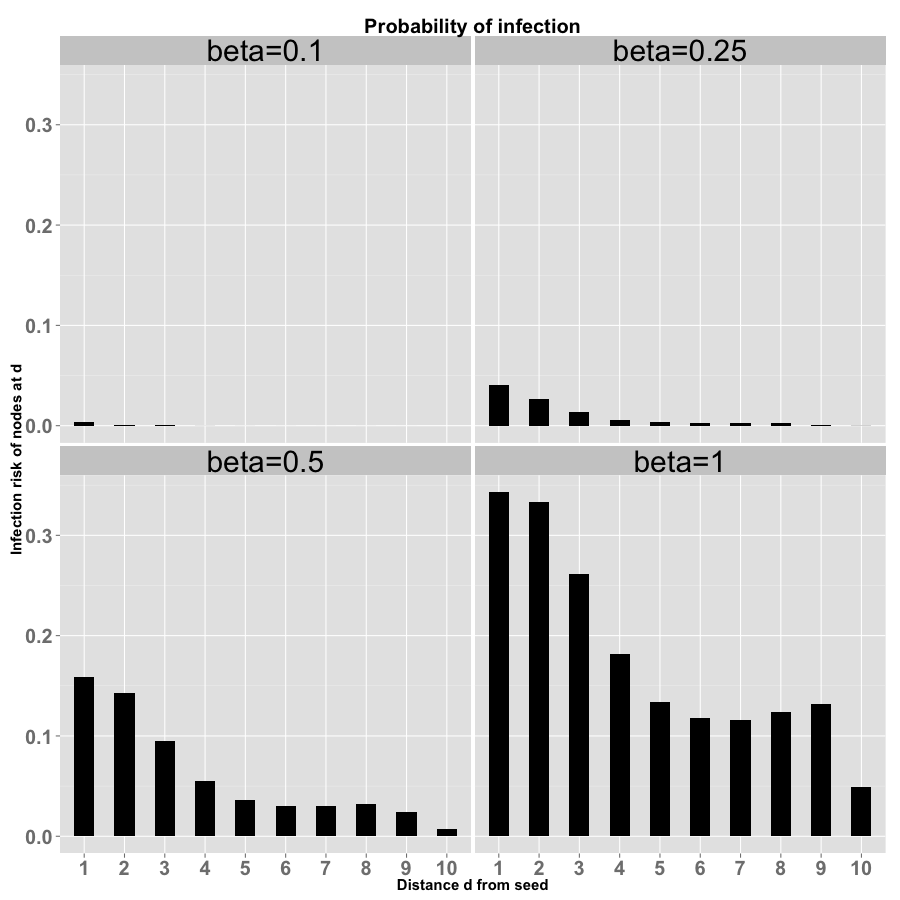}
\caption{\textbf{Epidemic risk with respect to distance on the friendship network.}
Empirical estimates of $\bar r_d(\beta)$ for $\beta\in\{0.1,0.25,0.5,1\}$ and $1\le d\le 10$.
For each value of the infection rate $\beta$, $10000$ simulations are run, each initiated at a random seed.}
\label{fig:friend_distance_risk}
\end{figure}

As the spreading of an infection process depends on the infection rate $\beta$, we write $\bar r_d(\beta)$ to compare infection processes with different infection rate.
Given a node $s$ in the encounter network, we recall that $t_0(\{s\})$ is the first time period in which $s$ has an edge (that is, the smallest $t$ such that $E_u(t)>0$).
As we consider infections spreading on the encounter network and distance on the friendship network, we consider seeds that are present in both networks.
In each simulation, a single seed is selected uniformly at random between all nodes $s\in \vert U_F\cap U_E \vert$ such that $t_0(\{s\})\le 500$ (as infections on time-varying networks spread for a limited number of time steps, we require them to start early enough).
For each $\beta\in\{0,0.1,0.25,0.5\}$ we run $10,000$ simulations.
The empirical estimates of $\bar r_d(\beta)$ for $1\le d\le 10$ are shown in Figure~\ref{fig:friend_distance_risk} and Table~\ref{tab:friend_distance_risk}.

\begin{table}
\centering
\caption{\textbf{Epidemic risk with respect to distance on the friendship network.}}
\begin{tabular}{|c|c|c|c|c|c|c|c|c|c|c|}
\hline
$\beta$ &
 $\bar r_1(\beta)$ & $\bar r_2(\beta)$ & $\bar r_3(\beta)$ & $\bar r_4(\beta)$ & $\bar r_5(\beta)$ &
 $\bar r_6(\beta)$ & $\bar r_7(\beta)$ & $\bar r_8(\beta)$ & $\bar r_9(\beta)$ & $\bar r_{10}(\beta)$ \\
\hline
0.10 & 3.9$\cdot 10^{-3}$ & 7.1$\cdot 10^{-4}$ & 2.1$\cdot 10^{-4}$ & 7.01$\cdot 10^{-05}$ & 3.2$\cdot 10^{-05}$ & 2.3$\cdot 10^{-05}$ & 1.3$\cdot 10^{-05}$ & 2.1$\cdot 10^{-05}$ & 1.1$\cdot 10^{-05}$ & 0 \\
0.25 & 0.041 & 0.027 & 0.014 & 0.006 & 0.003 & 0.003 & 0.002 & 0.003 & 0.001 & 1.6$\cdot 10^{-4}$ \\
0.50 & 0.159 & 0.143 & 0.095 & 0.055 & 0.036 & 0.031 & 0.030 & 0.032 & 0.025 & 0.007 \\
1.00 & 0.343 & 0.333 & 0.262 & 0.182 & 0.133 & 0.118 & 0.116 & 0.123 & 0.131 & 0.049 \\
\hline
\end{tabular}
\label{tab:friend_distance_risk}
\end{table}

\FloatBarrier
\section{The limits of the friendship network}
\label{yelp:sec:encounter_friendship_statics}

In this section, we consider SI processes on the the (time-varying) encounter network $\{N_E(t)\}_{t\in T}$, the static version of the encounter network $N_E$ and (static) friendship network $N_F$, initiated at single seeds (i.e., $\I(0)=\{s\}$).

As mentioned above, we identify the friendship network and the static encounter network with their giant components. We refer to the corresponding sets of nodes as $U_F$, with cardinality $n_F=\vert U_F\vert = 168,923$ users, and $U_E^{(s)}$, and $n_E^{(s)}=\vert U_E^{(s)}\vert = 113,187$.
Similarly, for the encounter network, we only consider users who had at least an encounter during the period of observation, that is, $u$ such that $E_u(t)>0$ for some $t$.
We refer to the set of these users as $U_E^{(t)}$, and $n_E=\vert U_E^{(t)}\vert = 133,038$.

Our objective is to compare the infection processes on the three different networks at a microscopic level, with the goal of evaluating both the friendship network and the static encounter network as predictors of epidemic risk on the (time-varying) encounter network.
In order to do that, we compare the sets of nodes that become infected on the three networks during independent infection processes starting at the same seed.
We therefore consider infection seeds that are present in all networks.
Given a node $s$ in the encounter network, we recall that $t_0(\{s\})$ is the first time period in which $s$ has an edge (that is, the smallest $t$ such that $E_u(t)>0$).
In each simulation, a single seed is selected uniformly at random between all nodes $s\in U_F\cap U_E^{(s)}\cap U_E^{(t)}$ such that $t_0(\{s\})\le 500$ (as infections on time-varying networks spread for a limited number of time steps, we require them to start early enough).

By considering both certain infection processes ($\beta=1$) and stochastic infection processes ($\beta<1$), we characterize how predictions of epidemic risk are affected by the structural differences between the networks, but their time-varying or static nature, and by the randomness of the infection processes.
To take into account the different edge density (and therefore the different speed of the infection process) on the encounter and the friendship network, we allow for different infection rates: $\beta_F$ on the friendship network, $\beta_{E^{(t)}}$ on the  encounter network, and $\beta_{E^{(s)}}$  on the static encounter network.
In Section~\ref{yelp:sec:encounter_friendship_statics_certain}, we consider the case of $\beta_F=\beta_{E^{(t)}}=\beta_{E^{(s)}}=1$ (certain infection), and show that the friendship network provides less accurate prediction of epidemic risk than the static encounter network.
In this case, given a seed, the differences between epidemic processes spreading on the three networks are solely determined by structural differences.
Our analyses suggest that the limits of the friendship network in predicting epidemic risk are not only due to its static nature as opposed to the time-varying nature of the encounter network, but also to the topological differences arising from the different semantic of the edges.
In Section~\ref{yelp:sec:encounter_friendship_statics_stochastic}, we set $beta_{E^{(t)}}=\beta_{E^{(s)}}=0.5$ and $\beta_F=0.01$ (stochastic infection), and show that also in this case the friendship network provides less accurate prediction of epidemic risk than the static encounter network.
Our analyses show that structural differences between friendship and encounter networks introduce more unpredictability than the randomness of the infection process.
Randomness introduces a certain amount of unpredictability in the spread of the infection, and two runs of the process on the same network starting from the same seed can result in different sets of infected nodes.
However, we observe that the unpredictability within a given network is substantially lower than the unpredictability between the two different networks.
Moreover, this unpredictability is not attributable only to the static nature of the friendship network as opposed to the time-varying nature of the encounter network, as the static version of the encounter network provides more accurate prediction of epidemic risk than the friendship network.
That is, the limits of the friendship network in predicting epidemic risk are primarily due to the structural differences between the two networks.

\subsection{Metrics}
\label{yelp:sec:encounter_friendship_statics_metrics}

Fixed a seed $s_i$, let $\I_{E^{(t)}_1}(t;s_i)$ and $\I_{E^{(t)}_2}(t;s_i)$ denote the set of infected nodes at time $t$ in two independent infection processes on the encounter network starting at $s_i$. $\I_{E^{(s)}_1}(t;s_i)$ and $\I_{E^{(s)}_2}(t;s_i)$ (resp. $\I_{F_1}(t;s_i)$ and $\I_{F_2}(t;s_i)$) are similarly defined by considering the static encounter (resp. friendship) network.
Let $I_{E^{(t)}_1}(t;s_i)$, $I_{E^{(t)}_2}(t;s_i)$, $I_{E^{(s)}_1}(t;s_i)$, $I_{E^{(s)}_2}(t;s_i)$, $I_{F_1}(t;s_i)$, $I_{F_2}(t;s_i)$ be their cardinality.
For $j=1,2$, let
$$
t_{E^{(t)}_j}(m;s_i) = \min\{t\in T: I_{E^{(t)}_j}(t;s_i)\ge m\}
$$
$$
t_{E^{(s)}_j}(m;s_i) = \min\{t\in T: I_{E^{(s)}_j}(t;s_i)\ge m\}
$$
$$
t_{F_j}(m;s_i) = \min\{t\in T: I_{F_j}(t;s_i)\ge m\}
$$
be the minimum time at which at least $m$ nodes are infected in the corresponding process.
$t_{E^{(t)}_j}(m;s_i)$ is undefined if $m$ nodes never get infected in the corresponding process (on the encounter network), and similar for the other processes.

If $t_{E^{(t)}_j}(m;s_i)$ is defined, then the corresponding infected set is 
$$
\I^*_{E^{(t)}_j}(m;s_i) = \I_{E^{(t)}_j}(t_{E^{(t)}_j}(m;s_i)).
$$ 
Instead, $t_{E^{(s)}_j}(m)$ and $t_{F_j}(m)$ are always defined on the static encounter network and on the friendship network (on which the infection process continues until the entire population is infected), and the corresponding infected sets are 
$$
\I^*_{E^{(s)}_j}(m;s_i) = \I_E^{(s)}(t_{E^{(s)}_j}(m;s_i)).
$$
$$
\I^*_{F_j}(m;s_i) = \I_F(t_{F_j}(m;s_i)).
$$

When the relevant values $\I^*_{E^{(t)}_j}(m;s_i)$, $\I^*_{E^{(s)}_j}(m;s_i)$ and $\I^*_{F_k}(m;s_i)$ for $j,k\in\{1,2\}$ are defined, we define the following measures of Jaccard similarity,
\begin{align*}
J_{E^{(t)}_1,E^{(t)}_2}(m;s_i) &= \frac{\vert \I^*_{E^{(t)}_1}(m;s_i) \cap \I^*_{E^{(t)}_2}(m;s_i)\vert}{\vert\I^*_{E^{(t)}_1}(m;s_i) \cup \I^*_{E^{(t)}_2}(m;s_i)\vert},\\
J_{E^{(t)}_j,F_k}(m;s_i) &= \frac{\vert \I^*_{E^{(t)}_j}(m;s_i) \cap \I^*_{F_k}(m;s_i)\vert}{\vert\I^*_{E^{(t)}_j}(m;s_i) \cup \I^*_{F_k}(m;s_i)\vert},\\
J_{E^{(s)}_j,F_k}(m;s_i) &= \frac{\vert \I^*_{E^{(s)}_j}(m;s_i) \cap \I^*_{F_k}(m;s_i)\vert}{\vert\I^*_{E^{(s)}_j}(m;s_i) \cup \I^*_{F_k}(m;s_i)\vert}.
\end{align*}
$J_{E^{(t)}_j,F_k}(m;s_i)$ and $J_{E^{(t)}_j,E^{(s)}_k}(m;s_i)$ are the similarities between the infected sets (for a target $m$) in two infection processes initiated at the same seed but evolving on the two different networks.
$J_{E^{(t)}_1,E^{(t)}_2}(m;s_i)$ is the similarity between the infected sets (for a target $m$) in the two independent processes on the encounter network.
In the case of $\beta_E=1$, the process on the encounter network is deterministic and $J_{E^{(t)}_1,E^{(t)}_2}(m;s_i)$ is not considered.

When the relevant values $\I^*_{E^{(t)}_j}(m;s_i)$, $\I^*_{E^{(s)}_j}(m;s_i)$ and $\I^*_{F_k}(m;s_i)$ for $j,k\in\{1,2\}$ are defined, we also define the following measures of precision,
\begin{align*}
P_{E^{(t)}_1,E^{(t)}_2}(m;s_i) &= \frac{\vert\I^*_{E^{(t)}_1}(m;s_i) \cap \I^*_{E^{(t)}_2}(m;s_i)\vert}{\vert \I^*_{E^{(t)}_1}(m;s_i)\vert},\\
P_{E^{(t)}_j,F_k}(m;s_i) &= \frac{\vert\I^*_{E^{(t)}_j}(m;s_i) \cap \I^*_{F_k}(m;s_i)\vert}{\vert \I^*_{E^{(t)}_j}(m;s_i)\vert},\\
P_{E^{(t)}_j,E^{(s)}_k}(m;s_i) &= \frac{\vert\I^*_{E^{(t)}_j}(m;s_i) \cap \I^*_{E^{(s)}_k}(m;s_i)\vert}{\vert \I^*_{E^{(t)}_j}(m;s_i)\vert},\\
P_{E^{(s)}_j,E^{(t)}_k}(m;s_i) &= \frac{\vert\I^*_{E^{(s)}_j}(m;s_i) \cap \I^*_{E^{(t)}_k}(m;s_i)\vert}{\vert \I^*_{E^{(s)}_j}(m;s_i)\vert},\\
P_{F_j,E^{(t)}_k}(m;s_i) &= \frac{\vert\I^*_{F_j}(m;s_i) \cap \I^*_{E^{(t)}_k}(m;s_i)\vert}{\vert \I^*_{F_j}(m;s_i)\vert}.\\
\end{align*}
For target $m$, $P_{E^{(t)}_j,F_k}(m;s_i)$ is the fraction of nodes infected in the process with index $j$ in the encounter network that are also infected in the process with index $k$ in the encounter network (started at the same seed).
The other quantities are similarly interpreted.

A comparison between $J_{E^{(t)}_1,F_1}(m;s_i)$ and $J_{E^{(t)}_1,E^{(t)}_2}(m;s_i)$ is not straightforward for the lack of an upper bound for $J_{E^{(t)}_1,F_1}(m;s_i)$.
There are $n_I=76,933$ nodes in the intersection of the friendship and encounter network and $n_U=225,028$ nodes in their union.
Therefore, for large values of target $m$, $J_{E^{(t)}_1,F_1}(m;s_i)$ is upper bounded by $n_I/n_U=0.3419$.
A bound that is independent of $s_i$ cannot be derived for general values of $m$, for which $J_{E^{(t)}_1,F_1}(m;s_i)$ is not constrained to have small values.
However, $J_{E^{(t)}_1,E^{(t)}_2}(m;s_i)$ can be as large as $1$ for all values of $m$.
To take this into account, we also define a rescaled version of the Jaccard similarity,
$$
\bar J_{E^{(t)}_1,F_1}(m;s_i) = \frac{J_{E^{(t)}_1,F_1}(m;s_i)}{J^U_{E^{(t)}_1,F_1}(m)},
$$
where $J^U_{E^{(t)}_1,F_1}(m)=\max_{s_i}J_{E^{(t)}_1,F_1}(m;s_i)$ is the empirical upper bound for $J_{E^{(t)}_1,F_1}(m;\cdot)$ (computed over all simulations).
We similarly define rescaled versions of the other similarity measures, considering the unions and intersections of the relevant sets of nodes.

The same argument hold for the precision measures for the lack of a straightforward upper bound for $P_{E^{(t)}_1,F_1}(m;s_i)$ and $P_{F_1,E^{(t)}_1}(m;s_i)$.
For large values of $m$, $P_{E^{(t)}_1,F_1}(m;s_i)$ and $P_{F_1,E^{(t)}_1}(m;s_i)$ are upper bounded by $n_I/n_E=0.5782$ and $n_I/n_F=0.4554$, respectively.
Bounds that are independent of $s_i$ cannot be derived for general values of $m$.
However, $P_{E^{(t)}_1,E^{(t)}_2}(m;s_i)$ can be as large as $1$ for all values of $m$.
To take this consideration into account, we define rescaled version of the precision measures, for example,
$$
\bar P_{E^{(t)}_1,F_1}(m;s_i) = \frac{P_{E^{(t)}_1,F_1}(m;s_i)}{P^U_{E^{(t)}_1,F_1}(m)},
$$
where $P^U_{E^{(t)}_1,F_1}(m)$ is an empirical upper bound obtained taking the maximum over all simulations.
We similarly define rescaled versions of the other similarity measures, considering  the intersections of the relevant sets of nodes.

\FloatBarrier
\subsection{Case 1: certain infection}
\label{yelp:sec:encounter_friendship_statics_certain}

We ran $5000$ groups of simulations of the SI process with $\beta_F=\beta_{E^{(t)}}=\beta_{E^{(s)}}=1$.
For each group of simulations, a single seed is selected uniformly at random among all nodes $s\in U_F\cap U_E^{(s)}\cap U_E^{(t)}$ (present in all three networks) such that $t_0(s_i)\le 500$ (that is, we consider nodes that have an encounter by time $t=500$).
For each choice of the seed, we separately run one infection process on each network.
Therefore, each seed selection is associated to three simulations: one on the encounter network ($E_1^{(t)}$), one on the static encounter network ($E_1^{(s)}$), one on the friendship network ($F_1$).
For target set size $m\in\{500, 1000, 2000, 5000, 10000, 20000\}$ and each of the $5000$ seeds $s_i$, we consider the metrics above when they are defined.
In particular, we consider the similarity metrics $J_{E_1^{(t)},E_1^{(s)}}(m;s_i)$, $J_{E_1^{(t)},F_1}(m;s_i)$,
and the precision metrics
$P_{E_1^{(t)},E_1^{(s)}}(m;s_i)$, $P_{E_1^{(t)},F_1}(m;s_i)$,
$P_{E_1^{(s)},E_1^{(t)}}(m;s_i)$, $P_{F_1,E_1^{(t)}}(m;s_i)$.
That is, fixed a seed $s_i$, we compare the infection processes on the encounter network with those on each static network.

\begin{figure}
\centering
\includegraphics[scale=0.50]{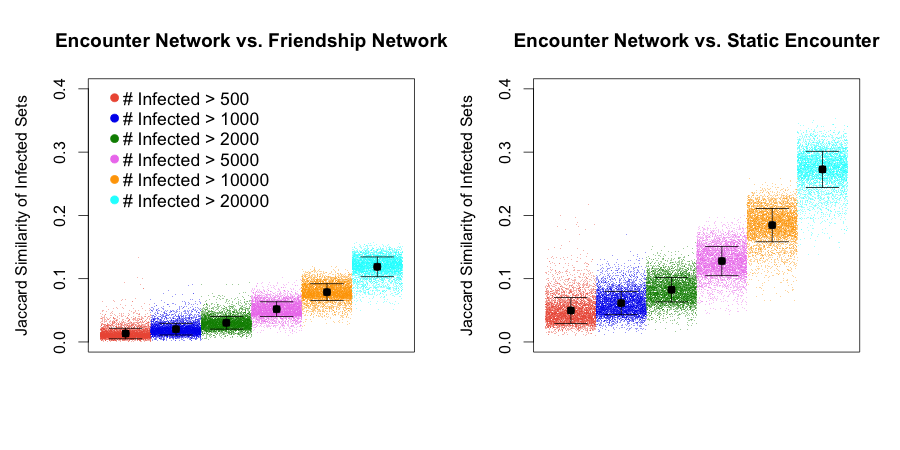}
\caption{\textbf{Single seed infection on encounter network and the static (encounter and friendship) networks -- Certain infection -- Similarity measures.} The two panels show the metrics $J_{E_1^{(t)},F_1}(m;s_i)$ and $J_{E_1^{(t)},E_1^{(s)}}(m;s_i)$, for $5000$ random choices of a single seeds, and different values of the target set size $m$.
For each seed, one simulation on the friendship network, one simulation on the encounter network and  one simulation on the static encounter network are run separately.
Each panel considers, for each of the $5000$ seeds, a pair of simulations on two different networks.
On the $x$- axis, observations for a given value of $m$ form a block with a constant color (within the block, the $x$ position is irrelevant).
We only consider pairs $(m,s_i)$ for which the metrics are defined.
For a given metric and each value $m$, the black point represents the average of the metric over all observations such that the metric is defined, and the bars represent standard deviations.
}
\label{diff_EFvsEE_Jaccard}
\end{figure}

Figure~\ref{diff_EFvsEE_Jaccard} plots the measures $J_{E_1^{(t)},F_1}(m;s_i)$ and $J_{E_1^{(t)},E_1^{(s)}}(m;s_i)$ in the left and right panels respectively.
Observations for a given value of $m$ constitute a block on the $x$-axis (larger values of $m$ correspond to $x$ positions on the right) and are represented with the same color.
For a fixed value of $m$, relative $x$ positions are irrelevant.
For a given metric and each value $m$, the black point represents the average of the metric over all observations such that the metric is defined, and the bars represent standard deviations.

Table~\ref{tab:EFvsEE_Jaccard} reports the averages of the measures $J_{E_1^{(t)},F_1}(m;s_i)$ and $J_{E_1^{(t)},E_1^{(s)}}(m;s_i)$, denoted by
$\langle J_{E_1^{(t)},F_1}(m)\rangle$  and $\langle J_{E_1^{(t)},E_1^{(s)}}(m)\rangle$,
together with their normalized versions
$\langle \bar J_{E_1^{(t)},F_1}(m)\rangle$  and $\langle \bar J_{E_1^{(t)},E_1^{(s)}}(m)\rangle$
and their empirical upper bounds
$J^U_{E_1^{(t)},F_1}(m)$  and $J^U_{E_1^{(t)},E_1^{(s)}}(m)$.

For all values of $m$, two-sample t-tests support the hypotheses that $J_{E_1^{(t)},E_1^{(s)}}(m)$ has larger average than $J_{E_1^{(t)},F_1}(m;s_i)$ (p-values$<2.2\cdot 10^{-16}$).
For all values of $m$, two-sample t-tests support the hypotheses that $\bar J_{E_1^{(t)},E_1^{(s)}}(m)$ has larger average than $\bar J_{E_1^{(t)},F_1}(m;s_i)$ (p-values$=0.0116$ for $m=20000$, p-values$<2.2\cdot 10^{-16}$ for other values of $m$).
That is, the similarity between the sets of infected nodes on the encounter network and on the static encounter network
is larger than the similarity between the sets of infected nodes on the encounter network and on the friendship network.

\begin{table}
\centering
\caption{\textbf{Single seed infection on the encounter network and the static (encounter and friendship) networks.} Certain infection - Similarity measures.}
\begin{tabular}{|c|c|c|c|c|c|c|}
\hline
$m$ &
$\langle J_{E_1^{(t)},F_1}(m)\rangle$  & $\langle J_{E_1^{(t)},E_1^{(s)}}(m)\rangle$ &
$\langle \bar J_{E_1^{(t)},F_1}(m)\rangle$  & $\langle \bar J_{E_1^{(t)},E_1^{(s)}}(m)\rangle$ &
$J^U_{E_1^{(t)},F_1}(m)$  & $J^U_{E_1^{(t)},E_1^{(s)}}(m)$ \\
\hline
500 & 0.013 & 0.050 & 0.100 & 0.228 & 0.133 & 0.218\\
1000 & 0.020 & 0.062 & 0.221 & 0.361 & 0.091 & 0.170\\
2000 & 0.030 & 0.083 & 0.322 & 0.525 & 0.094 & 0.157\\
5000 & 0.052 & 0.128 & 0.506 & 0.641 & 0.103 & 0.200\\
10000 & 0.079 & 0.185 & 0.660 & 0.720 & 0.120 & 0.258\\
20000 & 0.119 & 0.273 & 0.764 & 0.779 & 0.157 & 0.354\\
\hline
\end{tabular}
\label{tab:EFvsEE_Jaccard}
\end{table}

Table~\ref{tab:EFvsEE_Precision1} reports the averages of the precision measures $P_{E_1^{(t)},F_1}(m)$  and $P_{E_1^{(t)},E_1^{(s)}}(m)$, their empirical upper bounds, and the averages of the rescaled measures.
Table~\ref{tab:EFvsEE_Precision2} reports the averages of the precision measures $P_{F_1,E_1^{(t)}}(m)$  and $P_{E_1^{(s)},E_1^{(t)}}(m)$, their empirical upper bounds, and the averages of the rescaled measures.
For all values of $m$, two-sample t-tests support the hypotheses that $P_{E_1^{(t)},E_1^{(s)}}(m)$ has larger average than $P_{E_1^{(t)},F_1}(m;s_i)$, and that $P_{E_1^{(s)},E_1^{(t)}}(m)$ has larger average than $P_{F_1,E_1^{(t)}}(m;s_i)$(p-values$<2.2\cdot 10^{-16}$).
For all values of $m$, two-sample t-tests support the hypotheses that $\bar P_{E_1^{(t)},E_1^{(s)}}(m)$ has larger average than $\bar P_{E_1^{(t)},F_1}(m;s_i)$, and that $\bar P_{E_1^{(s)},E_1^{(t)}}(m)$ has larger average than $\bar P_{F_1,E_1^{(t)}}(m;s_i)$(p-values$<2.2\cdot 10^{-16}$).
That is, infections on the encounter network are better approximated by infections on the static encounter network than by infection on the friendship network.

\begin{table}
\centering
\caption{\textbf{Single seed infection on the encounter network and the static (encounter and friendship) networks.} Certain infection - Precision measures.}
\begin{tabular}{|c|c|c|c|c|c|c|}
\hline
$m$ &
$\langle P_{E_1^{(t)},F_1}(m)\rangle$  & $\langle P_{E_1^{(t)},E_1^{(s)}}(m)\rangle$ &
$\langle \bar P_{E_1^{(t)},F_1}(m)\rangle$  & $\langle \bar P_{E_1^{(t)},E_1^{(s)}}(m)\rangle$ &
$P^U_{E_1^{(t)},F_1}(m)$  & $P^U_{E_1^{(t)},E_1^{(s)}}(m)$ \\
\hline
500 & 0.177 & 0.298 & 0.388 & 0.388 & 0.594 & 0.766\\
1000 & 0.211 & 0.340 & 0.435 & 0.435 & 0.654 & 0.781\\
2000 & 0.259 & 0.392 & 0.478 & 0.478 & 0.648 & 0.821\\
5000 & 0.345 & 0.498 & 0.570 & 0.570 & 0.684 & 0.875\\
10000 & 0.405 & 0.583 & 0.647 & 0.647 & 0.688 & 0.902\\
20000 & 0.446 & 0.668 & 0.714 & 0.714 & 0.686 & 0.937\\
\hline
\end{tabular}
\label{tab:EFvsEE_Precision1}
\end{table}

\begin{table}
\centering
\caption{\textbf{Single seed infection on the encounter network and the static (encounter and friendship) networks.} Certain infection - Precision measures.}
\begin{tabular}{|c|c|c|c|c|c|c|}
\hline
$m$ &
$\langle P_{F_1,E_1^{(t)}}(m)\rangle$  & $\langle P_{E_1^{(s)},E_1^{(t)}}(m)\rangle$ &
$\langle \bar P_{F_1,E_1^{(t)}}(m)\rangle$  & $\langle \bar P_{E_1^{(s)},E_1^{(t)}}(m)\rangle$ &
$P^U_{F_1,E_1^{(t)}}(m)$  & $P^U_{E_1^{(s)},E_1^{(t)}}(m)$ \\
\hline
500 & 0.016 & 0.064 & 0.070 & 0.197 & 0.230 & 0.322\\
1000 & 0.025 & 0.078 & 0.180 & 0.289 & 0.137 & 0.268\\
2000 & 0.038 & 0.104 & 0.224 & 0.437 & 0.168 & 0.238\\
5000 & 0.064 & 0.159 & 0.382 & 0.508 & 0.168 & 0.314\\
10000 & 0.097 & 0.225 & 0.481 & 0.602 & 0.202 & 0.376\\
20000 & 0.147 & 0.327 & 0.585 & 0.677 & 0.252 & 0.487\\
\hline
\end{tabular}
\label{tab:EFvsEE_Precision2}
\end{table}

\begin{figure}
\centering
\includegraphics[scale=0.50]{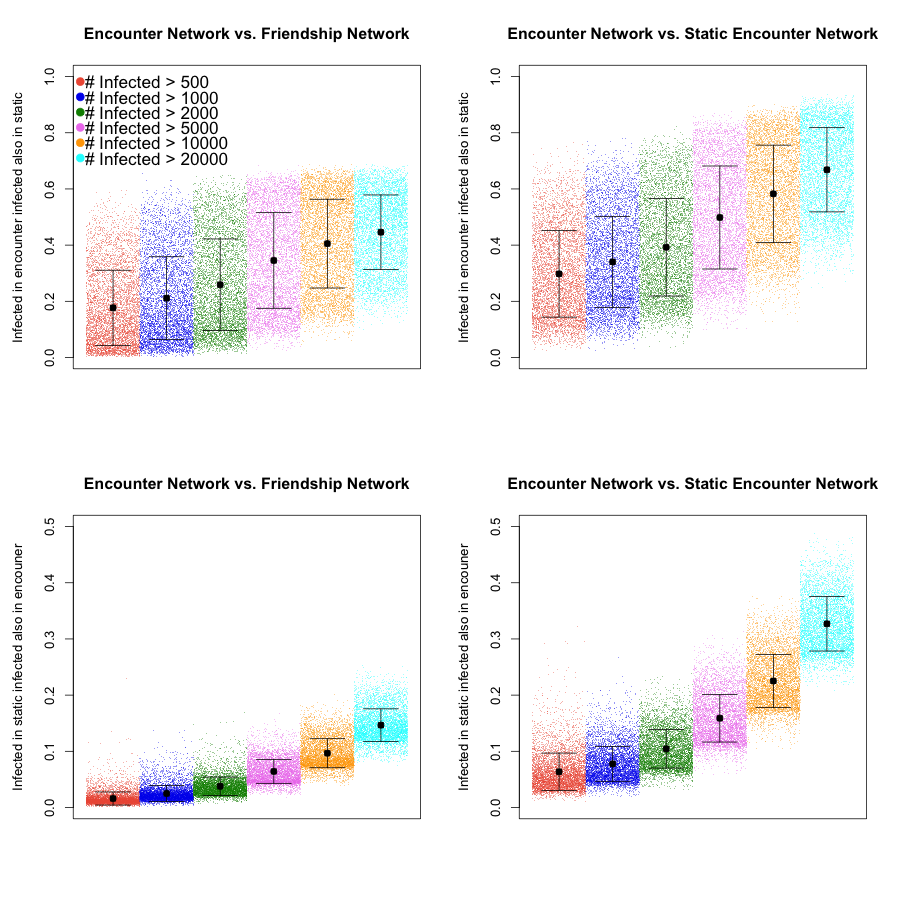}
\caption{\textbf{Single seed infection on encounter network and the static (encounter and friendship) networks -- Certain infection -- Precision measures.} The four panels show the metrics $P_{E_1^{(t)},F_1}(m)$ (top-left), $P_{E_1^{(t)},E_1^{(s)}}(m)$ (top-right), $P_{F_1,E_1^{(t)}}(m)$ (bottom-left) and $P_{E_1^{(s)},E_1^{(t)}}(m)$ (bottom-right), for $5000$ random choices of a single seeds, and different values of the target set size $m$.
For each seed, one simulation on the friendship network, one simulation on the encounter network and  one simulation on the static encounter network are run separately.
Each panel considers, for each of the $5000$ seeds, a pair of simulations on two different networks.
On the $x$- axis, observations for a given value of $m$ form a block with a constant color (within the block, the $x$ position is irrelevant).
We only consider pairs $(m,s_i)$ for which the metrics are defined.
For a given metric and each value $m$, the black point represents the average of the metric over all observations such that the metric is defined, and the bars represent standard deviations.
}
\label{diff_EFvsEE_Precision}
\end{figure}

\FloatBarrier
\subsection{Case 2: stochastic infection}
\label{yelp:sec:encounter_friendship_statics_stochastic}

We ran $5000$ groups of simulations of the SI process with $\beta_F=0.01$ and $\beta_{E^{(t)}}=\beta_{E^{(s)}}=0.5$.
For each group of simulations, a single seed is selected uniformly at random among all nodes $s\in U_F\cap U_E^{(s)}\cap U_E^{(t)}$ (present in all three networks) such that $t_0(s_i)\le 500$ (that is, we consider nodes that have an encounter by time $t=500$).
For each choice of the seed, we run two independent infection processes on each network.
Therefore, each seed selection is associated to six simulations: one on the encounter network ($E_1^{(t)}$, $E_2^{(t)}$), one on the static encounter network ($E_1^{(s)}$, $E_1^{(s)}$), one on the friendship network ($F_1$, $F_2$).
For target set size $m\in\{500, 1000, 2000, 5000, 10000, 20000\}$ and each of the $5000$ seeds $s_i$, we consider the metrics above when they are defined.
In particular, we consider the similarity metrics $J_{E_1^{(t)},E_2^{(t)}}(m;s_i)$, $J_{E_1^{(t)},E_1^{(s)}}(m;s_i)$, $J_{E_1^{(t)},F_1}(m;s_i)$,
and the precision metrics
$P_{E_1^{(t)},E_2^{(t)}}(m;s_i)$, $P_{E_1^{(t)},E_1^{(s)}}(m;s_i)$, $P_{E_1^{(t)},F_1}(m;s_i)$,
$P_{E_1^{(t)},E_2^{(t)}}(m;s_i)$, $P_{E_1^{(s)},E_1^{(t)}}(m;s_i)$, $P_{F_1,E_1^{(t)}}(m;s_i)$.
That is, fixed a seed $s_i$, we compare the two infection processes on the encounter network, and the the infections on the encounter network with those on each static network.

\begin{figure}
\centering
\includegraphics[scale=0.35]{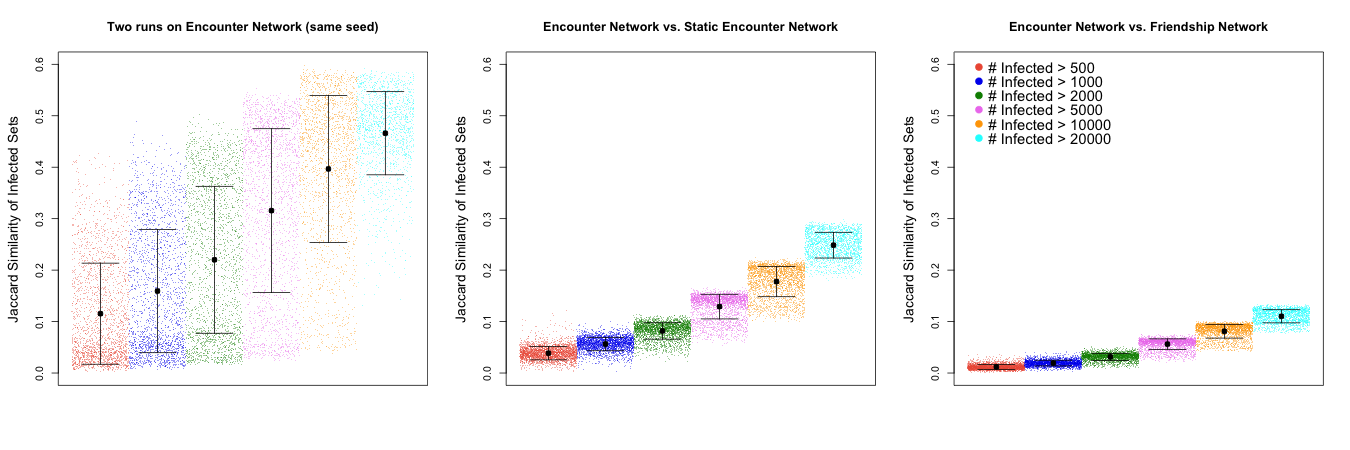}
\caption{\textbf{Single seed infection on encounter network and the static (encounter and friendship) networks -- Stochastic infection -- Similarity measures.} 
The three panels show the metrics $J_{E_1^{(t)},E_2^{(t)}}(m;s_i)$, $J_{E_1^{(t)},E_1^{(s)}}(m;s_i)$ and $J_{E_1^{(t)},F_1}(m;s_i)$, for $5000$ random choices of a single seeds, and different values of the target set size $m$.
For each seed, one simulation on the friendship network, two simulations on the encounter network, two on the static encounter network, and two on the friendship network are run separately.
Each panel considers, for each of the $5000$ seeds, a pair of simulations on two different networks.
On the $x$- axis, observations for a given value of $m$ form a block with a constant color (within the block, the $x$ position is irrelevant).
We only consider pairs $(m,s_i)$ for which the metrics are defined.
For a given metric and each value $m$, the black point represents the average of the metric over all observations such that the metric is defined, and the bars represent standard deviations.
}
\label{diff_Jaccard_stochastic}
\end{figure}

Figure~\ref{diff_Jaccard_stochastic} plots the measures $J_{E_1^{(t)},E_2^{(t)}}(m;s_i)$, $J_{E_1^{(t)},E_1^{(s)}}(m;s_i)$ and $J_{E_1^{(t)},F_1}(m;s_i)$ in the left, middle and right panels respectively.
Observations for a given value of $m$ constitute a block on the $x$-axis (larger values of $m$ correspond to $x$ positions on the right) and are represented with the same color.
For a fixed value of $m$, relative $x$ positions are irrelevant.
For a given metric and each value $m$, the black point represents the average of the metric over all observations such that the metric is defined, and the bars represent standard deviations.

Table~\ref{tab:EFvsEEvsE2_Jaccard} reports the averages of the measures $J_{E_1^{(t)},E_2^{(t)}}(m;s_i)$, $J_{E_1^{(t)},E_1^{(s)}}(m;s_i)$ and $J_{E_1^{(t)},F_1}(m;s_i)$ denoted by
$\langle J_{E_1^{(t)},E_2^{(t)}}(m)\rangle$, $\langle J_{E_1^{(t)},E_1^{(s)}}(m)\rangle$ and $\langle J_{E_1^{(t)},F_1}(m)\rangle$,
together with their normalized versions
$\langle \bar J_{E_1^{(t)},E_2^{(t)}}(m)\rangle$, $\langle \bar J_{E_1^{(t)},E_1^{(s)}}(m)\rangle$, $\langle \bar J_{E_1^{(t)},F_1}(m)\rangle$
and their empirical upper bounds
$J^U_{E_1^{(t)},E_2^{(t)}}(m)$, $J^U_{E_1^{(t)},E_1^{(s)}}(m)$ and $J^U_{E_1^{(t)},F_1}(m)$.

For all values of $m$, two-sample t-tests support the hypotheses that $J_{E_1^{(t)},E_2^{(t)}}(m)$ has larger average than $J_{E_1^{(t)},E_1^{(s)}}(m)$ and $J_{E_1^{(t)},F_1}(m;s_i)$, and that $J_{E_1^{(t)},E_1^{(s)}}(m)$ has larger average than $J_{E_1^{(t)},F_1}(m;s_i)$ (p-values$<2.2\cdot 10^{-16}$).
The similarity of the infected sets on two independent runs of the infection process within the encounter network is larger than the similarities of the infected sets between different networks.
In addition, the similarity between the sets of infected nodes on the encounter network and on the static encounter network
is larger than the similarity between the sets of infected nodes on the encounter network and on the friendship network.
For all values of $m$, two-sample t-tests support the hypotheses that $\bar J_{E_1^{(t)},E_2^{(t)}}(m)$ has smaller average than $\bar J_{E_1^{(t)},E_1^{(s)}}(m)$ and $\bar J_{E_1^{(t)},F_1}(m;s_i)$ (p-values$<2.2\cdot 10^{-16}$).
The hypotheses that $\bar J_{E_1^{(t)},E_1^{(s)}}(m)$ has larger average than $\bar J_{E_1^{(t)},F_1}(m;s_i)$ is supported for $m\in\{1000,2000,5000\}$ (p-values$<0.0248$) and the null hypothesis of equal mean cannot be rejected for the other values of $m$.
These analyses support the idea that topological differences accentuate the unpredictability of epidemic risk using the static networks, particularly in the case of the friendship network

\begin{table}
\centering
\caption{\textbf{Single seed infection on the encounter network and the static (encounter and friendship) networks.} Stochastic infection - Similarity measures.}
\begin{tabular}{|c|c|c|c|c|c|c|}
\hline
$m$ &
$\langle J_{E_1^{(t)},E_2^{(t)}}(m)\rangle$ & $\langle J_{E_1^{(t)},E_1^{(s)}}(m)\rangle$ & $\langle J_{E_1^{(t)},F_1}(m)\rangle$  & 
 $\langle \bar J_{E_1^{(t)},E_2^{(t)}}(m)\rangle$ &  $\langle \bar J_{E_1^{(t)},E_1^{(s)}}(m)\rangle$ & $\langle \bar J_{E_1^{(t)},F_1}(m)\rangle$ \\
\hline
500 & 0.115 & 0.039 & 0.012 & 0.270 & 0.315 & 0.323\\
1000 & 0.159 & 0.056 & 0.019 & 0.325 & 0.561 & 0.454\\
2000 & 0.220 & 0.082 & 0.031 & 0.438 & 0.716 & 0.615\\
5000 & 0.316 & 0.129 & 0.056 & 0.571 & 0.776 & 0.744\\
10000 & 0.397 & 0.178 & 0.081 & 0.664 & 0.790 & 0.806\\
20000 & 0.466 & 0.249 & 0.110 & 0.788 & 0.835 & 0.830\\
\hline
\end{tabular}
\label{tab:EFvsEEvsE2_Jaccard}
\end{table}

Table~\ref{tab:EFvsEEvsE2_Precision1} reports the averages of the precision measures $P_{E_1^{(t)},F_1}(m)$, $P_{E_1^{(t)},E_1^{(s)}}(m)$ and $P_{E_1^{(t)},E_2^{(t)}}(m)$, their empirical upper bounds, and the averages of the rescaled measures.
Table~\ref{tab:EFvsEEvsE2_Precision2} reports the averages of the precision measures $P_{F_1,E_1^{(t)}}(m)$, $P_{E_1^{(s)},E_1^{(t)}}(m)$ and $P_{E_1^{(2)},E_2^{(t)}}(m)$, their empirical upper bounds, and the averages of the rescaled measures (note that $P_{E_1^{(2)},E_2^{(t)}}(m)$ and $P_{E_2^{(2)},E_1^{(t)}}(m)$ are practically the same quantity).
For all values of $m$, two-sample t-tests support the hypotheses that $P_{E_1^{(t)},E_2^{(t)}}(m)$ has larger average than $P_{E_1^{(t)},E_1^{(s)}}(m)$, $P_{E_1^{(t)},F_1}(m;s_i)$, $P_{E_1^{(s)},E_1^{(t)}}(m)$, $P_{F_1,E_1^{(t)}}(m;s_i)$, that $P_{E_1^{(t)},E_1^{(s)}}(m)$ has larger average than $P_{E_1^{(t)},F_1}(m;s_i)$, and that $P_{E_1^{(s)},E_1^{(t)}}(m)$ has larger average than $P_{F_1,E_1^{(t)}}(m;s_i)$ (p-values$<2.2\cdot 10^{-16}$).
For the rescaled measures,
For all values of $m$, two-sample t-tests support the hypotheses that $\bar P_{E_1^{(t)},E_2^{(t)}}(m)$ has smaller average than $\bar P_{E_1^{(t)},E_1^{(s)}}(m)$, $\bar P_{E_1^{(t)},F_1}(m;s_i)$, $\bar P_{E_1^{(s)},E_1^{(t)}}(m)$, $\bar P_{F_1,E_1^{(t)}}(m;s_i)$ (p-values$<2.2\cdot 10^{-16}$),
that $\bar P_{E_1^{(t)},E_1^{(s)}}(m)$ has larger average than $\bar P_{E_1^{(t)},F_1}(m;s_i)$ (p-values$<0.00679$),
and, for $m\neq 10000$, that $\bar P_{E_1^{(s)},E_1^{(s)}}(m)$ has larger average than $\bar P_{F_1,E_1^{(t)}}(m;s_i)$ (p-values$<0.00679$), 
For all values of $m$, two-sample t-tests support the hypotheses that $\bar P_{E_1^{(t)},E_1^{(s)}}(m)$ has larger average than $\bar P_{E_1^{(t)},F_1}(m;s_i)$, and that $\bar P_{E_1^{(s)},E_1^{(t)}}(m)$ has larger average than $\bar P_{F_1,E_1^{(t)}}(m;s_i)$(p-values$<5.8e\cdot 10^{-7}$).
As above, these analyses support the idea that topological differences accentuate the unpredictability of epidemic risk using the static networks, particularly in the case of the friendship network

\begin{table}
\centering
\caption{\textbf{Single seed infection on the encounter network and the static (encounter and friendship) networks.} Stochastic infection - Precision measures.}
\begin{tabular}{|c|c|c|c|c|c|c|}
\hline
$m$ &
$\langle P_{E_1^{(t)},E_2^{(t)}}(m)\rangle$ & $\langle P_{E_1^{(t)},E_1^{(s)}}(m)\rangle$ & $\langle P_{E_1^{(t)},F_1}(m)\rangle$  &
$\langle \bar P_{E_1^{(t)},E_2^{(t)}}(m)\rangle$ & $\langle \bar P_{E_1^{(t)},E_1^{(s)}}(m)\rangle$ & $\langle \bar P_{E_1^{(t)},F_1}(m)\rangle$\\
\hline
500 & 0.194 & 0.078 & 0.034 & 0.323 & 0.350 & 0.305\\
1000 & 0.257 & 0.112 & 0.055 & 0.391 & 0.591 & 0.407\\
2000 & 0.338 & 0.158 & 0.083 & 0.506 & 0.720 & 0.515\\
5000 & 0.456 & 0.235 & 0.129 & 0.641 & 0.782 & 0.637\\
10000 & 0.551 & 0.306 & 0.167 & 0.737 & 0.804 & 0.754\\
20000 & 0.631 & 0.402 & 0.209 & 0.849 & 0.861 & 0.810\\
\hline
\end{tabular}
\label{tab:EFvsEEvsE2_Precision1}
\end{table}

\begin{table}
\centering
\caption{\textbf{Single seed infection on the encounter network and the static (encounter and friendship) networks.} Stochastic infection - Precision measures.}
\begin{tabular}{|c|c|c|c|c|c|c|}
\hline
$m$ &
$\langle P_{E_1^{(t)},E_2^{(t)}}(m)\rangle$ & $\langle P_{E_1^{(s)},E_1^{(t)}}(m)\rangle$ & $\langle P_{F_1,E_1^{(t)}}(m)\rangle$  &
$\langle \bar P_{E_1^{(t)},E_2^{(t)}}(m)\rangle$ & $\langle \bar P_{E_1^{(s)},E_1^{(t)}}(m)\rangle$ & $\langle \bar P_{F_1,E_1^{(t)}}(m)\rangle$\\
\hline
500 & 0.194 & 0.078 & 0.034 & 0.323 & 0.330 & 0.299\\
1000 & 0.257 & 0.112 & 0.055 & 0.392 & 0.572 & 0.417\\
2000 & 0.338 & 0.158 & 0.083 & 0.505 & 0.728 & 0.577\\
5000 & 0.456 & 0.235 & 0.129 & 0.640 & 0.777 & 0.706\\
10000 & 0.551 & 0.306 & 0.167 & 0.737 & 0.827 & 0.802\\
20000 & 0.631 & 0.402 & 0.209 & 0.849 & 0.869 & 0.833\\
\hline
\end{tabular}
\label{tab:EFvsEEvsE2_Precision2}
\end{table}

\begin{figure}
\centering
\includegraphics[scale=0.5]{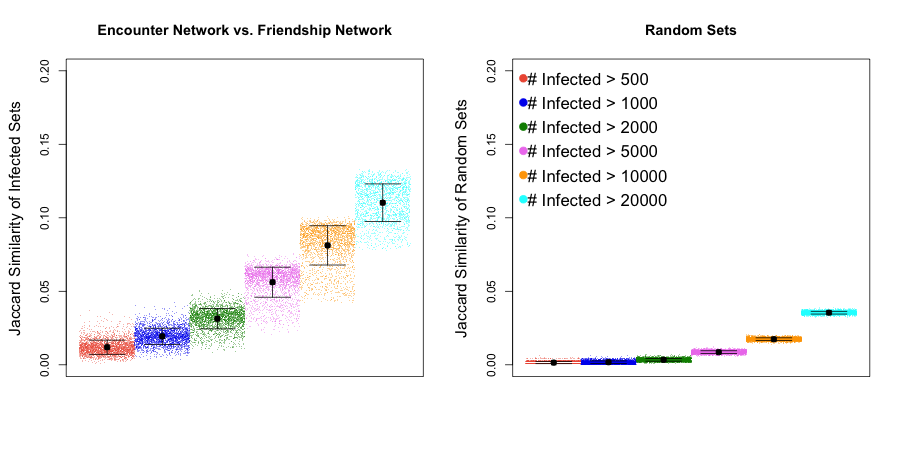}
\caption{\textbf{Comparison with the intersection of random sets -- Stochastic infection -- Similarity measures.} 
The left panels shows the metrics $J_{E_1^{(t)},F_1}(m;s_i)$, for $5000$ random choices of a single seeds, and different values of the target set size $m$.
For each seed, one simulation on the friendship network and one on the encounter network are run independently.
The right panel shows the same metric for pairs of random sets sampled from the two networks ($5000$ pairs for each target set size).
On the $x$- axis, observations for a given value of $m$ form a block with a constant color (within the block, the $x$ position is irrelevant).
For a given metric and each value $m$, the black point represents the average of the metric over all observations, and the bars represent standard deviations.
}
\label{diff_Jaccard_random}
\end{figure}

The intersection between the infected sets in the friendship network and the encounter network (considering infection started at the same seed) is much larger than the intersection of random sets, for each target set size (two-sample t-tests, p-values$<2.2\cdot 10^{-16}$).
Figure~\ref{diff_Jaccard_random} shows the Jaccard similarity of the infected sets on the encounter and friendship networks (left, for the $5000$ simulations considered above) and of random sets of the given target size sampled from the two networks (right, $5000$ pairs of random sets for each target size).
Table~\ref{tab:Jaccard_random} shows the averages of the metrics $J_{E_1^{(t)},F_1}(m;s_i)$, $P_{E_1^{(t)},F_1}(m;s_i)$ and $P_{F_1,E_1^{(t)}}(m;s_i)$ for the $5000$ pairs of simulations and the averages of the corresponding metrics for the $5000$ pairs of random sets.

\begin{table}
\centering
\caption{\textbf{Comparison with intersection of random sets.} Stochastic infection - Similarity and precision measures.}
\begin{tabular}{|c|c|c|c|c|c|c|}
\hline
$m$ &
 $\langle J_{E_1^{(t)},F_1}(m)\rangle$  & $\langle J^{rand}_{E_1^{(t)},F_1}(m)\rangle$  &
 $\langle P_{E_1^{(t)},F_1}(m)\rangle$  & $\langle P^{rand}_{E_1^{(t)},F_1}(m)\rangle$ &
 $\langle P_{F_1,E_1^{(t)}}(m)\rangle$  & $\langle P^{rand}_{F_1,E_1^{(t)}}(m)\rangle$ \\
\hline
500 & 0.012 & 0.001 & 0.020 & 0.002 & 0.011 & 0.002 \\
1000 & 0.019 & 0.002 & 0.031 & 0.003 & 0.017 & 0.003 \\
2000 & 0.031 & 0.003 & 0.047 & 0.007 & 0.028 & 0.007 \\
5000 & 0.056 & 0.009 & 0.071 & 0.017 & 0.050 & 0.017 \\
10000 & 0.081 & 0.017 & 0.086 & 0.034 & 0.071 & 0.034 \\
20000 & 0.110 & 0.035 & 0.092 & 0.069 & 0.084 & 0.069 \\
\hline
\end{tabular}
\label{tab:Jaccard_random}
\end{table}


\begin{figure}
\centering
\includegraphics[scale=0.35]{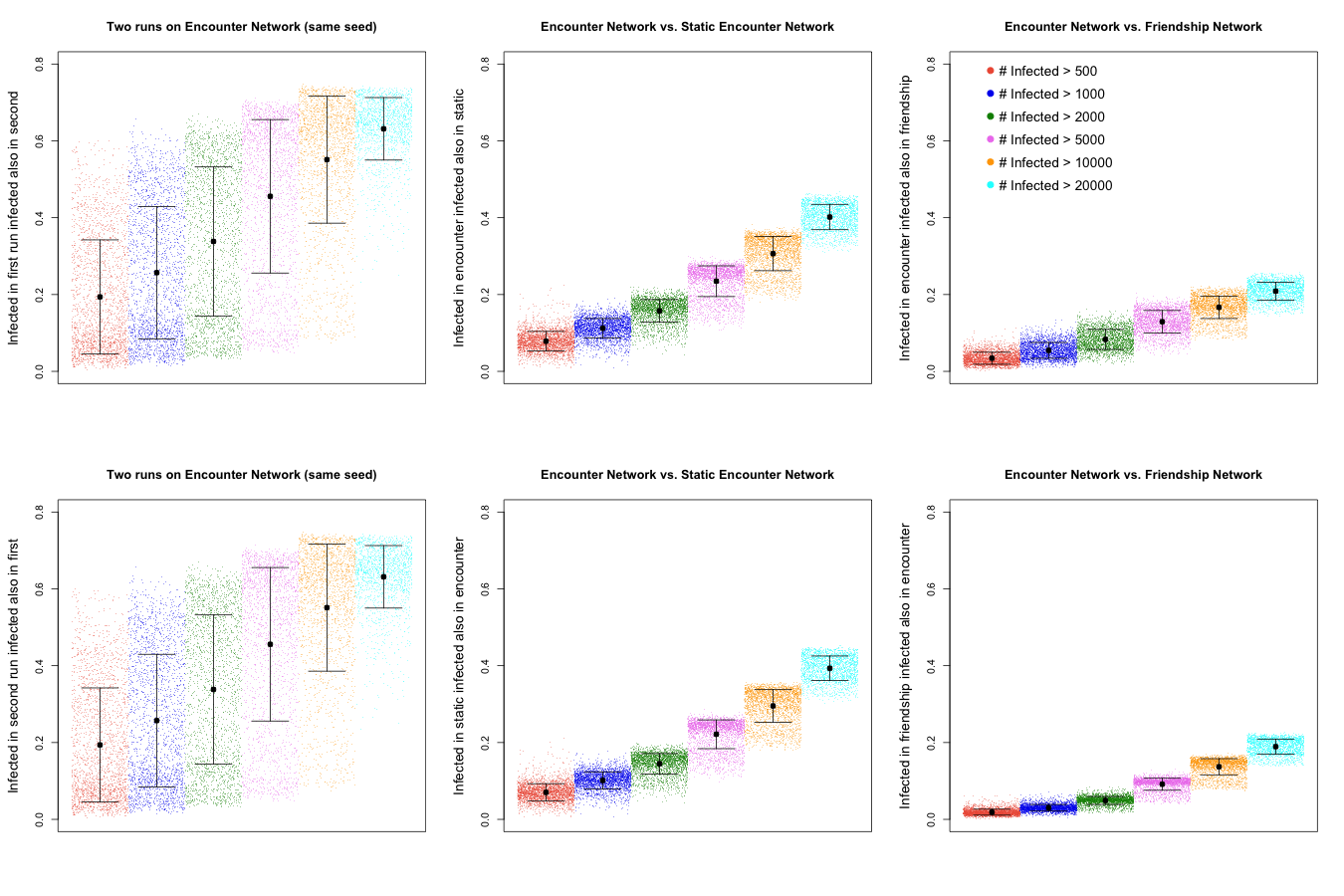}
\caption{\textbf{Single seed infection on encounter network and the static (encounter and friendship) networks -- Stochastic infection -- Precision measures.} 
The six panels show the metrics $P_{E_1^{(t)},E_2^{(t)}}(m)$ (top-left), $P_{E_1^{(t)},E_1^{(s)}}(m)$ (top-center), $P_{E_1^{(t)},F_1}(m)$ (top-right), $P_{E_2^{(t)},E_1^{(t)}}(m)$ (bottom-left), $P_{E_1^{(s)},E_1^{(t)}}(m)$ (bottom-center), $P_{F_1,E_1^{(t)}}(m)$ (bottom-right),
for $5000$ random choices of a single seeds, and different values of the target set size $m$.
For each seed, one simulation on the friendship network, two simulations on the encounter network, two on the static encounter network, and two on the friendship network are run separately.
Each panel considers, for each of the $5000$ seeds, a pair of simulations on two different networks.
On the $x$- axis, observations for a given value of $m$ form a block with a constant color (within the block, the $x$ position is irrelevant).
We only consider pairs $(m,s_i)$ for which the metrics are defined.
For a given metric and each value $m$, the black point represents the average of the metric over all observations such that the metric is defined, and the bars represent standard deviations.
}
\label{diff_Precision_stochastic}
\end{figure}

\FloatBarrier
\section{Epidemic risk: comparison between the time-varying networks}
\label{yelp:sec:time_population}
To argue that our results are not driven by the static nature of the friendship network as opposed to the time-varying nature of the encounter network, in this section we compare the encounter network with the time-varying friendship network defined in Section~\ref{yelp:sec:other_networks}.
In Section~\ref{yelp:sec:static_population}, we compare the friendship network with the static encounter network defined in Section~\ref{yelp:sec:other_networks}.
In both cases, the sets of individuals predicted to be at risk by friendship appear a poor approximation of those at risk in a process spreading according to physical encounter.
As before, we consider seed nodes that are present in both the friendship and the encounter network, and we compare the sets of nodes that become infected in independent processes on the two different networks initiated at the same seed.

We ran $5000$ groups of simulations of the SI process with $\beta=0.5$.
For each group of simulations, a single seed is selected at random among all nodes $s_i$ such that $t_0(s_i)\le 500$ in both the encounter and the time-varying friendship networks.
For each choice of the seed, we separately run two infection processes on the encounter network and two infection processes on the time-varying friendship network.
Therefore, each seed selection is associated to four simulations (referred to as $E_1$, $E_2$, $F_1$, $F_2$).
For target set size $m\in\{500, 1000, 2000, 5000, 10000, 20000\}$ and each of the $5000$ seeds $s_i$, we consider the similarity and precision metrics defined above.

\begin{figure}
\centering
\includegraphics[scale=0.50]{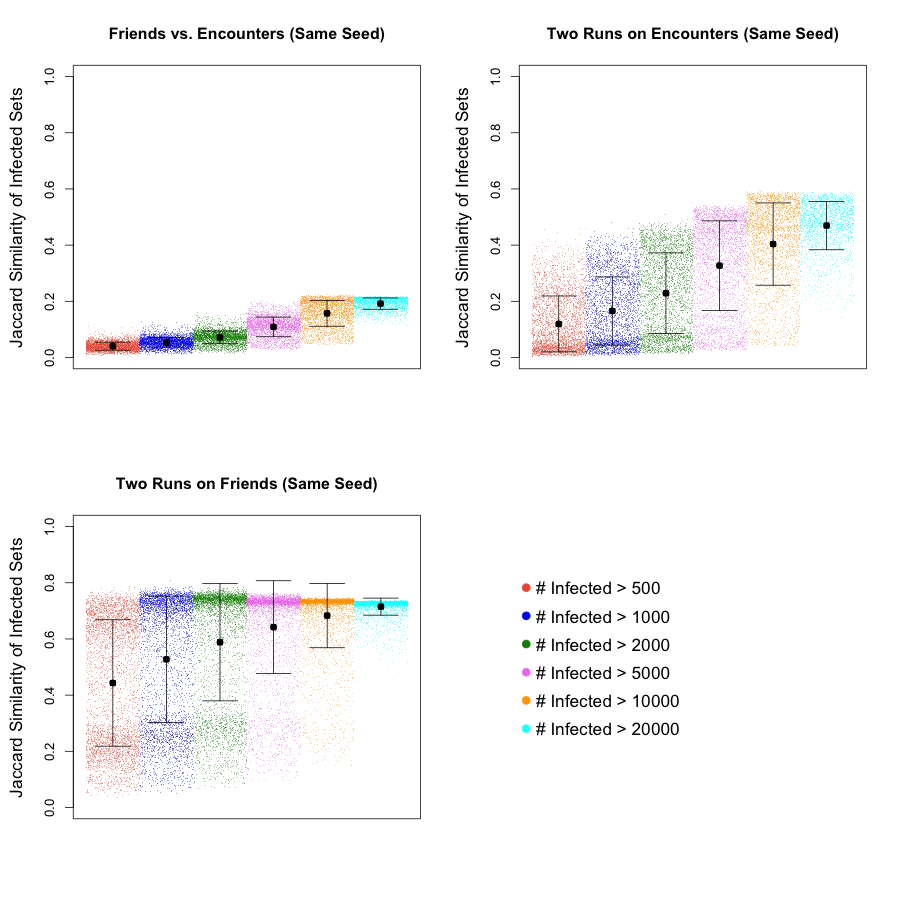}
\caption{\textbf{Single seed infection on time-varying networks -- Jaccard similarity.} The three panels show the metrics $J_{E_1,F_1}(m;s_i)$ (top-left), $J_{E_1,E_2}(m;s_i)$ (top-right) and $J_{F_1,F_2}(m;s_i)$ (bottom), for $5000$ random choices of a single seeds, and different values of the target set size $m$. For each seed, two simulations on the time-varying friendship network and two simulations on the encounter network are run separately.
The top-left panel considers, for each of the $5000$ seeds, a pair of simulations on the two different networks.
The top-right panel considers the $5000$ pairs of simulations ran on the encounter network.
The bottom panel considers the $5000$ pairs of simulations ran on the time-varying friendship network.
On the $x$- axis, observations for a given value of $m$ form a block with a constant color (within the block, the $x$ position is irrelevant).
We only consider pairs $(m,s_i)$ for which the metrics are defined.
For a given metric and each value $m$, the black point represents the average of the metric over all observations such that the metric is defined, and the bars represent standard deviations.
}
\label{fig:friVSenc_stochastic_jaccard}
\end{figure}

Figure~\ref{fig:friVSenc_stochastic_jaccard} plots the Jaccard similarity measures $J_{E_1,F_1}(m;s_i)$, $J_{E_1,E_2}(m;s_i)$, $J_{F_1,F_2}(m;s_i)$ in the top-left, top-right and bottom panels respectively.
Observations for a given value of $m$ constitute a block on the $x$-axis (larger values of $m$ correspond to $x$ positions on the right) and are represented with the same color.
For a fixed value of $m$, relative $x$ positions are irrelevant.
For a given metric and each value $m$, the black point represents the average of the metric over all observations such that the metric is defined, and the bars represent standard deviations.

For all values of $m$, two-sample t-tests support the hypotheses that $J_{E_1,F_1}(m;s_i)$ has smaller average than $J_{E_1,E_2}(m;s_i)$ and  $J_{F_1,F_2}(m;s_i)$, and that $J_{E_1,E_2}(m;s_i)$ has smaller average than $J_{F_1,F_2}(m;s_i)$ (p-values$<2.2\cdot 10^{-16}$).
A comparison between $J_{E_1,F_1}(m;s_i)$, $J_{E_1,E_2}(m;s_i)$, and $J_{F_1,F_2}(m;s_i)$ is not straightforward for the lack of an upper bound for $J_{E_1,F_1}(m;s_i)$.
There are $n_I=31,735$ nodes in the intersection of the time-varying friendship and encounter network and $n_U=142,967$ nodes in their union.
Therefore, for large values of target $m$, $J_{E_1,F_1}(m;s_i)$ is upper bounded by $n_I/n_U=0.2219$.
A bound that is independent of $s_i$ cannot be derived for general values of $m$, for which $J_{E_1,F_1}(m;s_i)$ is not constrained to have small values.
However, $J_{E_1,E_2}(m;s_i)$ and $J_{F_1,F_2}(m;s_i)$ can be as large as $1$ for all values of $m$.
As before, we consider rescaled versions of the Jaccard similarity.
Table~\ref{tab:J_upper_bound} reports the averages of the original and rescaled measures of Jaccard similarity.
Two-sample t-tests support the hypothesis that $\bar J_{E_1,F_1}(m;s_i)$ has a larger average than $\bar J_{E_1,E_2}(m;s_i)$ for $m\in\{500,1000,5000,10000,20000\}$ (p-values smaller that $0.0078$), whereas the null hypothesis of equal mean is not rejected for $m=2000$.
For all values of $m$, two-sample t-tests support the hypotheses that $\bar J_{E_1,F_1}(m;s_i)$ and $\bar J_{E_1,E_2}(m;s_i)$ have a smaller average than $\bar J_{F_1,F_2}(m;s_i)$ (p-values$<2.2\cdot 10^{-16}$).
The rescaled versions of the similarity measures suggest that the differences in local connectivity between the two networks play a major role in the inability of friendship to predict individuals at risk given a process driven by physical encounter.

\begin{table}
\centering
\caption{\textbf{Single seed infection on the time-varying networks.} Jaccard similarity measures: average of original measures, average of rescaled measures.}
\begin{tabular}{|c|c|c|c|c|c|c|}
\hline
$m$ &
$\langle\bar J_{E_1,F_1}(m)\rangle$ & $\langle\bar J_{E_1,E_2}(m)\rangle$ & $\langle\bar J_{F_1,F_2}(m)\rangle$ &
$\langle J_{E_1,F_1}(m)\rangle$ & $\langle J_{E_1,E_2}(m)\rangle$ & $\langle J_{F_1,F_2}(m)\rangle$\\
\hline
500  & 0.3656 & 0.2597 & 0.5636 & 0.0403 & 0.1194 & 0.4432 \\ 
1000  & 0.4177 & 0.3437 & 0.6526 & 0.0539 & 0.1655 & 0.5273 \\
2000  & 0.4425 & 0.4504 & 0.7377 & 0.0715 & 0.2287 & 0.5882 \\
5000  & 0.5449 & 0.5936 & 0.8390 & 0.1088 & 0.3270 & 0.6418 \\
10000 & 0.6978 & 0.6813 & 0.91091 & 0.1571 & 0.4037 & 0.6829 \\
20000 & 0.8765 & 0.7951 & 0.9668 & 0.19181 & 0.4695 & 0.7149 \\
\hline
\end{tabular}
\label{tab:J_upper_bound}
\end{table}

\begin{figure}
\centering
\includegraphics[scale=0.50]{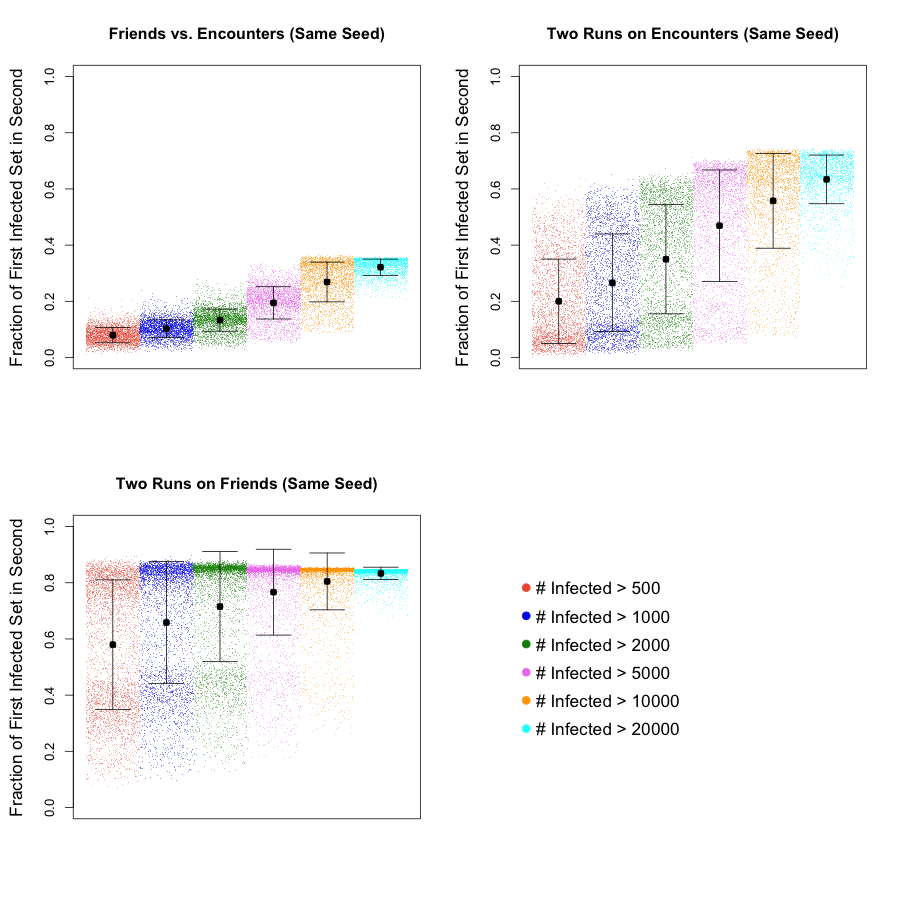}
\caption{\textbf{Single seed infection on time-varying networks -- Jaccard similarity.} The three panels show the metrics $P_{E_1,F_1}(m;s_i)$ (top-left), $P_{E_1,E_2}(m;s_i)$ (top-right) and  $P_{F_1,F_2}(m;s_i)$ (bottom), for $5000$ random choices of a single seeds, and different values of the target set size $m$. For each seed, two simulations on the time-varying friendship network and two simulations on the encounter network are run separately.
The top-left panel considers, for each of the $5000$ seeds, a pair of simulations on the two different networks.
The top-right panel considers the $5000$ pairs of simulations ran on the encounter network.
The bottom panel considers the $5000$ pairs of simulations ran on the time-varying friendship network.
On the $x$- axis, observations for a given value of $m$ form a block with a constant color (within the block, the $x$ position is irrelevant).
We only consider pairs $m$ of $s_i$ for which the metrics are defined.
For a given metric and each value $m$, the black point represents the average of the metric over all observations such that the metric is defined, and the bars represent standard deviations.
}
\label{fig:friVSenc_stochastic_fraction}
\end{figure}

Figure~\ref{fig:friVSenc_stochastic_fraction} plots the precision measures $P_{E_1,F_1}(m;s_i)$, $P_{E_1,E_2}(m;s_i)$, $P_{F_1,F_2}(m;s_i)$ in the top-left, top-right and bottom panels respectively.
Observations for a given value of $m$ constitute a block on the $x$-axis (larger $m$ correspond to $x$ positions on the right) and are represented with the same color.
For a fixed value of $m$, relative $x$ positions are irrelevant.
For a given metric and each value $m$, the black point represents the average of the metric over all observations such that the metric is defined, and the bars represent standard deviations.

Table~\ref{tab:P_upper_bound} reports the averages of the original and rescaled precision metrics.
For all values of $m$, two-sample t-tests support the hypotheses that both $P_{E_1,F_1}(m;s_i)$ and $P_{F_1,E_1}(m;s_i)$ have smaller average than both  $P_{E_1,E_2}(m;s_i)$ and  $P_{F_1,F_2}(m;s_i)$, and that $P_{E_1,E_2}(m;s_i)$ has smaller average than $P_{F_1,F_2}(m;s_i)$ (p-values$<2.2\cdot 10^{-16}$).
%
For all vales of $m$, two-sample t-tests support the hypothesis that $\bar P_{F_1,F_2}(m;s_i)$ has a larger average than all other precision measures.
For $m\in\{500,1000,20000\}$, two-sample t-tests support the hypotheses that $\bar P_{E_1,F_1}(m;s_i)$ and $\bar P_{F_1,E_1}(m;s_i)$ have larger average than $\bar P_{E_1,E_2}(m;s_i)$ (all p-values$<0.001$).
For $m\in\{500,1000,20000\}$, two-sample t-tests support the hypotheses that $\bar P_{E_1,F_1}(m;s_i)$ and $\bar P_{F_1,E_1}(m;s_i)$ have smaller average than $\bar P_{E_1,E_2}(m;s_i)$ (all p-values$<0.002$).
The null hypothesis that $\bar P_{E_1,F_1}(m;s_i)$ and $\bar P_{F_1,E_1}(m;s_i)$ have equal average is rejected only for $m\in\{500,1000\}$, for which the former has larger average (p-values$<1e-10$).
The rescaled versions of the precision measures stress the importance of the local connectivity properties between the two networks.

\begin{table}
\centering
\caption{\textbf{Single seed infection on the time-varying networks.} Precision measures: average of original and rescaled measures.}
\begin{tabular}{|c|c|c|c|c|c|c|c|c|}
\hline
$m$ &
$\langle\bar P_{E_1,F_1}(m)\rangle$ & $\langle\bar P_{F_1,E_1}(m)\rangle$ & $\langle\bar P_{E_1,E_2}(m)\rangle$ & $\langle\bar P_{F_1,F_2}(m)\rangle$ &
$\langle P_{E_1,F_1}(m)\rangle$ & $\langle P_{E_1,F_1}(m)\rangle$ & $\langle P_{E_1,E_2}(m)\rangle$ & $\langle P_{F_1,F_2}(m)\rangle$\\
\hline
500 & 0.3770 & 0.3996 & 0.3149 & 0.6484 & 0.07945 & 0.07527 & 0.2002 & 0.5797 \\
1000 & 0.4288 & 0.4610 & 0.4075 & 0.7339 & 0.1026 & 0.1010 & 0.2660 & 0.6585 \\
2000 & 0.4760 & 0.4775 & 0.5187 & 0.8067 & 0.1329 & 0.1325 & 0.3497 & 0.7150 \\
5000 & 0.5828 & 0.5840 & 0.6604 & 0.8837 & 0.1944 & 0.1945 & 0.4691 & 0.7665 \\
10000 & 0.7315 & 0.7308 & 0.7490 & 0.9395 & 0.2687 & 0.2688 & 0.5577 & 0.8047 \\
20000 & 0.8950 & 0.8951 & 0.853 & 0.9804 & 0.3213 & 0.3214 & 0.6339 & 0.8334 \\
\hline
\end{tabular}
\label{tab:P_upper_bound}
\end{table}

\FloatBarrier
\section{Epidemic risk: comparison between the static networks}
\label{yelp:sec:static_population}
In this section, we compare the friendship network with the static encounter network defined in Section~\ref{yelp:sec:other_networks}, in order to argue that our results are not driven by the static nature of the friendship network as opposed to the time-varying nature of the encounter network.
Also in this case, by comparing several independent runs of the infection process starting at each seed, we will observe that the unpredictability within a given network is substantially lower than the unpredictability between the two different networks.

We ran $10,000$ groups of simulations of the SI process with $\beta=0.01$ (stochastic infection).
For each group of simulations, a single seed is selected at random among all nodes $s_i$ in the intersection of the two networks ($u_I= \vert U_E\cap U_F \vert=71,211$).
For each choice of the seed, we separately run two infection processes on the static encounter network and two infection processes on the friendship network (denoted respectively by $E_1$, $E_2$, $F_1$, $F_2$).
For target set size $m\in\{500, 1000, 2000, 5000, 10000, 20000\}$ and each of the $10000$ seeds $s_i$, we consider the similarity and precision metrics defined above.
Observe that, as all nodes eventually become infected in a SI process on a static network, these quantities are defined for all choices of $s_i$ and $m\le n$ (where $n$ is the number of nodes in the network).

\begin{figure}
\centering
\includegraphics[scale=0.50]{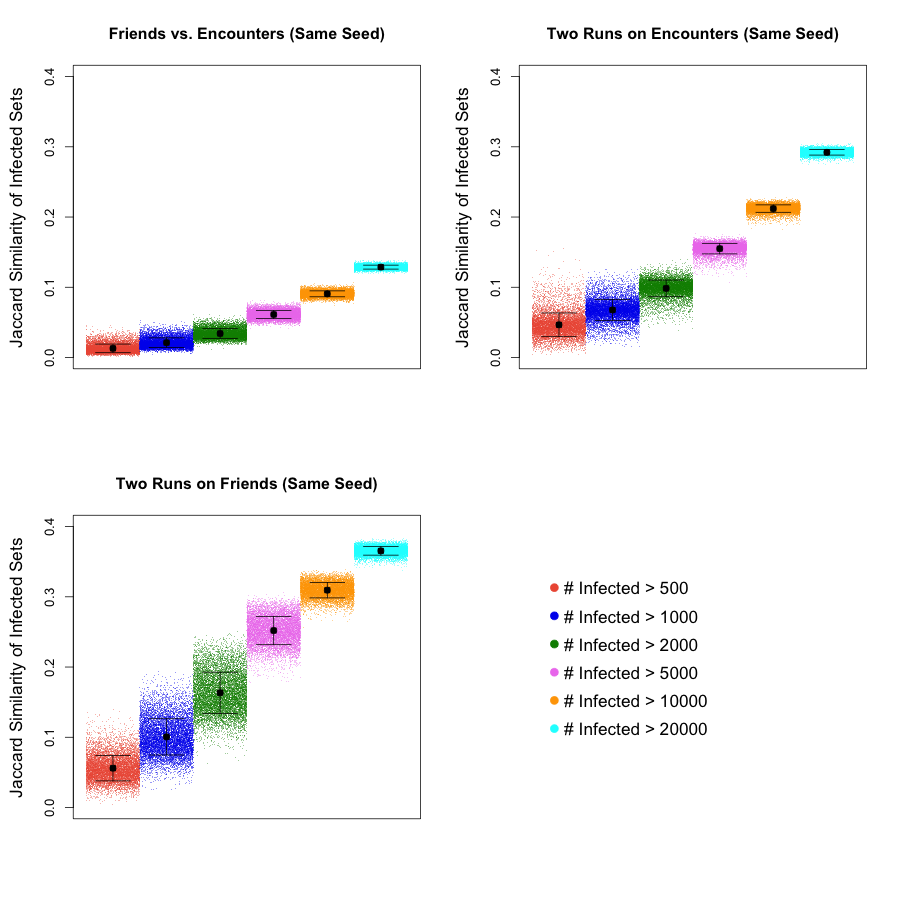}
\caption{\textbf{Single seed infection on the static networks -- Jaccard similarity.} The three panels show the metrics $J_{E_1,F_1}(m;s_i)$ (top-left), $J_{E_1,E_2}(m;s_i)$ (top-right) and $J_{F_1,F_2}(m;s_i)$ (bottom), for $10,000$ random choices of a single seeds, and different values of the target set size $m$. For each seed, two simulations on the friendship network and two simulations on the static encounter network are run separately.
The top-left panel considers, for each of the $10,000$ seeds, a pair of simulations on the two networks.
The top-right panel considers the $10,000$ pairs of simulations ran on the static encounter network.
The bottom panel considers the $10,000$ pairs of simulations ran on the friendship network.
On the $x$- axis, observations for a given value of $m$ form a block with a constant color (within the block, the $x$ position is irrelevant).
For a given metric and each value $m$, the black point represents the average of the metric over all the observations and bars represent standard deviations.
}
\label{fig:friVSenc_stochastic_jaccard_static}
\end{figure}

\begin{figure}
\centering
\includegraphics[scale=0.50]{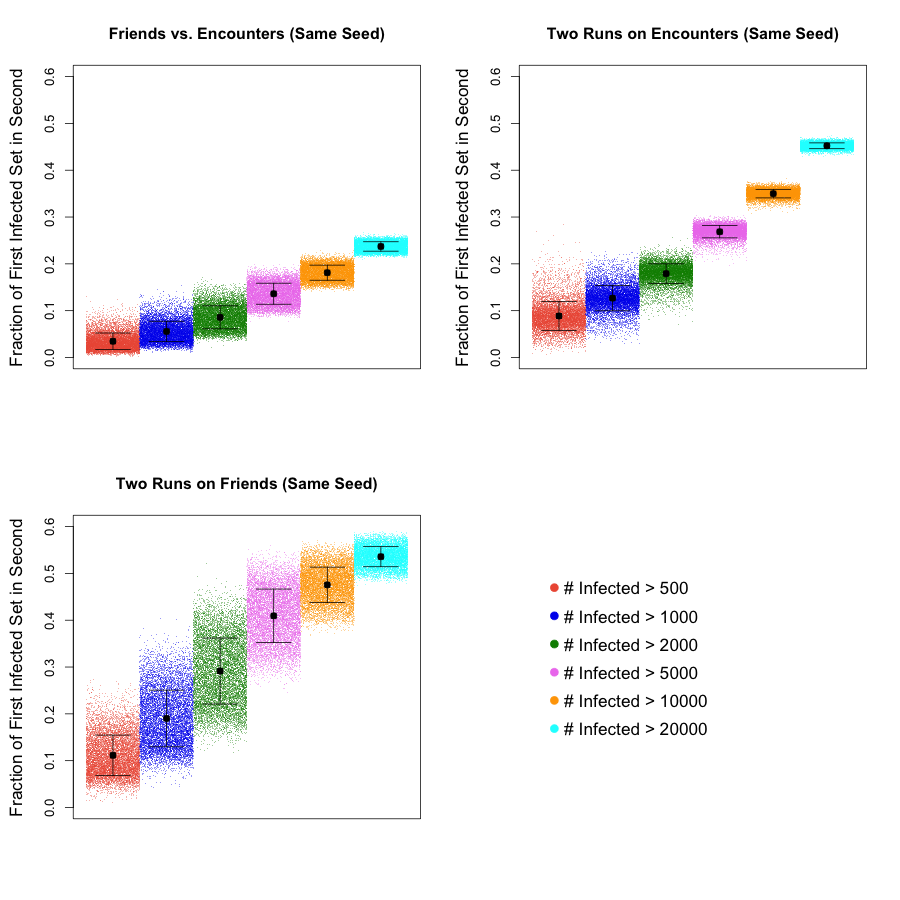}
\caption{\textbf{Single seed infection on the static networks -- precision.} The three panels show the metrics $P_{E_1,F_1}(m;s_i)$ (top-left), $P_{E_1,E_2}(m;s_i)$ (top-right) and  $P_{F_1,F_2}(m;s_i)$ (bottom), for $10,000$ random choices of a single seeds, and different values of the target set size $m$. For each seed, two simulations on the friendship network and two simulations on the static encounter network are run separately.
The top-left panel considers, for each of the $10,000$ seeds, a pair of simulations on the two networks.
The top-right panel considers the $10,000$ pairs of simulations ran on the static encounter network.
The bottom panel considers the $10,000$ pairs of simulations ran on the friendship network.
On the $x$- axis, observations for a given value of $m$ form a block with a constant color (within the block, the $x$ position is irrelevant).
For a given metric and each value $m$, the black point represents the average of the metric over all the observations and bars represent standard deviations.
}
\label{fig:friVSenc_stochastic_fraction_static}
\end{figure}

Figure~\ref{fig:friVSenc_stochastic_jaccard_static} plots the Jaccard similarity measures $J_{E_1,F_1}(m;s_i)$, $J_{E_1,E_2}(m;s_i)$, $J_{F_1,F_2}(m;s_i)$ in the top-left, top-right and bottom panels respectively.
Figure~\ref{fig:friVSenc_stochastic_fraction_static} plots the precision measures $P_{E_1,F_1}(m;s_i)$, $P_{E_1,E_2}(m;s_i)$, $P_{F_1,F_2}(m;s_i)$ in the top-left, top-right and bottom panels respectively.
Observations for a given value of $m$ constitute a block on the $x$-axis (larger $m$ corresponds to $x$ positions on the right) and are represented with the same color.
For a fixed value of $m$, relative $x$ positions are irrelevant.
For a given metric and each value $m$, the black point represents the average of the metric over all the observations and bars represent standard deviations.

$J_{E_1,F_1}(m;s_i)$ has smaller average than  $J_{E_1,E_2}(m;s_i)$, $J_{F_1,F_2}(m;s_i)$, and for $m>500$, $J_{E_1,E_2}(m;s_i)$ has larger average than $J_{F_1,F_2}(m;s_i)$ (two-paired t-tests, p-values$<2.2\cdot 10^{-16}$).
Similarly, $P_{E_1,F_1}(m;s_i)$ and $P_{F_1,E_1}(m;s_i)$ have smaller average than $P_{E_1,E_2}(m;s_i)$, $P_{F_1,F_2}(m;s_i)$, and for $J_{E_1,E_2}(m;s_i)$ has smaller average than $J_{F_1,F_2}(m;s_i)$ (two-paired t-tests, p-values$<2.2\cdot 10^{-16}$).

As before, it is not straightforward to rigorously compare the quantities for all values of $m$.
The metrics $J_{E_1,E_2}(m;s_i)$, $J_{F_1,F_2}(m;s_i)$, $P_{E_1,E_2}(m;s_i)$ and $P_{F_1,F_2}(m;s_i)$ can be as large as $1$ for all values of $m$.
Instead, for large $m$, $J_{E_j,F_k}(m;s_i)$ is upper bounded by $u_I/u_U=0.338$,
$P_{E_j,F_k}(m;s_i)$ is upper bounded by $u_I/u_E=0.629$, and $P_{F_j,E_k}(m;s_i)$ is upper bounded by $u_I/u_F=0.422$.
For general values of $m$, tight upper bounds for these quantities depend on $s_i$ and therefore on the network structure.
Therefore, we consider the rescaled version of the similarity and precision measures defined above.

Table~\ref{tab:J_upper_bound_static} reports the averages of the original and rescaled Jaccard similarity measures.
Table~\ref{tab:P_upper_bound_static} reports the averages of the original and rescaled precision measures.
For all values of $m$, $\bar J_{E_1,F_1}(m;s_i)$ has smaller average than  $\bar J_{E_1,E_2}(m;s_i)$ and $\bar J_{F_1,F_2}(m;s_i)$, and for $m>500$, $\bar J_{E_1,E_2}(m;s_i)$ has larger average than $\bar J_{F_1,F_2}(m;s_i)$ (two-sample t-tests, p-values$<2.2\cdot 10^{-16}$).
For all values of $m$, $\bar P_{E_1,F_1}(m;s_i)$ has smaller average than $\bar P_{E_1,E_2}(m;s_i)$ and $\bar P_{F_1,F_2}(m;s_i)$, whereas
$\bar P_{F_1,E_1}(m;s_i)$ has smaller average than  $\bar P_{F_1,F_2}(m;s_i)$ for $m\in\{500, 1000, 2000, 5000\}$ and larger for $m\in\{10000, 20000\}$  (two-sample t-tests, p-values$<2.2\cdot 10^{-16}$).
The rescaled measures suggest that the network structure has a large impact on the spread of the infection between the friendship and static encounter networks.

\begin{table}
\centering
\caption{\textbf{Single seed infection on the static networks.} Jaccard similarity measures: empirical upper bounds, average of original measures, average of the rescaled measures.}
\begin{tabular}{|c|c|c|c|c|c|c|c|c|c|}
\hline
$m$ &
$\langle\bar J_{E_1,F_1}(m)\rangle$ & $\langle\bar J_{E_1,E_2}(m)\rangle$ & $\langle\bar J_{F_1,F_2}(m)\rangle$ &
$\langle J_{E_1,F_1}(m)\rangle$ & $\langle J_{E_1,E_2}(m)\rangle$ & $\langle J_{F_1,F_2}(m)\rangle$\\
\hline
500 & 0.28350 & 0.3004 & 0.4029 & 0.01296 & 0.04653 & 0.05597 \\
1000 & 0.4047 & 0.5387 & 0.5175 & 0.02113 & 0.06772 & 0.1005 \\
2000 & 0.5531 & 0.7045 & 0.6509 & 0.03415 & 0.09841 & 0.1633 \\
5000 & 0.7493 & 0.8779 & 0.8234 & 0.06123 & 0.1550 & 0.2519 \\
10000 & 0.8521 & 0.9310 & 0.9116 & 0.09064 & 0.2120 & 0.30944 \\
20000 & 0.9290 & 0.9568 & 0.9527 & 0.1286 & 0.29213 & 0.36542 \\
\hline
\end{tabular}
\label{tab:J_upper_bound_static}
\end{table}

\begin{table}
\centering
\caption{\textbf{Single seed infection on the static networks.} Precision measures: average of original and rescaled measures.}
\begin{tabular}{|c|c|c|c|c|c|c|c|c|}
\hline
$m$ &
$\langle\bar P_{E_1,F_1}(m)\rangle$ & $\langle\bar P_{F_1,E_1}(m)\rangle$ & $\langle\bar P_{E_1,E_2}(m)\rangle$ & $\langle\bar P_{F_1,F_2}(m)\rangle$ &
$\langle P_{E_1,F_1}(m)\rangle$ & $\langle P_{E_1,F_1}(m)\rangle$ & $\langle P_{E_1,E_2}(m)\rangle$ & $\langle P_{F_1,F_2}(m)\rangle$\\
\hline
500 & 0.2672 & 0.2500 & 0.3122 & 0.40680 & 0.03462 & 0.02112581 & 0.08868 & 0.1114 \\
1000 & 0.3592 & 0.3707 & 0.5596 & 0.4717 & 0.05561 & 0.03419 & 0.12674 & 0.1901 \\
2000 & 0.5034 & 0.4567 & 0.7011 & 0.5584 & 0.08577 & 0.05527 & 0.1791 & 0.2914 \\
5000 & 0.6867 & 0.6927 & 0.8838 & 0.7317 & 0.1362 & 0.1014 & 0.2685 & 0.4093 \\
10000 & 0.7889 & 0.8461 & 0.9167 & 0.8330 & 0.1811 & 0.1543 & 0.3500 & 0.4755 \\
20000 & 0.8853 & 0.9207 & 0.9553 & 0.9083 & 0.2371 & 0.2198 & 0.4521 & 0.5356 \\
\hline
\end{tabular}
\label{tab:P_upper_bound_static}
\end{table}

\FloatBarrier
\section{Overcoming the limits of the friendship networks: correction}
\label{yelp:sec:window}

In the previous sections, in order to evaluate the friendship network as a predictor of epidemic risk on the encounter network, we initiated epidemic processes at a seed present on both networks and let them spread independently on the two networks.
This corresponds to a case in which the researcher has access neither to the contacts between individuals nor to the infected population (on the encounter network) and relies exclusively on the information provided by the friendship network.
In this section, we consider a less extreme scenario in which the researcher has still knowledge of the friendship network, but, in addition, is able to monitor the infected population (on the encounter network) at given times.
In such a situation, the infection propagation can be predicted according to the friendship network as long as information about the real infected population is unavailable.
When such information becomes available, the estimated set of infected individuals (on the friendship network) can be updated to the real set of infected individuals (on the encounter network).
As we show below, the ability to monitor the infection over time and correct the set of infected individuals overcomes the limits of the friendship networks in predicting epidemic risk highlighted in the previous sections.
In particular, we compare the sets of infected individuals on the two networks right before each correction and show that a good level of prediction accuracy is established early in the process and maintained over time.
Despite the level of accuracy decreases with larger window size, even relatively infrequent correction overcomes the limits of the friendship networks in predicting epidemic risk.

We proceed as follow.
Given a seed $s$ the is present in both the encounter and the friendship network, we consider two SI processes spreading on the two networks.
Let $\I^E(t)$ and $\I^F(t)$ be the sets of infected nodes on the two networks at time $t$, and let $I^E(t)$ and $I^F(t)$ be their cardinality.
We have that $\I^E(0)=\I^F(0)=\{s\}$.
We assume that every $W$ time steps the set $\I^E(t)$ is available and therefore $\I^F(t)$ can be corrected accordingly.
That is, we consider a ``corrected'' version of the infection process on the friendship network, whose set of infected nodes satisfies the relationship
$$
\I^F(kW) = \I^E(kW), \text{ for each } k>0.
$$ 
Between time $kW$ and $(k+1)W-1$ the set $\I^F(t)$ grows according to the ties of the friendship network.

We are interested in comparing the sets $\I^E(t)$ and $\I^F(t)$ at times $t=kW-1$, that is, right before each correction.
Let 
$$
J_{E,F}(k;s,W) = \frac{\I^E(kW-1) \cap \I^F(kW-1)}{\I^E(kW-1) \cup \I^F(kW-1)},
$$
be the Jaccard similarity of the infected sets on the two networks right before a correction.
Similarly, let
Let 
\begin{align*}
P_{F,E}(k;s,W) &= \frac{\I^E(kW-1) \cap \I^F(kW-1)}{\I^F(kW-1)},\\
P_{E,F}(k;s,W) &= \frac{\I^E(kW-1) \cap \I^F(kW-1)}{\I^E(kW-1)}.
\end{align*}
$P_{F,E}(k;s,W)$ represents the fraction of infected nodes before a correction on the friendship network that are also infected in the encounter network (precision).
$P_{E,F}(k;s,W)$ represents the fraction of infected nodes in the encounter network which were correctly predicted to be infected before a correction on the friendship network (recall).
In addition we consider the relative size of the infected sets on the two networks,
$$
r_{E,F}(k;s,W) = \frac{I^F(kW-1)}{\I^E(kW-1)},
$$
which compares the two infection from a more coarse point of view.
All quantities above depend on the window size $W$.

For window size $W\in\{10,20,50\}$, we ran $6000$ groups of simulations of the SI process with $\beta_F=0.0001$ on the friendship network and $\beta_{E}=0.5$ on the encounter network (we allow for different infection rates on the two network in order to compensate for their different degree distributions).
For each group of simulations, a single seed is selected uniformly at random among all nodes $s\in U_F\cap U_E$ (present in both networks) such that $t_0(s_i)\le 900$ (that is, we consider nodes that have an encounter by time $t=900$).
For each choice of the seed, we run one infection process on the encounter network for $T=500$ time steps (that is, from $t=t_0(s_i)$ to $t=t_0(s_i)+500$).
On the friendship network, the infection process is initiated at the same seed $s_i$ and spreads according to the ties of the friendship network for $T=500$ time steps (at each time $t=kW$, it is set $\I^F(kW)=\I^E(kW)$).

\begin{figure}
\centering
\includegraphics[scale=0.25]{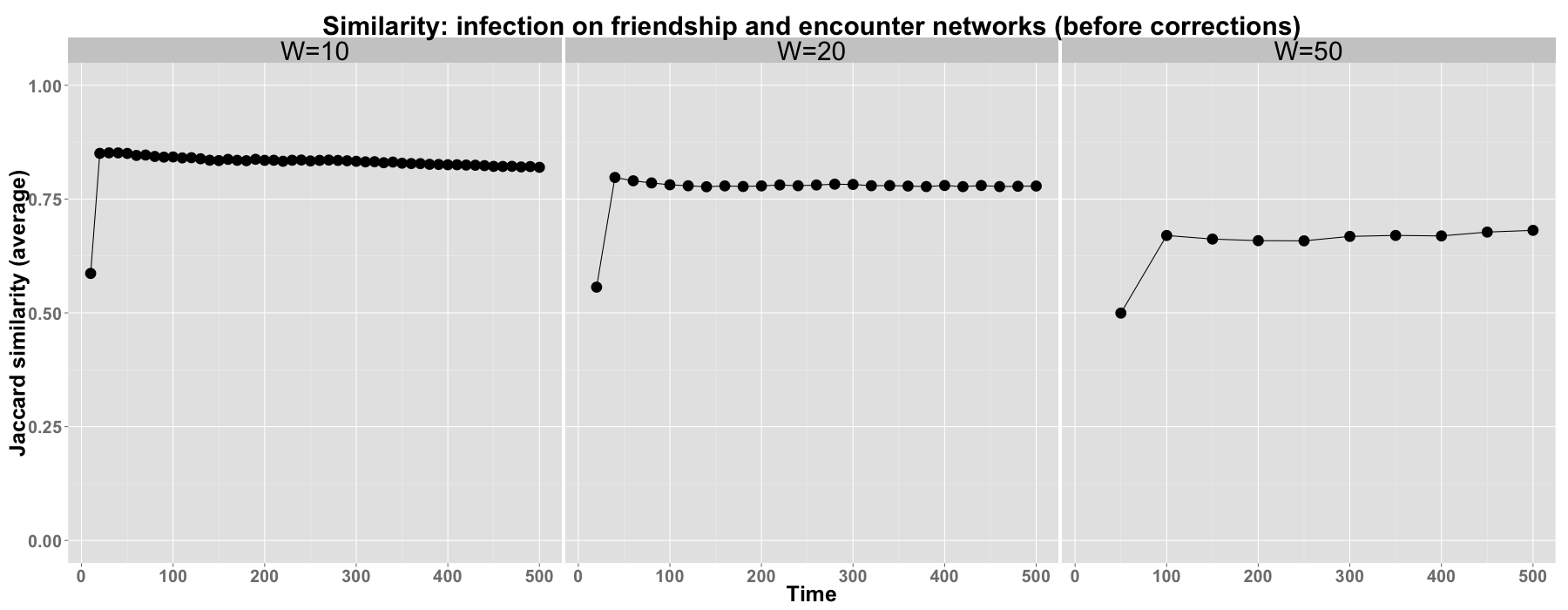}
\caption{Jaccard similarity of infected sets before corrections.
The $x$-axis shows time.
The $y$-axis shows the measure $J_{E,F}(k;s,W)$ averaged over $6000$ pairs of simulations
(each associated to an independent choice of the seed).
Subplots consider different window sizes $W$.}
\label{fig:window_jaccard}
\end{figure}

\begin{figure}
\centering
\includegraphics[scale=0.25]{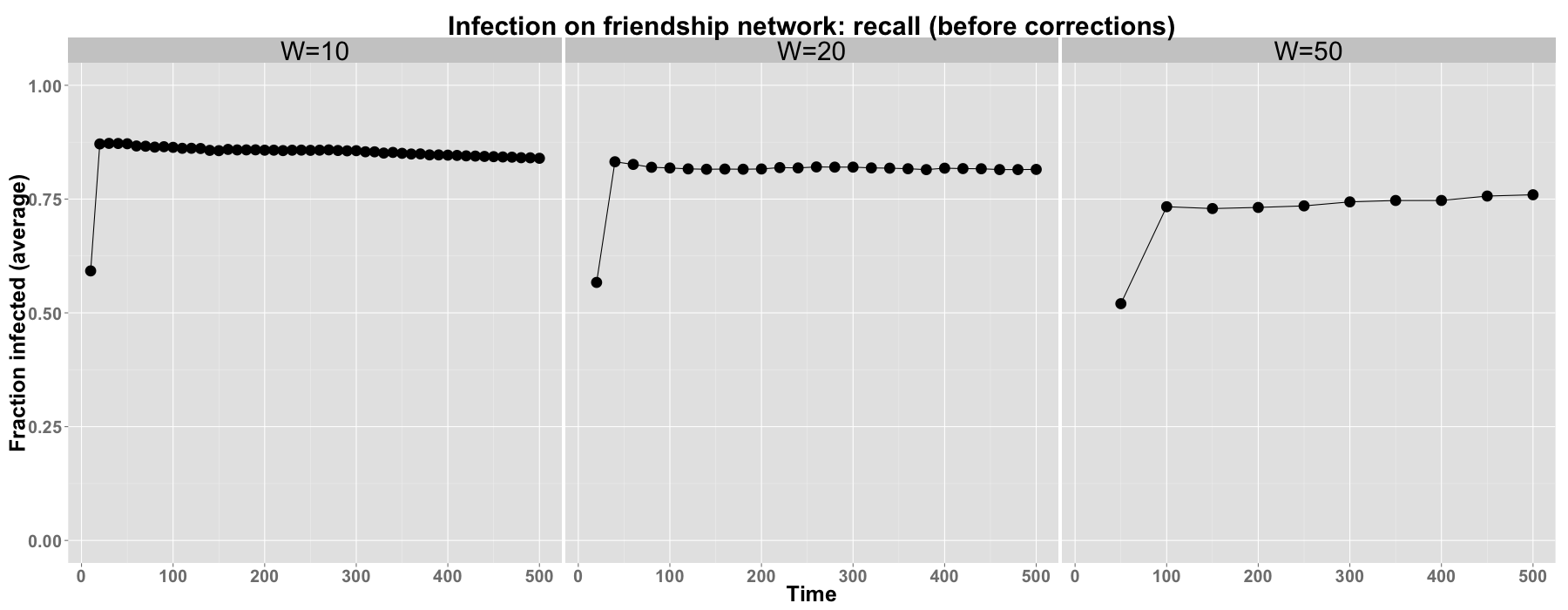}
\caption{Fraction of infected nodes in the encounter network that are predicted to be infected in the friendship network, before corrections.
The $x$-axis shows time.
The $y$-axis shows the measure $P_{E,F}(k;s,W)$ averaged over $6000$ pairs of simulations
(each associated to an independent choice of the seed).
Subplots consider different window sizes $W$.}
\label{fig:window_enc_prec}
\end{figure}

\begin{figure}
\centering
\includegraphics[scale=0.25]{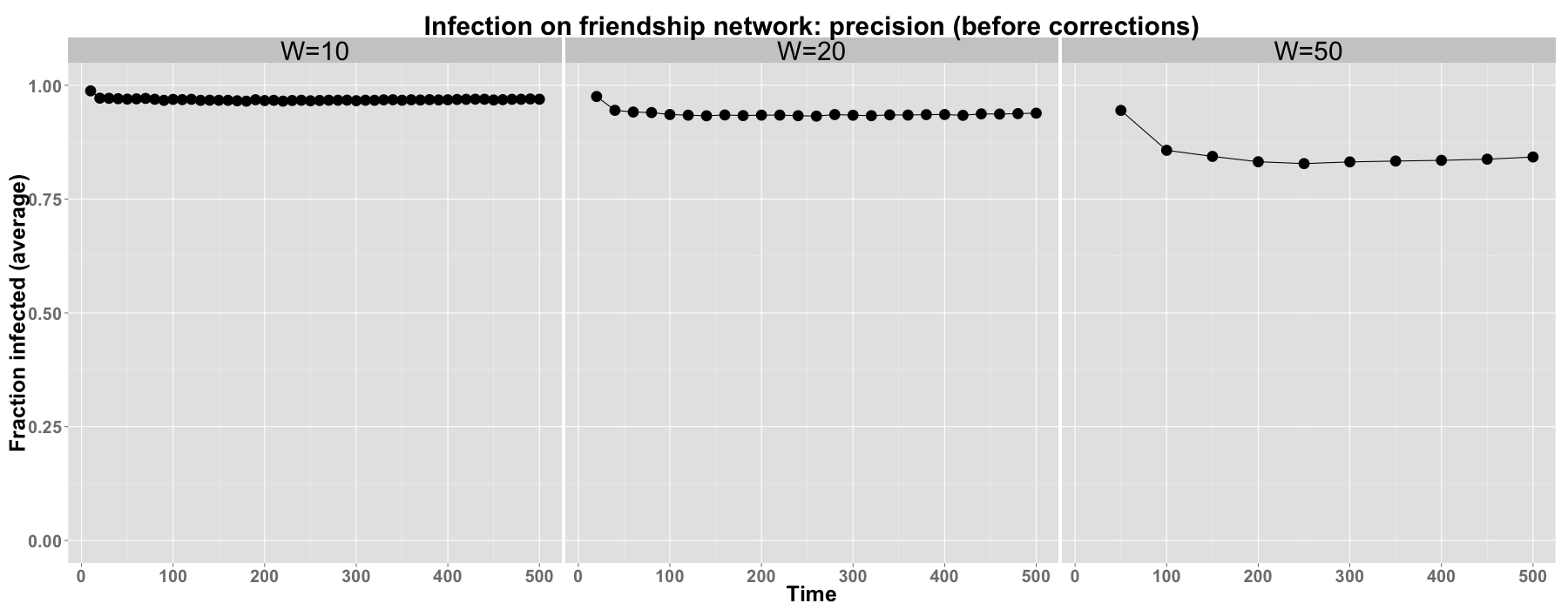}
\caption{Fraction of predicted infected nodes on the friendship network that are infected in the encounter network, before corrections.
The $x$-axis shows time.
The $y$-axis shows the measure $P_{F,E}(k;s,W)$ averaged over $6000$ pairs of simulations
(each associated to an independent choice of the seed).
Subplots consider different window sizes $W$.}
\label{fig:window_fri_prec}
\end{figure}

\begin{figure}
\centering
\includegraphics[scale=0.25]{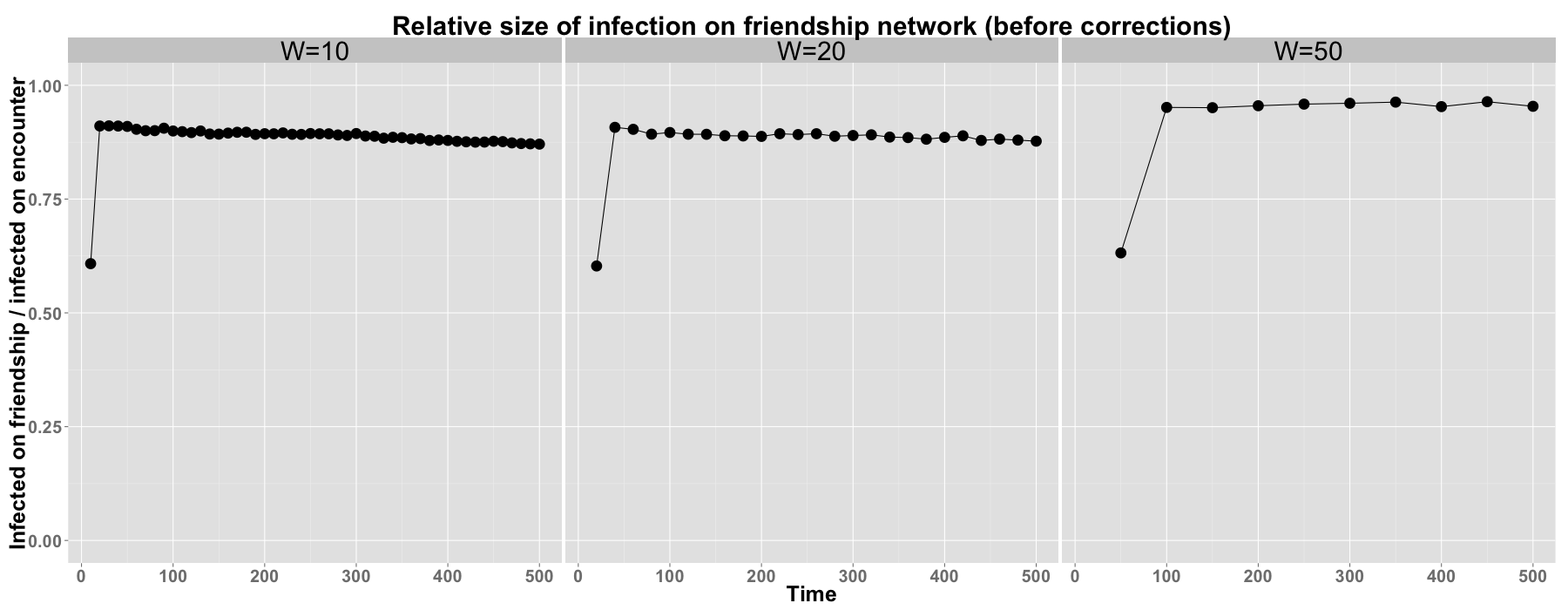}
\caption{Relative size of infected sets before corrections.
The $x$-axis shows time.
The $y$-axis shows the measure $r_{E,F}(k;s,W)$ averaged over $6000$ pairs of simulations
(each associated to an independent choice of the seed).
Subplots consider different window sizes $W$.}
\label{fig:window_rel_size}
\end{figure}

Figures~\ref{fig:window_jaccard} to~\ref{fig:window_rel_size} show the average of the defined metrics over all simulations as a function of time and for all choices of $W$.
Note that, as each infection process is run for $T=500$ time steps, the number of corrections (and therefore the number of points in the plots) depends on the choice of $W$ and equals $T/W$.
The plots show that the ability to periodically observe the infected sets on the encounter network (that is, to correct the set $\I^F(kW)$ at each time window) overcomes the limitations of the friendship network in predicting epidemic risk that was highlighted in the previous sections.
Interestingly, good accuracy of the prediction (through the friendship network) emerges early in the process (after the first correction) and is maintained over time with relatively few observations (with only slight degrade or improvement over time).
The accuracy decreases with larger window size.
However, even the largest considered window size ($W=50$) guarantees a good prediction accuracy that slowly increases over time.
The particular value of the obtained results might partially depend on the choice of the infection rates on the two networks.

In order to compare window sizes $W=10$ and $W=20$, we consider all time steps corresponding to a correction for both choices of $W$ and ignore the first correction (i.e., we consider times $20k$ for $k>1$).
The trend of the average of the Jaccard similarity $J_{E,F}(k;s,W)$ with respect to time $t$ and window size $W$ is captured by a linear relationship.
OLS with interaction between $t$ and $W$ shows that the Jaccard similarity is lower in the case of $W=20$ then $W=10$ ($-0.0623$, p-value$<2\cdot 10^{-16}$) and slowly decreases over time ($-1.126\cdot 10^{-3}$ every $20$ time steps for $W=10$, p-value$=1.13\cdot 10^{-15}$, $-3.73\cdot 10^{-3}$ every $20$ time steps for $W=20$, p-value$=7.77\cdot 10^{-7}$).
Similarly, the trend of the average of the precision measure $P_{E,F}(k;s,W)$ with respect to time $t$ and window size $W$ is captured by a linear relationship.
OLS with interaction between $t$ and $W$ shows that the Jaccard similarity is lower in the case of $W=20$ ($-0.0478$, p-value$<2\cdot 10^{-16}$) and slowly decreases over time ($-1.117\cdot 10^{-3}$ every $20$ time steps for $W=10$, p-value$<2\cdot 10^{-16}$, $-3.04\cdot 10^{-4}$ every $20$ time steps for $W=20$, p-value$=2.67\cdot 10^{-9}$).
The average of the precision measure $P_{F,E}(k;s,W)$ is lower in the case of $W=20$ ($-0.0308$, p-value$<2\cdot 10^{-16}$) and present not statistically significant trend with respect to the time $t$.
The trend of the average of size ratio $r_{E,F}(k;s,W)$ with respect to time $t$ and window size $W$ is captured by a linear relationship.
OLS with interaction between $t$ and $W$ shows that the ratio is lower in the case of $W=20$ ($-7.66\cdot 10^{-3}$, p-value$=8.4\cdot 10^{-4}$) and slowly decreases over time ($-1.45\cdot 10^{-3}$ every $20$ time steps for $W=10$, p-value$<2\cdot 10^{-16}$, $-8.65\cdot 10^{-4}$ every $20$ time steps for $W=20$, p-value$=1.52\cdot 10^{-4}$).

In order to compare all window sizes $W\in\{10,20,50\}$, we consider all time steps corresponding to a correction for all choices of $W$ and ignore the first correction (i.e., we consider times $100k$ for $1\le k\le 5$).
The trends of the average of all defined measures with respect to time $t$ and window size $W$ are captured by a linear relationships.
In the case of Jaccard similarity $J_{E,F}(k;s,W)$, the metric is lower in the case of $W=50$ ($-0.188$ with respect to $W=10$, p-value$=2.74\cdot 10^{-10}$), value for which it increases over time ($3.29\cdot 10^{-3}$ every $100$ time steps, p-value$=1.27\cdot 10^{-3}$).
Similar trends as the ones above are found in the case of the precision measures $P_{E,F}(k;s,W)$ and $P_{F,E}(k;s,W)$.
In the case of the relative size of infected sets $r_{E,F}(k;s,W)$, the largest window size results in a more accurate prediction of the size of the infected set over time ($+0.049$ with respect to $W=10$, p-value$=2.55\cdot 10^{-5}$).

\FloatBarrier
\section{Containment of epidemic outbreaks using the friendship network}
\label{yelp:sec:immunization}

In this section, we show that the friendship network encodes useful information for the containment of epidemic outbreaks.
We consider a scenario in which a fixed budget is available for immunization, corresponding to the number of individuals that can be made immune to the infection.
This budget might represent the total amount of vaccine that is available.
Immune individuals do not get infected and do not infect other individuals (i.e., according to our framework, they are removed from the network).
Our goal is to spend the budget in an effective way, in order to contain the spread of the disease.
A simple, straightforward immunization strategy is to select individuals at random (random immunization).
This method is unlikely to target the most connected individuals and can result in inefficient allocation of the immunization budget.
We propose the strategy of selecting random friends of randomly chosen individuals (friend immunization).
Such strategy is motivated by the ``friendship paradox'', the network property for which the average friend of an individual is more connected than the average individual~\cite{feld1991your}, and has been proposed to predict the peak of an epidemic outbreak~\cite{christakis2010social} and the spread of information online~\cite{garcia2014using}.
Instead of selecting individuals for immunization at random, the method first selects random individuals and then gives immunization to a random friend of each selected individual, according to the friendship network.
The method is simple, as its implementation only requires individuals to name a friend, and is able to target individuals who are more connected on average.
In addition, we consider another benchmark, in which immunization is given to encounters of random individuals (encounter immunization).
This latter method is similar to the one just described (but selects individuals for immunization according to the static version of the encounter network rather than the friendship network) but requires knowledge of the encounters between individuals, that might be unavailable for the reasons discussed in the introduction.
However, given its potential to identify individuals who have a large number of encounters, it represents an upper bound for the capability of outbreak containment.
We do not consider more sophisticated methods that require the computation of quantities such as nodes degree or centrality.

We consider infection processes spreading on the encounter network
and an immunization budget $b$ representing the percentage of individuals who can receive immunization.
We refer to $b$ as the immunization rate.
The sets of immune individuals depend on the immunization method and on the randomness of the selection of individuals, friends and encounters.
Let $X_R(b)$, $X_F(b)$, $X_E(b)$ be respectively three immunization sets obtained with the three described methods (random, friend and encounter immunization).
In the implementation, we guarantee that the three sets have the same cardinality.
Obtaining sets of the same cardinality might require sampling more individuals in the case of friend and encounter immunization than random immunization (e.g., the same friend might be named multiple times).
However, we don't consider sampling as a cost and we focus our attention on the immunization rate.

We consider a wide range of immunization rates, $b\in\{1\%,2\%,5\%,10\%,15\%\}$ and compare them to the case of no immunization ($b=0\%$).
For each value of $b$, we run $5000$ groups of three simulations.
For each group of simulations, a seed $s_i$ such that $t_0(s_i)\le 500$ is selected uniformly at random (that is, we consider nodes that have an encounter by time $t=500$).
Then, three immunization sets $X_R(b,s_i)$, $X_F(b,s_i)$, $X_E(b,s_i)$ are built according to the three methods (with the constraint that the seed $s_i$ cannot receive immunization).
Then, three independent SI processes are initiated at $s_i$ and spread on the encounter network.
In the first process (denoted by $R$), individuals in $X_R(b;s_i)$ are immune to the infection.
In the second process (denoted by $F$), individuals in $X_F(b;s_i)$ are immune to the infection.
In the third process (denoted by $E$), individuals in $X_E(b;s_i)$ are immune to the infection.
Let
$$
r_R(b,s_i),\quad r_F(b,s_i),\quad r_E(b,s_i),
$$
be the final infection rates of the three processes, respectively.

\begin{figure}
\centering
\includegraphics[scale=0.4]{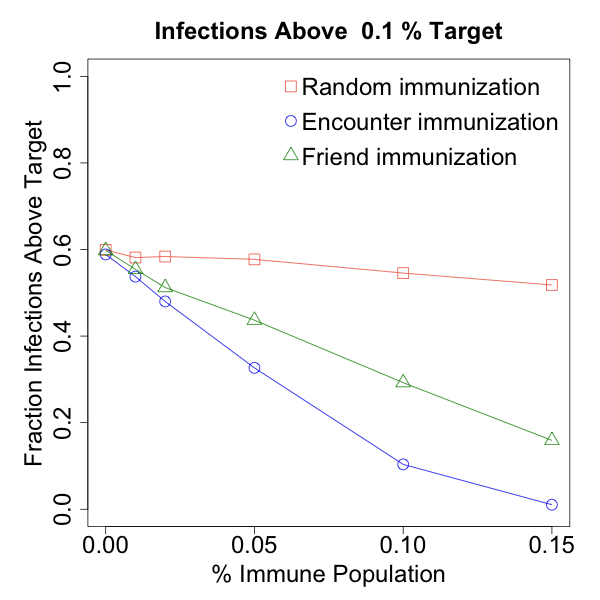}
\caption{Fraction of infections above $1\%$ final infection rate as a function of immunization rate and immunization method (random immunization: red squares, encounter immunization: blue circles, friend immunization: green triangles).
The $x$-axis shows the immunization rate $b$ (the fraction of immune individuals).
The $y$-axis shows the fraction of infections above the $1\%$ target ($5000$ simulations for each immunization method and value of $b$).
}
\label{fig:immunization_above1percent}
\end{figure}

Figure~\ref{fig:immunization_above1percent} shows the fraction of infections with final infection rate above $0.1\%$ as a function of the immunization rate $b$ and for all considered immunization methods (random immunization: red squares, encounter immunization: blue circles, friend immunization: green triangles).
We consider a $0.1\%$ target for the final infection rate as an indicator that the infection did not die out.
In the case of no immunization ($b=0\%$), we observe that only $60\%$ of infections hit the $0.1\%$ target.
The the remaining $40\%$ correspond to infections that die out in their early stage.
In the case of random immunization, the fraction of infections that die out is not very sensible to the immunization rate.
In both cases of friend immunization and encounter immunization, increasing the immunization rate substantially increases the fraction of infections that die out, suggesting that both methods are effective at preventing outbreaks.
The effect is stronger in the case of encounter immunization.
However, friend immunization provides a comparatively similar effect to encounter immunization, and a substantial improvement with respect to random immunization.
The trend in Figure~\ref{fig:immunization_above1percent} is captured by a linear model that considers the interaction between immunization type and immunization rate. 
In the case of random immunization, each $1\%$ increase of the immunization rate determines a $0.5\%$ decrease in the fraction of infections above the $0.1\%$ target (p-value$=0.0299$).
In the case of friend immunization, each $1\%$ increase of the immunization rate determines an additional $3.5\%$ (with respect to random immunization) decrease in the fraction of infections above the $0.1\%$ target (p-value$=4.03\cdot 10^{8}$).
In the case of encounter immunization, each $1\%$ increase of the immunization rate determines an additional $2.36\%$ (with respect to random immunization) decrease in the fraction of infections above the $0.1\%$ target (p-value$=2.77\cdot 10^{6}$).

\begin{figure}
\centering
\includegraphics[scale=0.3]{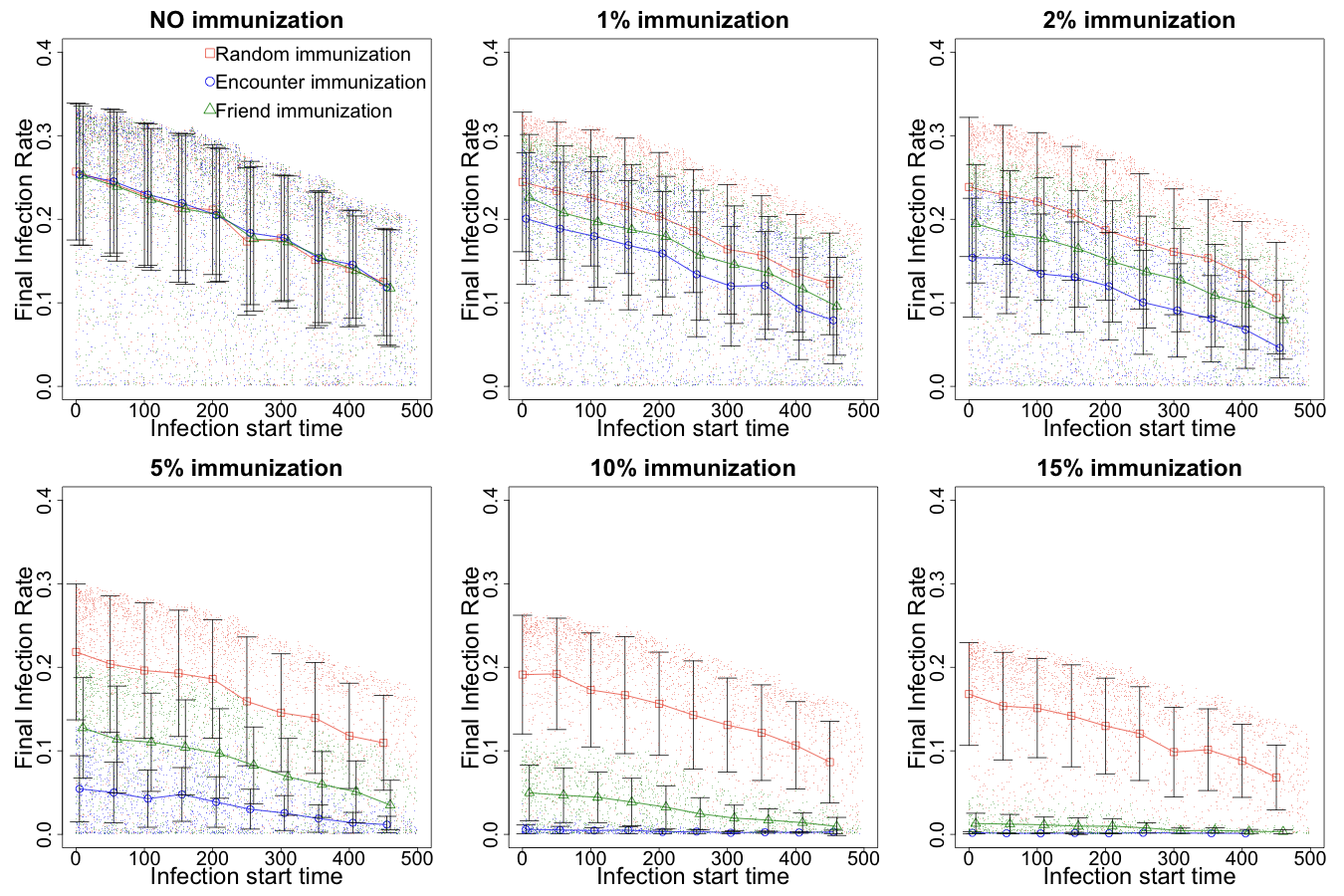}
\caption{Average final infection rate over infections that do not die out (i.e., final infection rate above $0.1\%$) as a function of the infection start time $t_0(s_i)$, for all immunization methods (random immunization: red squares, encounter immunization: blue circles, friend immunization: green triangles) and immunization rates $b$ (subplots)
The $x$-axis shows $t_0(s_i)$.
The $y$-axis shows average final infection rate.
Bars represent standard errors.
}
\label{fig:immunization_final}
\end{figure}

Figure~\ref{fig:immunization_final} shows the average final infection rate among all infections that do not die out (according to the $0.1\%$ target considered above) as a function of the infection start time $t_0(s_i)$ (i.e., the first time in which the seed is connected in the encounter network, grouped into bins of width equal to $50$ time steps), for all immunization methods (random immunization: red squares, encounter immunization: blue circles, friend immunization: green triangles) and immunization rates (subplots).
For each value of the immunization rate $b$, friend immunization provides a substantial reduction of the average final infection rate with respect to random immunization.
Encounter immunization results in the lowest infection rates.
To analyze the trends in Figure~\ref{fig:immunization_final}, we fit separate models to each subset of simulations with a given immunization rate $b$, as each $b$ results in a different number of infections above the $0.1\%$ target (see Figure~\ref{fig:immunization_above1percent}).
For example, in the case of $b=1\%$, friend immunization results in an average final infection rate $2.4\%$ lower than random immunization (p-value$<2.2\cdot 10^{16}$),
and encounter immunization results in an average final infection rate $4.4\%$ lower than random immunization (p-value$<2.2\cdot 10^{16}$).
A model considering the interaction of immunization type and infection start time $t_0(s_i)$ shows similar reduction effects of friend and encounter immunization as above (respectively $-2.45\%$, p-value$<2.14\cdot 10^{12}$, and $-4.65\%$ p-value$<2.2\cdot 10^{16}$) and a decreasing final infection rate with respect to $t_0(s_i)$ ($-0.027\%$ for each time step of delay, p-value$<2.2\cdot 10^{16}$), but slopes do not depend on the immunization type.
Analyses have a similar flavor for the different choices of the immunization rate $b$, and the infection containment effect of both friendship and encounter immunization increases for larger $b$ (fixed effects of linear models).
In addition, for larger $b$, the decrease of the average final infection rate with respect to $t_0(s_i)$ is less steep in the case of both friendship and encounter immunization than random immunization.
Interestingly, encounter immunization results in an almost null average final infection rate for immunization rate $b=10\%$,
and the same is obtained in the case of friend immunization for immunization rate $b=15\%$.
This highlights the effectiveness of friend immunization, which is able to obtain the same effect as encounter immunization at a small additional cost.

\begin{figure}
\centering
\includegraphics[scale=0.3]{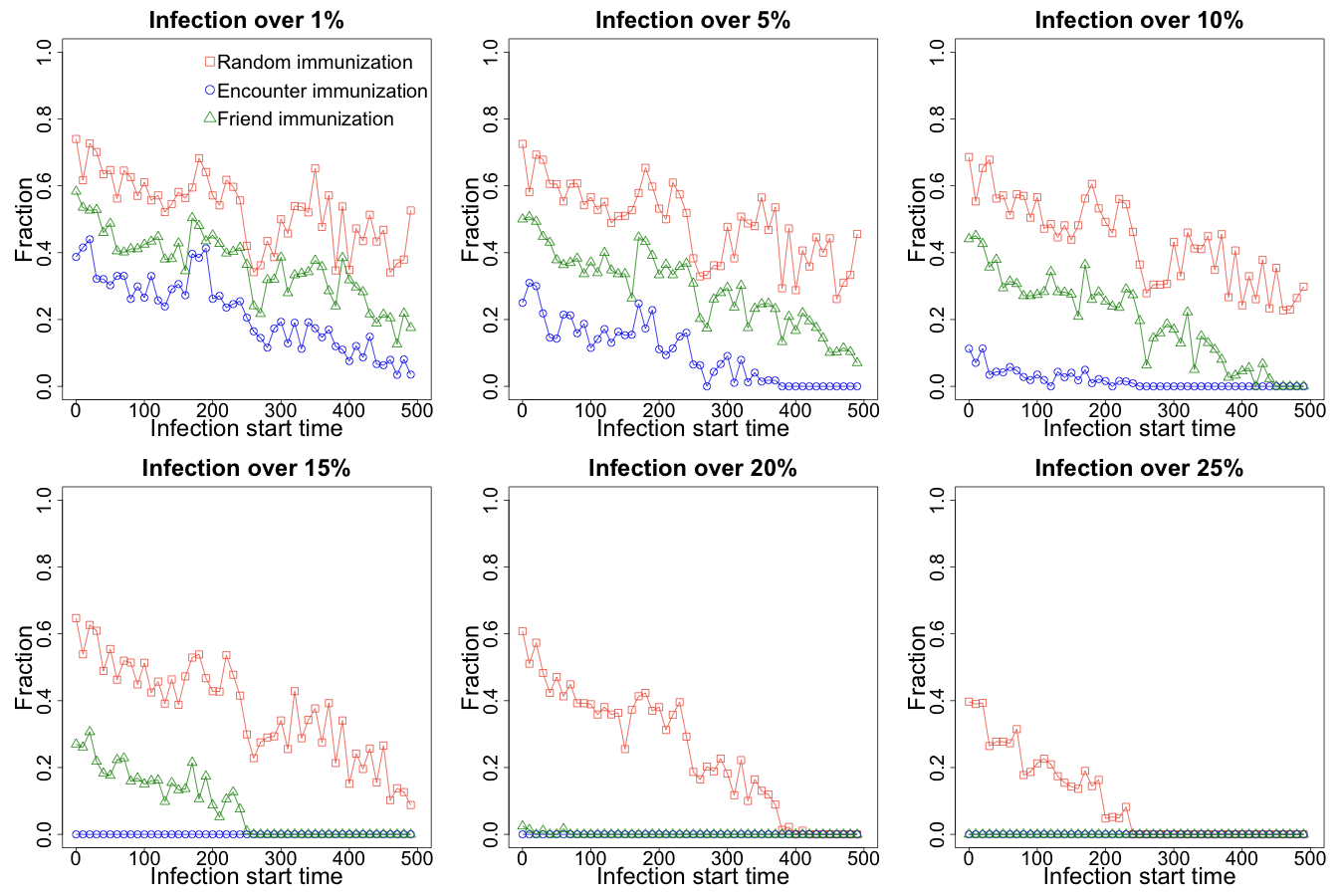}
\caption{Fraction of infections whose final rate is above given targets as a function of the infection start time $t_0(s_i)$, for all immunization methods (random immunization: red squares, encounter immunization: blue circles, friend immunization: green triangles) and immunization rate $b=5\%$.
Subplots consider targets of $r_R(b,s_i), r_F(b,s_i), r_E(b,s_i)\in\{1\%,5\%,10\%,15\%,20\%,25\%,\}$.
The $x$-axis shows $t_0(s_i)$.
The $y$-axis shows fraction of infection above the considered targets.
}
\label{fig:immunization_success_5percent}
\end{figure}

\begin{figure}
\centering
\includegraphics[scale=0.3]{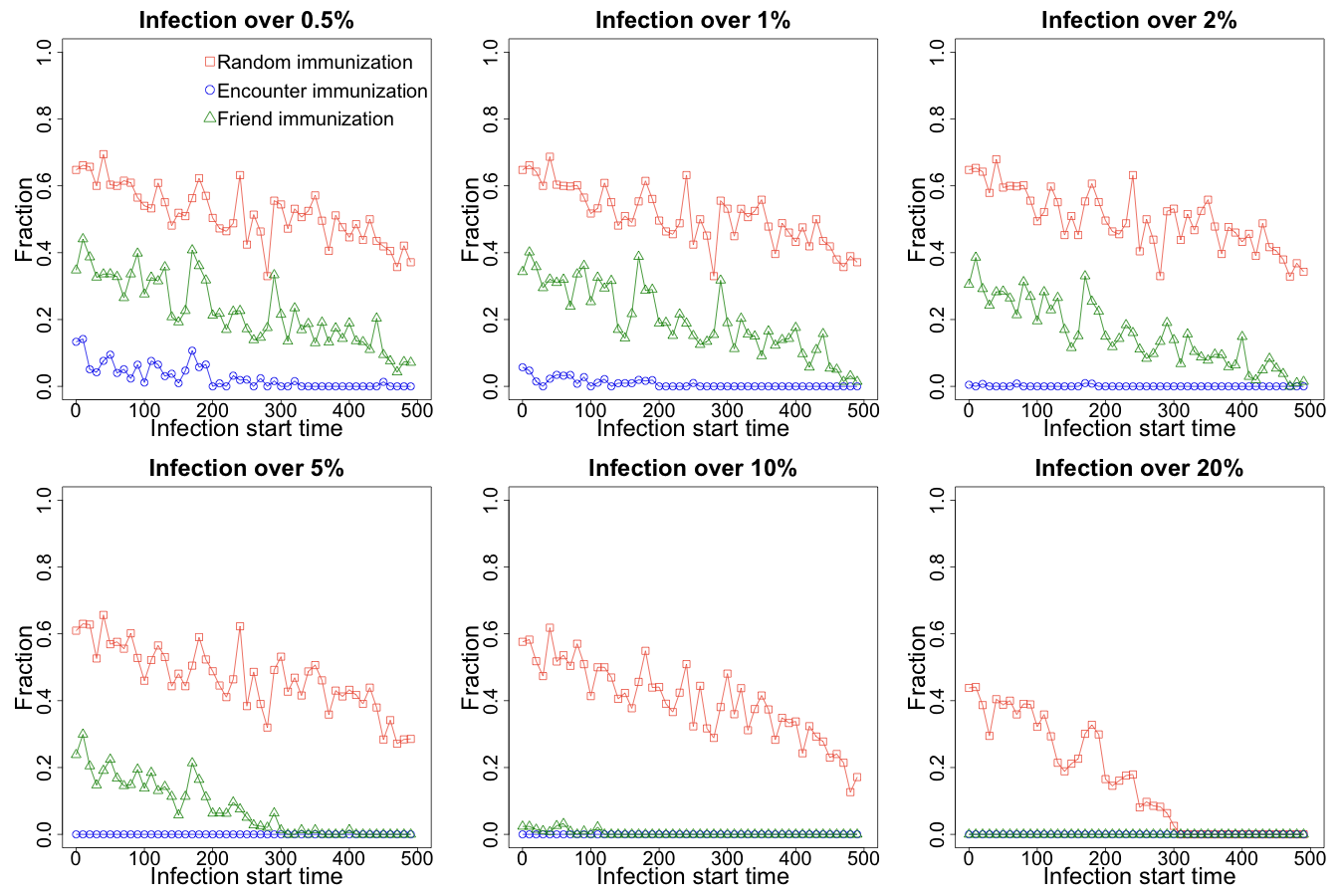}
\caption{Fraction of infections whose final rate is above given targets as a function of the infection start time $t_0(s_i)$, for all immunization methods (random immunization: red squares, encounter immunization: blue circles, friend immunization: green triangles) and immunization rate $b=10\%$.
Subplots consider targets of $r_R(b,s_i), r_F(b,s_i), r_E(b,s_i)\in\{1\%,5\%,10\%,15\%,20\%,25\%,\}$.
The $x$-axis shows $t_0(s_i)$.
The $y$-axis shows fraction of infection above the considered targets.
}
\label{fig:immunization_success_10percent}
\end{figure}

Figure~\ref{fig:immunization_success_5percent} and~\ref{fig:immunization_success_10percent} show (for immunization rate of $b=5\%$ and $b=10\%$, respectively) the fraction of infections with final rate ($r_R(b,s_i)$, $r_F(b,s_i)$, $r_E(b,s_i)$) above given targets as a function of the infection start time $t_0(s_i)$, for all immunization methods (random immunization: red squares, encounter immunization: blue circles, friend immunization: green triangles).
Each subplot considers a fixed target value of the final infection rate and, for each immunization method, plots the fraction of infections that hit that target.
As in the other figures, both friend and encounter immunization provide substantial improvement over random immunization, widely reducing the fraction of infections that hit the targets.
The improvement obtained with encounter immunization is larger than that obtained with friend immunization.

\FloatBarrier
\section{Epidemics at the macroscopic level: time-varying networks}
\label{yelp:sec:time_infection}
In this section and in Section~\ref{yelp:sec:static_infection}, we look at the epidemic processes on the different networks from a macroscopic point of view.
Rather than comparing the sets of individuals at risk according to the two spreading models (i.e., friendship and encounter).
we focus on quantities such as the size of the infected population and the infection detection time.
We also consider infection detection time through sensors, as defined in Section~\ref{yelp:sec:sensors}.

Our simulations confirm the idea that the dynamics on different networks present similarities. 
Both on static and time-varying networks, the fraction of infected nodes increases linearly over time after an initial period of incubation, during which the infected population is small.
In the case of time-varying networks (where the infection process runs for a finite number of time steps), we find an inverse relationship between the infection starting time and the final rate of infection, showing that earlier connectivity results in faster infection.
Final infection rates are higher on the friendship network, due to it larger density. However, infection rates evolve similarly on the two networks.
If we consider the probability that an infection hits a target $\alpha$-fraction of the population, some targets are never reached on the encounter network while they are on the friendship network (due to the different density), but the trends are similar on both networks.
In the case of static networks (where the infection runs until the entire population is infected), the time to infect a target $\alpha$-fraction of the nodes is smaller for seeds with larger degree, confirming that higher connectivity results in faster infection.
Even if the infection spreads faster on the friendship network, we observe similar trends on both networks.

In this section, we consider SI processes on the time-varying networks $\{N_F(t)\}_{t\in T}$ and $\{N_E(t)\}_{t\in T}$.
In Section~\ref{yelp:sec:static_infection}, we consider SI processes on the static networks $N_F=(U,F)$ and $N_E=(U,E)$.

\subsection{Infection Rate}
\label{yelp:sec:time_rate}
With $\beta=1$, we perform $10,000$ simulations on each time-varying network.
In each simulation, a single seed is selected uniformly at random between all nodes $s$ such that $t_0(\{s\})\le 500$ on the considered network.
That is, in the case of the friendship (respectively, encounter) network, we consider potential seeds that have an edge in $N_F(t)$ (respectively, $N_E(t)$) for some $t\le 500$.
As infections on time-varying networks spread for a limited number of time steps, we require them to start early enough.

Each simulation $i$ is therefore associated to a seed $s_i$ and, as $\beta=1$, the first time in which a node other than $s_i$ is infected is
$$
t_0(s_i) = \min\{t:\exists (s_i,v)\in\E(t) \text{ for some } v\neq s_i\} \in [1,500],
$$
We refer to $t_0(s_i)$ as the starting time of the infection.
Let $t_F(s_i)$ be the last time in which a node is infected in an infection starting from $s_i$ (i.e., the time after which the size of the infected population stops increasing).
It holds that $t_F(s_i)\le \max T$.
At time $t_F(s_i)$, the infection reaches its peak, infecting a fraction $r(s_i)\in[0,1]$ of the population.

The final infection $r(s_i)$ decreases with increasing infection starting time $t_0(s_i)$, for both the  time-varying friendship network
(OLS, coefficient $-4.255\cdot 10^{-4}$, p-value$<2\cdot 10^{-16}$, intercept $0.451$, p-value$<2\cdot 10^{-16}$)
and the encounter network
(OLS, coefficient $-3.922\cdot 10^{-4}$, p-value$<2\cdot 10^{-16}$, intercept $0.757$, p-value$<2\cdot 10^{-16}$).
Instead, $t_0(s_i)$ does not predict $t_F(s_i)$ for either the time-varying friendship network (OLS, coefficient $-0.03701$, p-value $0.376$) or the encounter network (OLS, coefficient $-0.03662$, p-value $0.381$).
This suggests that the networks remain connected over time and therefore infections that start earlier do not stop earlier.

\begin{figure}
\centering
\includegraphics[scale=0.50]{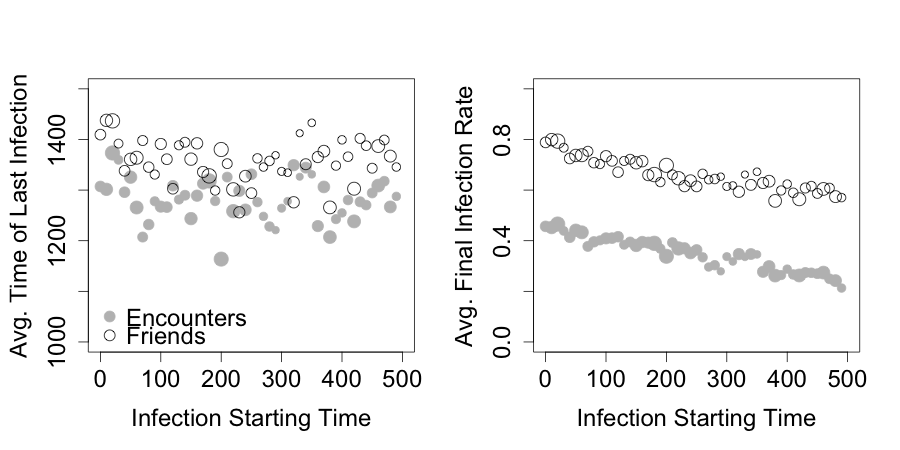}
\caption{SI process on the time-varying friendship network (white circles) and encounter network (grey circles), $\beta=1$ (certain infection). $10,000$ simulations are run on each network, each with a single seed $s_i$ selected at random among all nodes such that $t_0(s_i)\le 500$.
The $x$-axis represents the infection start time $t_0(s_i)$, rounded to the lower multiple of $10$. Point size is proportional to the number of observations for the corresponding value of the $x$-axis.
\textbf{Left:} Average of the last time of infection $t_F(s_i)$ (i.e., the time at which the peak of the infection is reached) with respect to $t_0(s_i)$, for both the friendship and encounter networks.
\textbf{Right:} Average of the final infection $r(s_i)$ with respect to $t_0(s_i)$, for both the friendship and encounters networks.
}
\label{fig:temp_singleSeed_lastFinal}
\end{figure}

Due to higher connectivity, the final rate of infection $r(s_i)$ is on average $31.5\%$ higher on the time-varying friendship network than on the encounter network (OLS, $0.3149$, p-value$<2\cdot 10^{-16}$, when controlling for $t_0(s_i)$), see Figure~\ref{fig:temp_singleSeed_lastFinal} (right panel).
Also, the time $t_F(s_i)$ of maximum infection is reached on average $79$ time steps later on the time-varying friendship network than on encounter network (OLS, $79.19$, p-value$<2\cdot 10^{-16}$, when controlling for $t_0(s_i)$), see Figure~\ref{fig:temp_singleSeed_lastFinal} (left panel).

\begin{figure}
\centering
\includegraphics[scale=0.50]{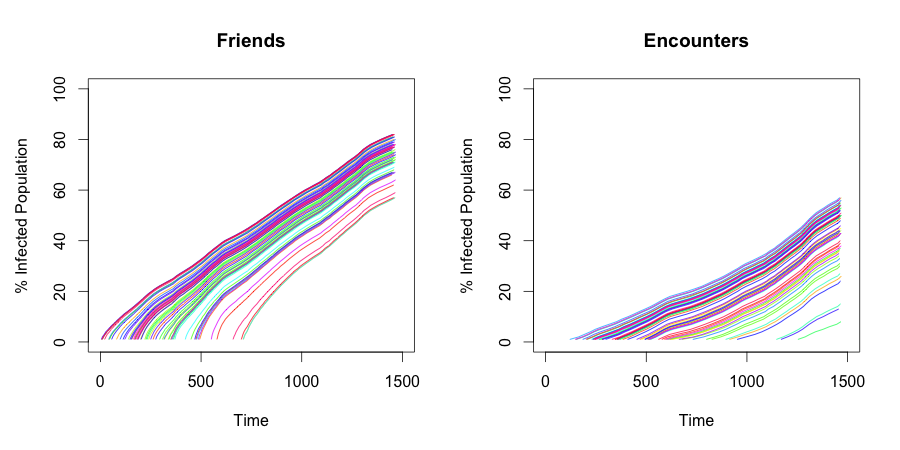}
\caption{Fraction of infected nodes over time, for the time-varying friendship network (left) and the encounter network (right).
Each SI process (with $\beta=1$) is started from a single seed $s_i$ selected at random among all nodes such that $t_0(s_i)\le 500$.
For each network, $60$ simulations that infected at least $1\%$ of the population are considered.
Colors are not meaningful.
}
\label{fig:linear_increase_infection}
\end{figure}

The fraction of infected nodes increases linearly over time in both networks (see Figure~\ref{fig:linear_increase_infection}).
In particular, we consider all infections that infected at least $1\%$ of the total population ($7,888$ out of $10,000$ simulations in the encounter network, and $9,100$ in the time-varying friendship network).
The infection spreads faster in the time-varying friendship network (OLS, slope $0.06209$, p-value$<2\cdot 10^{-16}$) than in the encounter network (OLS, slope $0.03187$, p-value$<2\cdot 10^{-16}$),
with a significantly different slope difference (OLS, interaction coefficient of $7.57\cdot 10^{-3}$, p-value$<2\cdot 10^{-16}$).
Moreover, even if an infection starts at time $t\le 500$, it still might take a while to infect a significant amount of the population (see Figure~\ref{fig:linear_increase_infection}). There is, therefore, a period of ``incubation'' during which the fraction of the infected population remains very low.

\subsection{Sensor monitoring}
\label{yelp:sec:time_sensor}

Instead of monitoring the entire population, in each run of the SI process, we consider a random set of sensors composed by $1\%$ of the population.
Sensors are selected in the two ways described above: random sensors and friend sensors (where the selection is based on friendship rather than encounter, even when considering a process spreading on the encounter network).
We perform $10,000$ simulations on each time-varying network and each sensor type, setting $\beta=1$ (i.e., infection is certain).
In each simulation, a single seed is selected uniformly at random between all nodes $s_i$ such that $t_0(s_i)\le 500$.

\begin{figure}
\centering
\includegraphics[scale=0.50]{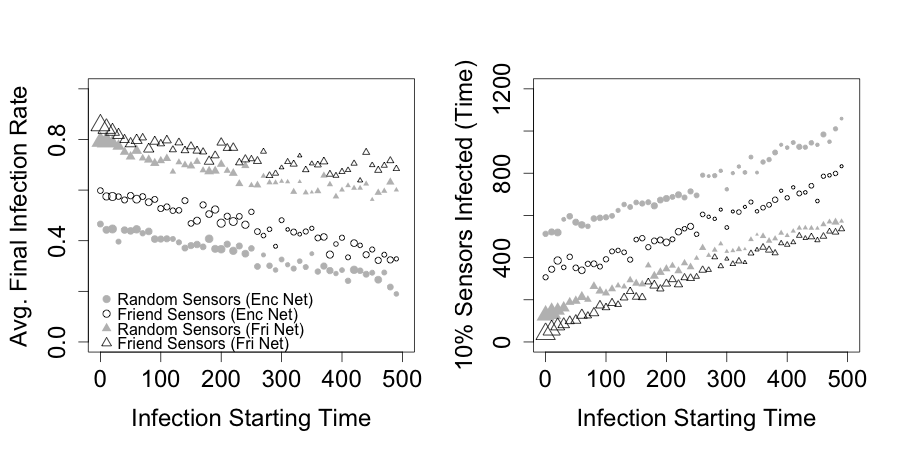}
\caption{
Infection detection with random sensors and friend sensors on the friendship and encounter time-varying networks.
$10,000$ simulations, with $\beta=1$, are run on each network and for each sensor type. Each simulation starts with a seed $s_i$ selected at random among all nodes such that $t_0(s_i)\le 500$. Sensor size is $1\%$ of the population.
The $x$-axis represents the infection start time $t_0(s_i)$, rounded to the lower multiple of $10$.
Point size proportional to the number of observations for the corresponding value of the $x$-axis.
\textbf{Left}: average final sensor infection versus infection start time, for the time-varying friendship network (triangles) and the encounter network (circles), with random sensors (grey) and friend sensors (white).
\textbf{Right}: average time to infect $10\%$ of the sensors versus infection start time, considering only the simulations in which at least $10\%$ of the sensors are infected.
}
\label{fig:sensor_type_final_10percent}
\end{figure}

Let $r_S(s_i)$ denote the final infection rate of the sensors (considered instead of $r(s_i)$, defined for the entire node set).
Also $r_S(s_i)$ linearly decreases with increasing infection start time (Figure~\ref{fig:sensor_type_final_10percent}, left).
On average, friend sensors predict an infection rate $9.5\%$ higher than random sensors (OLS, coefficient $0.0953$, p-value$<2\cdot 10^{-16}$,  controlling for infection starting time $t_0(s_i)$ and type of network).
As random sensor constitute a random sample of the population, their infection reflects the infection of the entire population.
Instead, friends sensors are more connected that average nodes (the \emph{friend paradox}) and therefore their larger infection constitutes an overestimation of the infection of the population.
Such overestimation can be beneficial for early detection of an outbreak.
The overestimation effect is larger on the encounter network (OLS, coefficient $0.1197$, p-value$<2\cdot 10^{-16}$,  controlling for infection starting time) than on the time-varying friendship network (OLS, coefficient $0.0709e$, p-value$<2\cdot 10^{-16}$, when controlling for infection starting time).
However, the sensor type does not significantly affect the slope of the observed linear decrease (OLS: interaction between infection starting time and sensor type, $2.299\cdot 10^{-5}$, p-value $0.21$).
We also observe that, on the time-varying friendship network, the $r_S(s_i)$ is on average $29\%$ higher than on the encounter network (OLS, coefficient $-0.2935$, p-value$<2\cdot 10^{-16}$,  controlling for infection starting time and type of network).
This effect is larger for random sensors (OLS, coefficient $0.2809$, p-value$<2\cdot 10^{-16}$,  controlling for infection starting time) than for friend sensors  (OLS, coefficient $0.2130$, p-value$<2\cdot 10^{-16}$,  controlling for infection starting time).

When restricting our attention to simulations which infected at least $10\%$ of the sensors (on the encounter network, $7,669$ with random sensors, $7,781$ with friend sensors, on the friendship network, $9,140$ with random sensors, $9,109$ with friend sensors), on average, the $10\%$ infection of friends sensors is reached $128$ time units earlier than the $10\%$ infection of random sensors (OLS, coefficient $-128.0$, p-value$<2\cdot 10^{-16}$,  controlling for infection starting time and type of network).
For the same consideration as above, friend sensors offer earlier detection with respect to the $10\%$ infection of the entire population.
This underestimation effect is larger on the encounter network (OLS, coefficient $-197.3$, p-value$<2\cdot 10^{-16}$,  controlling for infection starting time) than on the time-varying friendship network (OLS, coefficient $-69.2$, p-value$<2\cdot 10^{-16}$,  controlling for infection starting time).
Also in this case, the sensor type does not affect the slope of the observed linear increase (OLS: interaction between infection starting time and sensor type, $-2.302\cdot 10^{-3}$, p-value $0.892$).
We also observe that, on the time-varying friendship network, the infection of $10\%$ of the sensors requires on average $302$ units of time less than on the encounter network (OLS, coefficient $-302.6$, p-value$<2\cdot 10^{-16}$,  controlling for infection starting time and type of sensors). This effect is larger for random sensors (OLS, coefficient $-3.672$, p-value$<2\cdot 10^{-16}$,  controlling for infection starting time) than friend sensors  (OLS, coefficient $-2.384$, p-value$<2\cdot 10^{-16}$,  controlling for infection starting time).

\begin{figure}
\centering
\includegraphics[scale=0.40]{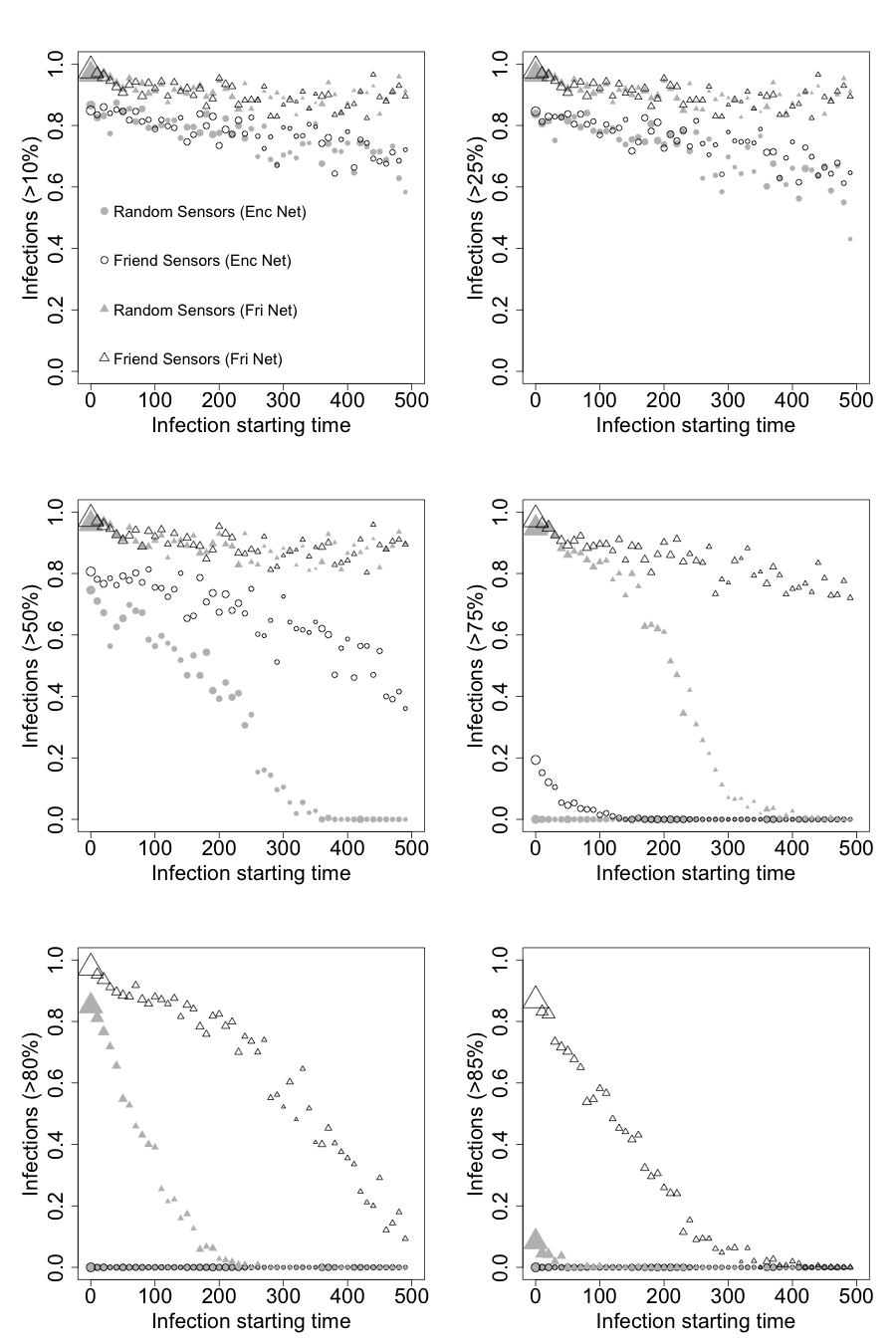}
\caption{Fraction of simulations that reached a target sensors' infection versus the infection starting time, for different targets, for the encounter (circles) and time-varying friendship networks (triangles), using random sensors (grey) and friend sensors (white).
$10,000$ simulations, with $\beta=1$, are run on each network and for each sensor type. Each simulation starts with a seed $s_i$ selected at random among all nodes such that $t_0(s_i)\le 500$. Sensor size is $1\%$ of the population.
The $x$-axis represents the infection start time $t_0(s_i)$, rounded to the lower multiple of $10$. Point size proportional to the number of observations for the corresponding value of the $x$-axis.
}
\label{fig:success_sensor_type}
\end{figure}

Figure~\ref{fig:success_sensor_type} plots the fraction of simulations that reached a target sensors' infection versus the infection starting time (values of the target: $10\%, 25\%, 50\%, 75\%, 80\%, 85\%$). We refer to the infections that reached a given target as successful (for the given target).
For targets of $10\%$ and $20\%$ (top plots) the observations are the same as above.
For a target of $50\%$ (middle left plot), on the encounter network (circles), the fraction of successful infection decreases more steeply for random sensors (grey) than friend sensors (white). with the former, the fraction of successful infections approaches zero for infection starting time above $t=350$. This effect is not observed in the case of the time-varying friendship network (triangles) for target of $50\%$.
For a target of $75\%$ (middle right plot), we observe a similar effect also on the time-varying friendship network, on which the success rate decreases faster with random sensors (approaching zero for infection starting times above $t=400$). On the encounter network, there is no successful infection of random sensors, whereas some successful infection of friends sensors happens for infection starting time before $t=100$.
For targets of $80\%$ and $85\%$ (bottom plots), the observations are similar.

\FloatBarrier
\section{Epidemics at the macroscopic level: static networks}
\label{yelp:sec:static_infection}

In this section, we consider SI processes on the static networks $N_F=(U,F)$ and $N_E=(U,E)$, started from a single seed, $\I(0)=\{s\}$.
As mentioned in Section~\ref{yelp:sec:dataset}, we consider the giant components of these networks,
consisting in $n_F=168,923$ nodes in the friendship network and $n_E=113,187$ nodes in the static encounter network (their union has cardinality $n_U= 210,899$).

\subsection{Infection Rate}
\label{yelp:sec:static_rate}
We perform $5,000$ simulations on each static network, setting $\beta=0.01$.
In each simulation, a single seed $s_i$ is selected uniformly at random between all nodes in the corresponding network.
Given that in a SI process nodes never recover from infection, the entire population eventually becomes infected for each $\beta>0$ and for each seed $s_i$.
Recall that, for $0\le\alpha\le 1$, $\tau(\alpha)$ represents the first time in which a $\alpha$-fraction of the population is infected (for ease of notation, we omit the dependency on $s_i$).
In this section, we study how the infection grows over time, that is, how $\tau(\alpha)$ grows with $\alpha$.

\begin{figure}
\centering
\includegraphics[scale=0.50]{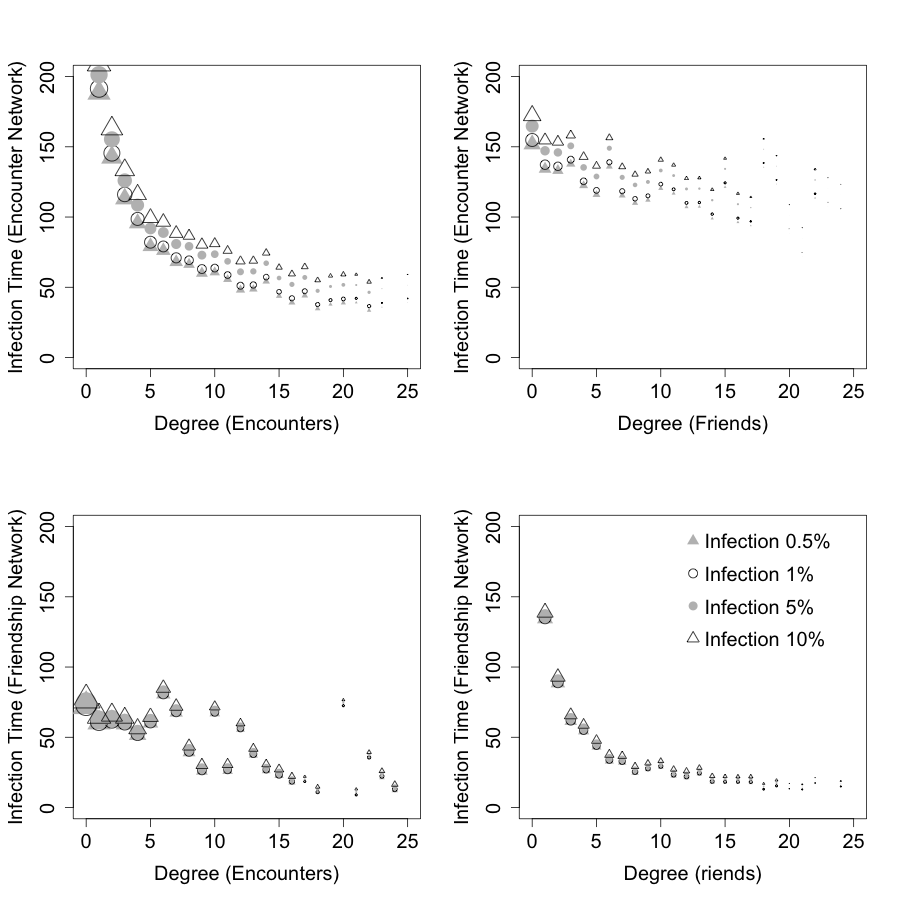}
\caption{\textbf{Infection speed versus degree - static networks.} The plots relate the encounter and friendship degree of a seed node with the infection speed for different target infections $\alpha$. $5000$ simulations with $\beta=0.01$ are run per network, selecting a seed uniformly at random for each simulation.
The left panels relate encounter degree of the seed with infection on the static encounter network (top) and friendship network (bottom), for degree of at most $25$.
The right panels relate friends degree of the seed with infection on the static encounter network (top) and friendship network (bottom), for degree of at most $25$.
Point size are proportional to the logarithm of the number of observations for each degree.
}
\label{fig:degree_vs_infection_static}
\end{figure}

Figure~\ref{fig:degree_vs_infection_static} relates the degree of the infection seed (i.e., encounter and friend degree) to the time $\tau(\alpha)$ to reach infection targets of $\alpha\in\{0.5\%, 1\%, 5\%, 10\%\}$.
Top and bottom panels consider the SI process on the static encounter network and the friendship network, respectively.
The $x$-axis show either the encounter degree (left panels) or the friend degree (right panels) of the seed (with degree at most $25$).

In general, for all targets $\alpha\in\{0.5\%, 1\%, 5\%, 10\%\}$, increasing encounter (reps. friendship) degree is related to an initial steep decrease in the infection time on the encounter (reps. friendship) network, that then smooths out when the degree surpasses a threshold.

In the static encounter network (compare Figure~\ref{fig:degree_vs_infection_static}, top-left panel), encounter degree larger than $10$ results in a four-fold decrease of the infection time with respect to degree one, for all values of $\alpha$
(two-sample t-tests, 
means $188$ and  $42$ for $\alpha=0.5\%$,
$191$ and  $45$ for $\alpha=1\%$,
$201$ and  $55$ for $\alpha=5\%$,
$209$ and  $62$ for $\alpha=10\%$,
p-value$<2.2\cdot 10^{-16}$ for all $\alpha$).
The decrease of the infection time with respect to seed degree is slow for degree larger than $15$
(OLS, restricted to seeds with encounter degree larger than $15$, degree coefficient $-0.300$ for all $\alpha$, p-value$<5.57e-11$).
The effect of the seed's friend degree on the infection speed on the encounter network is limited
(degree coefficient $-0.14$ for all $\alpha$, p-value$<2.48e-8$; compare Figure~\ref{fig:degree_vs_infection_static}, top-right panel).

In the friendship network (compare Figure~\ref{fig:degree_vs_infection_static}, bottom-right panel), friend degree larger than $5$ results in a six-fold decrease of the infection time with respect to degree one, for all values of $\alpha$
(two-sample t-tests,
means $134.97$ and  $20.63$ for $\alpha=0.5\%$,
$135.60$ and  $21.23$ for $\alpha=1\%$,
$137.48$ and  $23.11$ for $\alpha=5\%$,
$139.37$ and  $25.01$ for $\alpha=10\%$,
p-value$<2.2\cdot 10^{-16}$ for all $\alpha$).
The decrease of the infection time is slow for degree larger than $10$
(OLS, restricted to seed with encounter degree larger than $10$, degree coefficient $-0.0547$ for all $\alpha$, p-value$<9.63e-14$).
Larger encounter degree is not related to an equally steep decrease of the infection speed on the friendship network (compare Figure~\ref{fig:degree_vs_infection_static}, bottom-left panel),
despite its effect is somewhat
(degree coefficient $-1.270$ for all $\alpha$, p-value$<2.08e-7$),
likely due to the low average encounter degree (mean $2.594$). 

If we look at how the infection grows over time, we observe an initial ``incubation'' period, during which the infected population is very small, followed by an explosion of the infection.
Figure~\ref{fig:degree_vs_infection_static} plots the percentage of the infected population over time (up to $25\%$) for a sample of $60$ randomly selected seeds on the encounter (right panel) network and $60$ randomly selected seeds on the friendship network (left panel).
Overall, on the friendship network, an infection starting from a single seed takes on average $59.44$ time units to infect an initial $0.01\%$ of the population (about $17$ nodes), with more connected nodes requiring less time (OLS, degree coefficient $-0.273$, p-value$<2.2\cdot 10^{-16}$).
On the static encounter network, an infection starting from a single seed takes on average $107.78$ time units to infect an initial $0.01\%$ of the population (about $12$ nodes), with more connected nodes requiring less time (OLS, degree coefficient $-3.384$, p-value$<2.2\cdot 10^{-16}$).

The incubation period is in large part determined by the time required by the seed to infect its first neighbor (and thus depends on the infection rate $\beta$).
Indeed, the first infection happens, on average, after $39.68$ time units in the friendship network (decreasing with degree, OLS, $-0.1608$, p-value $4.46e-11$),
and after $53.07$ time units in the static encounter network (decreasing with degree, OLS, $-2.011$, p-value $<2\cdot 10^{-16}$).

\begin{figure}
\centering
\includegraphics[scale=0.50]{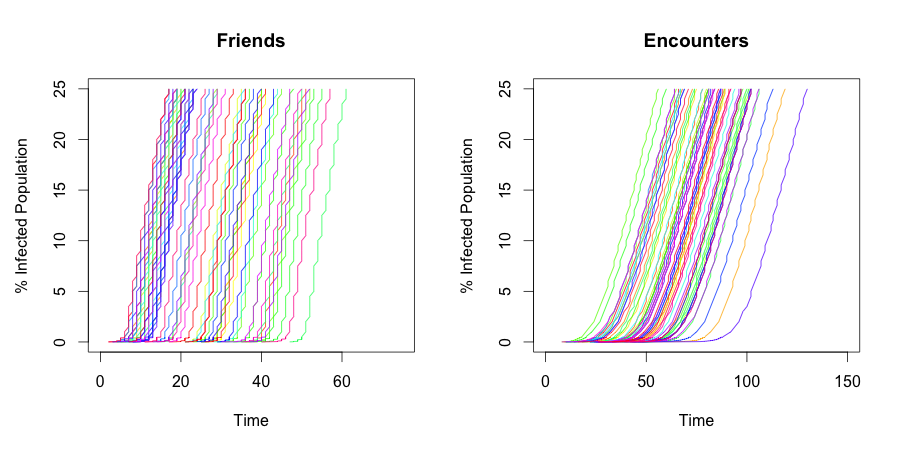}
\caption{\textbf{Growth of the infection over time - static networks.} $60$ simulations with $\beta=0.01$ are shown for the friendship network (left) and for the static encounter network (right).
For each simulation, a seed is selected uniform at random and the infection starts at time $t=0$.
Colors are not meaningful.
An initial ``incubation'' period, during which the infection spreads from the seed to its first neighbors, is followed by an explosion of the infection.
}
\label{fig:linear_infection_static}
\end{figure}

\subsection{Sensor Monitoring}
\label{yelp:sec:static_sensor}

Instead of monitoring the entire population, in each run of the SI process, we consider a random set of sensors composed by $1\%$ of the population.
Sensors are selected in the two ways described above: random sensors and friend sensors (where the selection is based on friendship rather than encounter, even when considering a process spreading on the encounter network).
We perform $5,000$ simulations on each static network and each sensor type, setting $\beta=0.01$ (stochastic infection).
In each simulation, a single seed is selected uniformly at random between all nodes in the network.

\begin{figure}
\centering
\includegraphics[scale=0.50]{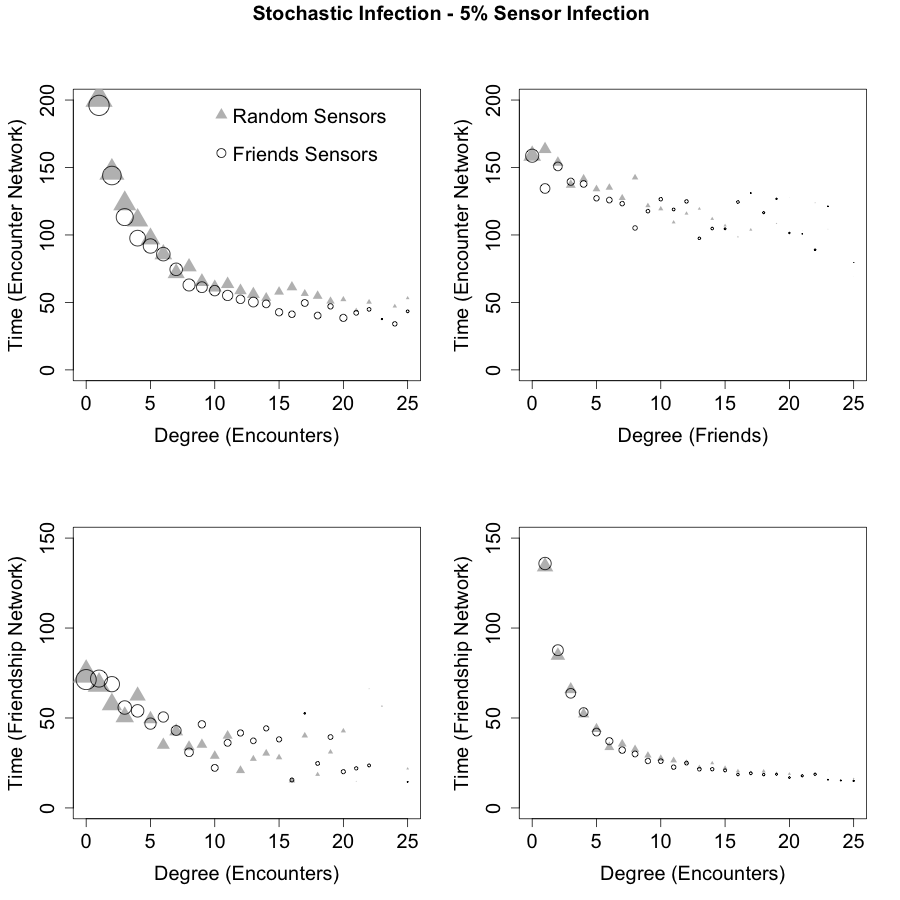}
\caption{\textbf{Sensor infection monitoring versus seed degree - static networks.}
The plots show the average time to infect $5\%$ of the sensors versus the degree of the infection seed.
$5000$ simulations with $\beta=0.01$ are run per network and per sensor type.
For each simulation, a seed is selected uniformly at random, and the sensor size is $1\%$ of the total population.
The left panels relate encounter degree of the seed with infection on the static encounter network (top) and friendship network (bottom), for degree of at most $25$.
The right panels relate friends degree of the seed with infection on the static encounter network (top) and friendship network (bottom), for degree of at most $25$.
Point size are proportional to the logarithm of the number of observations for each degree.
}
\label{fig:degree_vs_infection_static_sensor}
\end{figure}

Figure~\ref{fig:degree_vs_infection_static_sensor} plots the average time to detect a $5\%$ infection of the sensors versus the seed degree,
on the static encounter network (top panels) and friendship network (bottom panels).
The $x$-axis shows either the encounter degree (left panels) or the friend degree (right panels) of the seed (degree at most $25$).

On the static encounter network (compare Figure~\ref{fig:degree_vs_infection_static_sensor}, top panels), friends sensors guarantee earlier detection than random sensors.
The average detection time is smaller with friend sensors than with random sensors,
both for a $5\%$ infection ($135.36$ times units  versus $141.66$, t-test, p-value $0.00487$),
a $10\%$ infection ($139.96$ time units versus $149.13$, t-test, p-value $4.109\cdot 10^{-5}$),
and a $25\%$ infection ($151.17$ time units versus $168.51$, t-test, p-value $9.913e-15$).
The earlier detection provided by friend sensors over random sensors is not statistically significant for targets of $0.05\%$ and $1\%$ infection.

On the friendship network (compare Figure~\ref{fig:degree_vs_infection_static_sensor}, bottom panels), despite friend sensors provide a lower average detection time than random sensors, the difference is not statistically significant for any target infection rate.

The results above are driven by the stochastic incubation time needed to get the infection started, driven by the parameter $\beta$, as we observed in the previous section.
In order to control for such randomness, we perform $5,000$ additional simulations for each time-varying network and sensor type, setting $\beta=1$ (certain infection).
This choice allows to study the effect of the structural properties of the selected sensors on the infection detection time.
Friend sensors provide faster detection of the infection both on the friendship and the encounter network, and for all targets $\alpha$.

\begin{figure}
\centering
\includegraphics[scale=0.50]{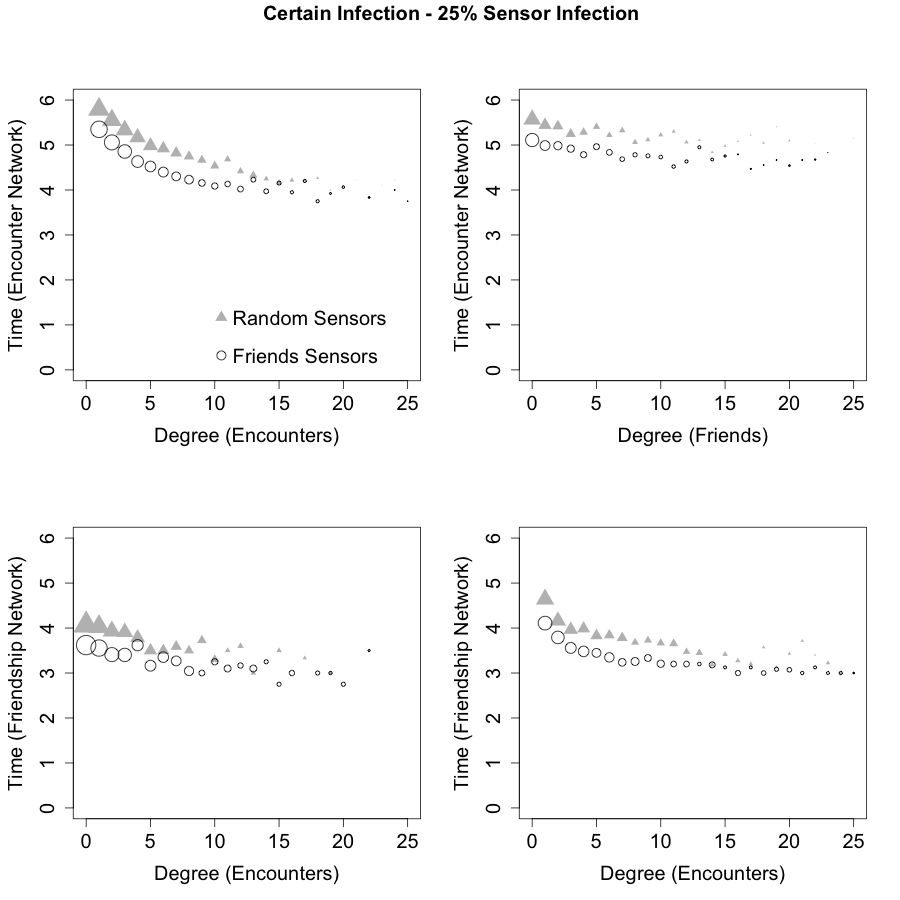}
\caption{\textbf{Sensor infection monitoring versus seed degree - static networks.}
The plots show the average time to infect $25\%$ of the sensors versus the degree of the infection seed.
$5000$ simulations with $\beta=1$ (certain infection) are run per network and per sensor type.
For each simulation, a seed is selected uniformly at random, and the sensor size is $1\%$ of the total population.
The left panels relate encounter degree of the seed with infection on the static encounter network (top) and friendship network (bottom), for degree of at most $25$.
The right panels relate friends degree of the seed with infection on the static encounter network (top) and friendship network (bottom), for degree of at most $25$.
Point size are proportional to the logarithm of the number of observations for each degree.
}
\label{fig:degree_vs_infection_static_sensor_beta1}
\end{figure}

Figure~\ref{fig:degree_vs_infection_static_sensor_beta1} plots the average time to detect a $25\%$ infection of the sensors versus the seed degree,
on the static encounter network (top panels) and friendship network (bottom panels).
The $x$-axis shows either the encounter degree (left panels) or the friend degree (right panels) of the seed (degree at most $25$).
On the encounter network (compare Figure~\ref{fig:degree_vs_infection_static_sensor_beta1}, top panels), the average detection time for a $0.5\%$ infection of friends sensors is $3.3516$ time units (versus $3.7426$ for random sensors),
for a $1\%$ infection is $3.5488$ (versus $3.9894$),
for a $5\%$ infection is $4.1192$ (versus $4.5660$),
for a $10\%$ infection is $4.4008$ (versus $4.8526$),
for a $25\%$ infection is $4.8922$ (versus $5.3340$), and all value are statistically significant (t-tests, p-values$<2\cdot 10^{-16}$).
On the friendship network (compare Figure~\ref{fig:degree_vs_infection_static_sensor_beta1}, bottom panels), the average detection time for a $0.5\%$ infection of friends sensors is $3.3516$ time units (versus $3.7426$ for random sensors),
for a $1\%$ infection is $2.4748$ (versus $2.8432$),
for a $5\%$ infection is $2.6004$ (versus $2.9804$),
for a $10\%$ infection is $2.9286$ (versus $3.6092$),
for a $25\%$ infection is $3.5142$ (versus $3.9056$), and all value are statistically significant (t-tests, p-values$<2\cdot 10^{-16}$).

\FloatBarrier

\section{Discussion}
\label{yelp:sec:conclusion}
This paper started from the observations that physical encounter is the most common vehicle for the spread of infectious diseases, but pervasive and detailed information about said encounters is often unavailable.
Therefore, given an infection driven by physical encounter, we explored the question of whether friendship ties successfully predict the individuals at risk.
Through computer simulation, we argued that this is not the case: friendship networks do not provide accurate prediction of epidemic risk.
In particular, building a \emph{friendship network} and an \emph{encounter network} between the same set of individuals, we showed that epidemic processes initiated at the same seed but spreading independently on the two networks infect very different sets of nodes,
even after controlling for the fact that individuals might be connected in one network and not in the other.
The difference is not determined by the randomness of the infection process, but by the differences in local connectivity between the two networks.
Also, the difference is not determined by the static nature of the friendship network, whose edges do not change over time,
as opposed to the time-varying nature of the encounter network, whose edges are activated when individuals encounter.
Our analyses reveal a striking contrast between the similarity at the macroscopic level of processes spreading on different networks (confirmed by our simulations) and the possibly misleading prediction of risk resulting from the chosen definition of edges.
However, despite the limits highlighted by our analyses, we show that periodical and relatively infrequent monitoring of the real infection on the encounter network allows to correct the predicted infection on the friendship network and to achieve satisfactory prediction accuracy.
In this sense, the ability to periodically monitor the infection on the encounter network is the key to overcome the limits of the friendship network in predicting epidemic risk.
In addition, the friendship network allows to effectively employ a given immunization budget (e.g., limited vaccine amount) for the containment of epidemic outbreaks.
A simple strategy that gives immunization to random friends of randomly chosen individuals substantially increases the probability that an infection dies out in its early stage and strongly reduces the expected final infection size with respect to purely random immunization.

When it is known who is infected or likely to become infected (e.g., individuals traveling to certain countries who might have come in contact with a pathogen),
accurate prediction of the individuals at risk of contagion would allow targeted monitoring and immunization.
Despite friendship and other social relationships might be informative about the encounters between individuals,
our work suggests that they do not always give a complete picture of the paths a pathogen might take.
Information about future encounters between individuals is likely to be unavailable, at least at a detailed level.
However, a feasible approach could use past encounter as a proxy of future encounter.
In fact, it is known that human mobility and encounter present high spatial and temporal regularity and predictability~\cite{brockmann2006scaling,gonzalez2008understanding,song2010limits,sun2013understanding}.
From a practical perspective, networks based on social relationships might be complemented by information about past encounter.
Our simulations are based on a dataset that allowed us to build a static friendship network and a time-varying encounter network that is a candidate vehicle for the spread of a pathogen.
The dataset considers a large number of individuals and spans several years of activity.
In general, other datasets might be available and allow similar analyses.
Friendship networks whose edges have a different semantic than that considered in the present work might lead to different observations.

%

\bibliographystyle{plain}
\bibliography{friendship_epidemic_risk}

\end{document}